\documentclass[12pt]{article}

\usepackage[title]{appendix}
\usepackage{lscape}
\usepackage{amsfonts}
\usepackage{amsmath}
\usepackage{amssymb}
\usepackage{epstopdf}
\usepackage[bottom,multiple]{footmisc}
\usepackage{float}
\usepackage{natbib}
\usepackage[margin=.75in]{geometry}
\usepackage[hang]{caption}
\usepackage[pdftex]{graphicx}
\usepackage{rotating}

\newcommand{\qed}{\hfill $\blacksquare$}
\numberwithin{equation}{section}
\renewcommand{\baselinestretch}{1.2}
\allowdisplaybreaks

\newtheorem{theorem}{Theorem} \numberwithin{theorem}{section}
\newtheorem{assumption}{Assumption}
\newtheorem{corollary}[theorem]{Corollary}
\newtheorem{lemma}[theorem]{Lemma}
\newtheorem{proposition}[theorem]{Proposition}

\begin{document}
\title{The realized empirical distribution function of stochastic variance with application to goodness-of-fit testing\thanks{
This paper was presented at the 10th International Conference on Computational and Financial Econometrics (CFE 2016) in Seville, Spain; the 2017 Humboldt-Aarhus-Xiamen Workshop in Aarhus, Denmark; the 10th annual SoFiE meeting in New York, USA; the European meeting of Statisticians in Helsinki, Finland; the 4th annual conference of the International Association for Applied Econometrics (IAAE) in Sapporo, Japan; and in lunch seminars at CREATES, Aarhus University, and Kellogg School of Management, Northwestern University. We thank the audience at these venues, and in addition Torben Andersen, Rolf Poulsen and Viktor Todorov, for their comments and insightful suggestions that helped to improve various aspects of previous drafts of this paper. The authors were partially funded by a grant from the Danish Council for Independent Research (DFF -- 4182-00050). This work was also supported by CREATES, which is funded by the Danish National Research Foundation (DNRF78). Corresponding author: thyrsgaard@econ.au.dk.}}
\author{Kim Christensen\thanks{CREATES, Department of Economics and Business Economics, Aarhus University, Fuglesangs All\'{e} 4, 8210 Aarhus V, Denmark.} \thanks{Research fellow at Danish Finance Institute.}
\and Martin Thyrsgaard\footnotemark[2]
\and Bezirgen Veliyev\footnotemark[2] \footnotemark[3]}

\date{April, 2019}
\maketitle
\vspace*{-1.0cm}
\begin{abstract}
We propose a nonparametric estimator of the empirical distribution function (EDF) of the latent spot variance of the log-price of a financial asset. We show that over a fixed time span our realized EDF (or REDF)---inferred from noisy high-frequency data---is consistent as the mesh of the observation grid goes to zero. In a double-asymptotic framework, with time also increasing to infinity, the REDF converges to the cumulative distribution function of volatility, if it exists. We exploit these results to construct some new goodness-of-fit tests for stochastic volatility models. In a Monte Carlo study, the REDF is found to be accurate over the entire support of volatility. This leads to goodness-of-fit tests that are both correctly sized and relatively powerful against common alternatives. In an empirical application, we recover the REDF from stock market high-frequency data. We inspect the goodness-of-fit of several two-parameter marginal distributions that are inherent in standard stochastic volatility models. The inverse Gaussian offers the best overall description of random equity variation, but the fit is less than perfect. This suggests an extra parameter (as available in, e.g., the generalized inverse Gaussian) is required to model stochastic variance.

 \vspace*{-0.1cm}

 \vspace*{-0.1cm}
\end{abstract}

\setlength{\baselineskip}{18pt}\setlength{\abovedisplayskip}{10pt} %
\belowdisplayskip \abovedisplayskip \setlength{\abovedisplayshortskip }{5pt} %
\abovedisplayshortskip \belowdisplayshortskip \setlength{%
\abovedisplayskip}{8pt} \belowdisplayskip \abovedisplayskip%
\setlength{\abovedisplayshortskip }{4pt}

\thispagestyle{empty}

\vfill

\pagebreak

\section{Introduction} \setcounter{page}{1} \renewcommand{\baselinestretch}{1.45} \normalsize

Stochastic volatility is a central concept in financial economics with numerous implications for capital allocation, asset- and derivatives pricing, or risk management. It is therefore essential to select a model for stochastic volatility that is able to reproduce the main features observed in data from financial markets, such as the implied volatility surface, as closely as possible. There is, however, no consensus about which model is ``best,'' as the extensive list of papers on stochastic volatility suggests \citep*[the literature is too big to enumerate, but a partial and necessarily incomplete set of proposed specifications can be found in][and the references therein]{andersen-benzoni-lund:02a, barndorff-nielsen-shephard:02a, chernov-gallant-ghysels-tauchen:03a, christoffersen-jacobs-mimouni:10a, comte-renault:98a, gatheral-jaisson-rosenbaum:18a, heston:93a, hull-white:87a}.

The goodness-of-fit of a stochastic volatility model can be assessed in several ways. If the parameters are estimated with a method of moment-based estimator, a standard diagnostic check is available via a test of overidentifying restrictions \citep*[see also, e.g.,][]{gallant-hsieh-tauchen:97a}. This approach can be highly inefficient, however, as it depends heavily on the selection of moment conditions and the associated estimation of the weighting matrix, which may produce significant size distortions in finite samples \citep*{andersen-sorensen:96a}.

Another idea is to compare the model-implied distribution of volatility with a nonparametric empirical measure of it \citep*[e.g.,][]{ait-sahalia:96a}. In a continuous-time setting, for instance, a model is often formulated via a stochastic differential equation for the variance process (or a transformation thereof, such as the square root or natural logarithm). This, in turn, implies that the marginal distribution of spot volatility (at least in the stationary case) belongs to a particular class of distributions. So if the underlying volatility was observed, a specification test based on the distance---with respect to a suitable norm---between the empirical distribution function (EDF) and the marginal distribution imposed by the model would be feasible. However, as spot volatility is not directly observed, this approach is not immediately applicable.

The recent access to financial high-frequency data has alleviated these concerns, as it enables the computation of an error-free measure of realized volatility, allowing for a more direct evaluation of the goodness-of-fit of stochastic volatility models. Inspired by the above, a common approach is to compare (conditional) moments of the integrated variance of a parametric model with a nonparametric estimator hereof \citep*[see, e.g.,][]{bollerslev-zhou:02a, corradi-distaso:06a, dette-podolskij:08a, dette-podolskij-vetter:06a, todorov:09a, todorov-tauchen-grynkiv:11a, vetter-dette:12a}. \citet*{zu:15a} proposes a test based on a de-convolution kernel density estimator of the distribution of the integrated variance, \citet*{lin-lee-guo:13a, lin-lee-guo:16a} resort to a characteristic function approach, while \citet*{bull:17a} proposes a wavelet-based test. As pointed out by \cite{todorov-tauchen:12a}, however, the mapping between the probability distribution of the spot and integrated variance is in general not one-to-one. Thus, goodness-of-fit tests based on the latter suffer from lack of power against some alternatives, due to the smoothing entailed by integrating spot volatility over a discrete time interval.

In this article, we therefore construct goodness-of-fit tests for stochastic volatility models that are based on a realized EDF (REDF hereafter) of spot volatility. We build on \citet*{li-todorov-tauchen:13a, li-todorov-tauchen:16a}, who show that inference about the EDF can be based on the volatility occupation measure \citep*[e.g.,][]{geman-horowitz:80a}. In their framework, the latent spot variance at any point in time is retrieved by employing a \citet*{foster-nelson:96a} rolling window-type estimator computed on small non-overlapping blocks of high-frequency data. In contrast, we employ overlapping blocks, which is theoretically more efficient, but it also induces further technical problems in the part of the proofs, where a uniform bound on the estimation errors of spot variance is derived.

In addition, we allow the asset price to be recorded at the tick-by-tick level, so that the instantaneous variance can be recovered as accurately as possible. This is important if the whole distribution, including the tails, is of interest and not only measures of central tendency. At that sampling frequency, however, the data are distorted by microstructure noise \citep*[see, e.g.][]{hansen-lunde:06b}, and this complicates the issue. The derived problem of estimating the integrated---or cumulative---variance in the presence of noise has received a lot of attention in the literature \citep*[see, e.g.][]{barndorff-nielsen-hansen-lunde-shephard:08a, jacod-li-mykland-podolskij-vetter:09a, zhang-mykland-ait-sahalia:05a}. We here adapt the pre-averaging approach of \citet*{jacod-li-mykland-podolskij-vetter:09a} and \citet*{podolskij-vetter:09a, podolskij-vetter:09b} to design a consistent spot volatility estimator from contaminated high-frequency data \citep*[][suggest an alternative procedure]{zu-boswijk:14a}. The REDF is then constructed from the recovered sample path of spot variance.

We show that under weak regularity conditions the REDF is a consistent noise- and jump-robust estimator of the EDF over any fixed time interval in the infill limit, thus extending \citet*{li-todorov-tauchen:13a, li-todorov-tauchen:16a} to a setting with overlapping blocks and microstructure noise. This is merely a stepping stone in our context, however, as the main goal is to do goodness-of-fit tests for stochastic volatility models. In the second part, we therefore prove a functional CLT for the REDF in a long-span asymptotic framework by letting time increase to infinity. This amounts to establish convergence of what \citet*{vaart-wellner:07a} call an empirical process indexed by an estimated function, but here time is continuous and the observations are not i.i.d.

The theoretical results are refined into two new goodness-of-fit tests for measuring the discrepancy between the marginal density implied by a candidate stochastic volatility model and the nonparametric gauge at the EDF represented by the REDF. The first is reminiscent to a Kolmogorov-Smirnov statistic for an observed process, while the second is based on a weighted $L^{2}$ norm. Both are trivial to compute once the REDF has been constructed, but there are a number of subtleties that render the asymptotic distribution hard to evaluate, making it difficult to set critical regions and determine p-values. This is to some extent related to the complexities encountered in a classical setup, when the parameters of the model under the null are estimated. We resolve the issue via a parametric bootstrap \citep*[based on, e.g.,][and described in Appendix \ref{appendix:critical-value}]{bull:17a}, which leads to fast evaluation of the $t$-statistic.

The paper is organized as follows. In Section \ref{section:theory}, we cover the setting, assumptions, and introduce the EDF of volatility. In Section \ref{section:redf}, the properties of the REDF are analyzed. The asymptotic theory for our goodness-of-fit tests is also developed here. Section \ref{section:simulation} is devoted to a Monte Carlo study of the REDF and an assessment of the proposed goodness-of-fit tests. An empirical application is conducted in Section 5, where we recover the REDF from stock market high-frequency data and test goodness-of-fit of several marginal distributions that are induced by mainstream stochastic volatility models. We conclude in Section 6. The proofs appear in Appendix \ref{appendix:proofs}.

\section{The setting} \label{section:theory}

Let $X = (X_{t})_{t\geq0}$ denote the efficient log-price process of a financial asset, which is defined on a filtered probability space $\big( \Omega,\mathcal{F}, ( \mathcal{F}_{t} )_{t \geq 0}, \mathbb{P} \big)$. As consistent with no-arbitrage \citep*[e.g.,][]{delbaen-schachermayer:94a}, we assume $X$ is an It\^{o} semimartingale:
\begin{equation} \label{equation:X}
X_{t} = X_{0} + \int_{0}^{t} b_{s}ds + \int_{0}^{t} \sigma_{s}dW_{s} + J_{t}, \quad \text{for} \quad t \in [0,T],
\end{equation}
where $X_{0}$ is $\mathcal{F}_{0}$-measurable, $(b_{t})_{t \geq 0}$ is a locally bounded, predictable drift, $( \sigma_{t})_{t \geq 0}$ is an adapted, c\`{a}dl\`{a}g volatility, $(W_{t})_{t \geq 0}$ a Brownian motion, and $(J_{t})_{t \geq 0}$ is a jump process:
\begin{equation}
J_{t} = \int_{0}^{t} \int_{ \mathbb{R}} \delta(s,z) 1_{ \{| \delta(s,z)| \leq 1 \}}(\mu - \nu)(d(s,z)) + \int_{0}^{t} \int_{ \mathbb{R}} \delta(s,z) 1_{ \{| \delta(s,z)| > 1 \}} \mu(d(s,z)),
\end{equation}
where $\mu$ is a Poisson random measure on $(\mathbb{R}_+,\mathbb{R})$, $\nu(d(s,z)) = ds \otimes \lambda(dz)$ is a compensator, $\lambda$ is a $\sigma$-finite measure on $\mathbb{R}$, and $\delta: \Omega \times \mathbb{R}_{+} \times \mathbb{R} \rightarrow \mathbb{R}$ is a predictable function. Furthermore, we assume that there exists a function $\Gamma$ such that $| \delta(\omega, s,z)|\leq \Gamma(z)$ for all $(\omega, s, z)$ and
\begin{equation} \label{equation:jump-activity}
\int_{ \mathbb{R}}(1 \wedge \Gamma(z))^{r} \lambda(dz) < \infty,
\end{equation}
for some $r \in [0,2]$.

The constant $r$ in \eqref{equation:jump-activity} controls the activity level of the price jumps and is an upper bound on the Blumenthal-Getoor index. When $r=0$, the jumps are of finite activity, whereas $r>0$ corresponds to the infinite-activity setting.

We collect assumptions about the drift, volatility, and jumps of $X$ below.

\begin{assumption} \label{assumption:volatility} We assume that for each $p \geq 1$ and $t \geq 0$, $\mathbb{E} \big [|b_{t}|^{p} \big] + \mathbb{E} \big[| \sigma_{t}|^{p}] + \int_{ \mathbb{R}}(1\vee \Gamma(z))^{p} \lambda( \text{d}z) \leq C$ for some constant $C$. In addition, $\sigma$ fulfills at least one of the following conditions: \\[0.25cm]
\textbf{\textnormal{(i)}} There exists $H \in (0, 1)$ such that for each $p \geq 1, s \geq 0, u > 0$, and a constant $C$ (dependent on $H$ and $p$):
\begin{equation}
\mathbb{E} \bigg[ \Big(\sup_{ s \leq t \leq s+u} | \sigma_{t} - \sigma_{s} | \Big)^{p}  \bigg] \leq C u^{p H}.
\end{equation}
\textbf{\textnormal{(ii)}} $\sigma$ has the representation:
\begin{equation}
\sigma_{t} = \sigma_{0} + \int_{0}^{t} \tilde{b}_{s}ds + \int_{0}^{t} \tilde{ \sigma}_{s}dW_{s} + \int_{0}^{t} \tilde{ \sigma}_{s}'dB_{s} + \int_{0}^{t} \int_{ \mathbb{R}} \tilde{ \delta}(s,z)( \mu - \nu)(d(s,z)),
\end{equation}
where $\sigma_{0}$ is $\mathcal{F}_{0}$-measurable,   $\mathbb{E} \big [|\tilde{b}_{t}|^{p} \big] + \mathbb{E} \big[| \tilde{\sigma}_{t}|^{p}]+
\mathbb{E} \big[|\tilde{ \sigma}'_{t}|^{p}] \leq C$ for some constant $C$ and for each $p\geq 1$ and $t \geq 0$, while $B$ is a Brownian motion independent of $W$, $\tilde{ \delta}$ is a predictable function, and $| \tilde{ \delta}( \omega,t,z)| \wedge1\leq\tilde{\Gamma}(z)$ for all $(\omega,t,z)$ and some $\tilde{\Gamma}:\mathbb{R} \rightarrow \mathbb{R}$ such that $\int_{ \mathbb{R}} \tilde{ \Gamma}(z)^{2} \lambda(dz) < \infty$.
\end{assumption}

This places some moment conditions on $b_{t}$, $\sigma_{t}$, and the large jumps. While it is general and encompasses many volatility models applied in practice, it may feel a bit restrictive. The assumption can be relaxed,  but this entails that a constant $\iota > 0$ (restricting the rate at which $T \rightarrow \infty$ and $\Delta_{n} \rightarrow 0$, cf. Section \ref{section:redf}) has to be bounded from below, while it can be made arbitrarily small within the above setup.

Assumption 1(i) implies $\sigma$ is continuous. This means it can be approximated by a locally constant process over short time intervals. The condition covers both ``rough'' and long-memory volatility processes \citep*[e.g.,][]{comte-renault:98a, gatheral-jaisson-rosenbaum:18a} with Hurst exponent different from one-half.

The conditions in Assumption 1(ii), which says $\sigma$ is an It\^{o} semimartingale, are common in the literature. It allows volatility to exhibit fairly unrestricted jump dynamics (at the expense of excluding fractional Brownian motion as in 1(i)) and also enables a leverage effect, as it does not restrict the correlation structure between the increments of $X$ and $\sigma$.

Next, we define the EDF of volatility, which in the continuous-time setting is given by the random function:
\begin{equation} \label{equation:EDF}
F_{T}(x) = \frac{1}{T} \int_{0}^{T} 1_{ \{ V_{t} \leq x \}} dt,
\end{equation}
where $V_{t} \equiv \sigma_{t}^{2}$.\footnote{In probability theory, $F_{T}(x) = \int_{0}^{T} 1_{ \{ Y_{t} \leq x \}} dt$ is called the occupation---or local---time of the stochastic process $Y$ \citep*[e.g.,][]{geman-horowitz:80a}. This convention was adopted by \citet*{li-todorov-tauchen:13a, li-todorov-tauchen:16a}, who applied it to high-frequency volatility estimation. We normalize $F_{T}$ by $T$ here, as we are heading toward a setting with stationary volatility and an asymptotic theory with $T \rightarrow \infty$.}
\begin{assumption} \label{assumption:pathwise}
$F_{T}$ is a.s. continuous.
\end{assumption}
This condition is fulfilled by most stochastic volatility models. It does not impose the existence of a density nor does it require $\sigma$ itself to be continuous.

The right-hand side of \eqref{equation:EDF} can be written $T^{-1} \int_{0}^{T} g(V_{t}) dt$, where $g(v) = 1_{\{ v \leq x\}}$.\footnote{Much work in the high-frequency literature studies integrals of the form $\int_{0}^{T} g( V_{t}) dt$, where $g$ is smooth \citep*[e.g.,][]{barndorff-nielsen-graversen-jacod-podolskij-shephard:06a, jacod-rosenbaum:13a}. Here, in contrast, $g$ is discontinuous, which makes the theory a lot more inaccessible.} Then, if $F_{T}$ is absolutely continuous with respect to the Lebesgue measure---which is stronger than Assumption \ref{assumption:pathwise}---and $g$ is a bounded or non-negative Borel function:
\begin{equation}
\frac{1}{T} \int_{0}^{T} g( V_{t}) dt = \frac{1}{T} \int_{ \mathbb{R}_{+}} g(x)dF_{T}(x) = \frac{1}{T} \int_{ \mathbb{R}_{+}} g(x)f_{T}(x)dx,
\end{equation}
where $f_{T}$ is the density of $F_T$ with respect to the Lebesgue measure. The EDF therefore encapsulates all the information about $(V_{t})_{t \geq 0}$ available in $[0,T]$ and can be viewed as a pathwise version of the distribution function of volatility.

\subsection{A discrete and noisy high-frequency record of $X$}

The main difficulty is that $F_{T}$ is latent, because $\sigma$ is not observable. The aim of this paper is therefore to construct a consistent estimator of $F_{T}$, while making the above minimal assumptions about $X$.

Of course, in an ideal world with no markets frictions and continuous trading---i.e. if the entire trajectory of $X$ is available---we can recover the volatility process perfectly and also the exact time and size of jumps in $X$ and $\sigma$ \citep*[in constrast to the drift, which cannot be consistently estimated in finite time, not even if it is constant, see, e.g.,][]{merton:80a, foster-nelson:96a}. This setup is not realistic, however. In practice, we operate with discrete high-frequency data, which we assume are available at times $i \Delta_{n}$, for $i = 0, 1, \ldots, n$, where $\Delta_{n}$ is the time gap between consecutive observations and $n = \lfloor T / \Delta_{n} \rfloor$ is the sample size (equidistant sampling is enforced here, but it can be weakened). In a near-ideal world, the database then constitutes a high-frequency record of $X$, i.e. $(X_{i \Delta_{n}})_{i=0}^{n}$, and the estimator proposed by \citet*{li-todorov-tauchen:13a, li-todorov-tauchen:16a} can be applied without further ado.

The microstructure of financial markets adds measurement error, however, which implies that we do not observe $X$ directly (e.g., due to bid-ask bounce or price discreteness). Instead, we record a contaminated log-price $Z$, which we assume is related to $X$ as follows:
\begin{equation} \label{equation:noisy-price}
Z_{i \Delta_{n}} = X_{i \Delta_{n}} + U_{i \Delta_{n}}, \quad \text{for} \quad i = 0, 1, \ldots, nT.
\end{equation}
\begin{assumption} \label{assumption:noise}
We have $U_{i \Delta_{n}} = \omega_{i \Delta_{n}} \varepsilon_{i \Delta_{n}}$, where $(\omega_{t})_{t \geq 0}$ is a c\`{a}dl\`{a}g stochastic process that is adapted to $\mathcal{F}_{t}$, and $(\varepsilon_{i \Delta_{n}})_{i \geq 0}$ is i.i.d. with mean zero, variance one, and $\mathbb{E}(| \varepsilon_{i \Delta_{n}}|^{p}) < \infty$ for every $p > 0$ and $i \geq 0$. In addition, $(\varepsilon_{i \Delta_{n}})_{i \geq 0}$ is independent of $\mathcal{F}$.
\end{assumption}
This setting is comparable to Assumption (K) in \citet*{jacod-li-mykland-podolskij-vetter:09a} or Assumption (4) in the Appendix of \citet*{li-todorov-tauchen:17a} \citep*[see, e.g.,][for a related approach]{bibinger-winkelmann:18a}. The noise is permitted to be both heteroscedastic and serially dependent via $(\omega_{t})_{t \geq 0}$, so the model is broad enough to replicate realistic structures in the microstructure of financial markets \citep*[e.g.,][]{diebold-strasser:13a, hansen-lunde:06b}. The moment condition is not standard, however, but it is merely made for technical convenience. In the proofs, we only require moments of $\varepsilon_{i\Delta_{n}}$ up to some order (for instance, in Lemma \ref{lemma:spot-variance} the 4th moment should exist).

Later, we impose a stronger smoothness assumption on $( \omega_{t})_{t \geq 0}$:
\begin{assumption} \label{assumption:noise-strong}
Assumption \ref{assumption:noise} is satisfied. In addition, for each $p \geq 1$, $s \geq 0$, and $u > 0$:
\begin{equation}
\mathbb{E} \bigg[ \Big(\sup_{ s \leq t \leq s+u} | \omega_{t} - \omega_{s} | \Big)^{p}  \bigg] \leq C u^{p}.
\end{equation}
and $\mathbb{E} \big[ | \omega_{t}|^{p} \big] \leq C$, for some constant $C$.
\end{assumption}


We define the noisy log-return:
\begin{equation} \label{equation:noisy-return}
\Delta_{i}^{n} Z = Z_{i \Delta_{n}} - Z_{(i-1) \Delta_{n}}, \quad \text{for} \quad i = 1, \ldots, nT.
\end{equation}
To deal with the noise, we pre-average the return series \citep*[see, e.g.,][]{jacod-li-mykland-podolskij-vetter:09a, podolskij-vetter:09a, podolskij-vetter:09b}:
\begin{equation} \label{equation:pre-averaged-return}
\bar{Z}_{i} = \sum_{j=1}^{k_{n}-1} g \left(\frac{j}{k_{n}}\right) \Delta_{i+j}^{n}Z, \quad \text{for} \quad i = 0, \ldots, Tn-k_{n}+1,
\end{equation}
where $k_{n}$ is a positive integer, while $g$ is a kernel.

Any $g:[0,1] \mapsto \mathbb{R}$ with $g$ continuous and piecewise $C^{1}$ with Lipschitz derivative $g'$, such that $g(0) = g(1) = 0$ and $\int_{0}^{1} g(x)^{2}dx > 0$ is permitted. We follow the default choice in the literature by setting $g(x) = \min(x,1-x)$.

The selection of $k_{n}$ entails a trade-off. The intuition is that while pre-averaging lessens the noise, it also smooths out the underlying volatility of $X$. This can render it hard to construct estimates of spot volatility that fit the tails of the distribution if a too wide pre-averaging window is applied. As shown by \citet*{jacod-li-mykland-podolskij-vetter:09a}, an optimal $k_{n}$ is achieved via:
\begin{equation} \label{equation:kn}
k_{n} = \frac{\theta}{ \sqrt{ \Delta_{n}}} + o( \Delta_{n}^{-1/4}),
\end{equation}
where $\theta > 0$ is a tuning parameter, which controls the balance struck between the above forces in small samples. As consistent with prior work, we base our analysis on $\theta = 1/3$ and $k_{n} = [\theta/ \sqrt{ \Delta_{n}}]$.

\subsection{The realized EDF}

We now exploit the pre-averaged high-frequency data to form local estimates of $V_{t}$. The estimator we propose is both jump- and noise-robust and can therefore be plugged into a ``realized'' version of the EDF (or REDF, as defined in \eqref{equation:redf}). It is computed on small blocks of pre-averaged returns. We denote the number of increments in a block by a sequence of positive integers $h_{n}$ with $h_{n} \Delta_{n} \rightarrow 0$ and $h_{n}/k_{n} \rightarrow \infty$, as $\Delta_{n} \rightarrow 0$, so the time span of the block is decreasing, but there is an increasing amount of data within it.

Then, we set:
\begin{equation} \label{equation:variance-estimator}
\tilde{V}_{i \Delta_{n}} = \frac{1}{h_{n} \sqrt{ \Delta_{n}}} \sum_{m=0}^{h_{n}-1} \big( \bar{Z}_{i+m} \big)^{2} 1_{ \left\{ |\bar{Z}_{i+m}| \leq v_n  \right\}}, \quad \text{for} \quad i = 0, 1, \ldots, Tn - h_{n} - k_{n} + 1,
\end{equation}
where $v_{n} = \alpha \Delta_{n}^{ \bar{ \omega}}$, for some $\bar{ \omega} \in (0, 1/4)$ and $\alpha>0$.

As the notation suggests, $\tilde{V}_{i \Delta_{n}}$ is an estimator of the spot variance at time $i \Delta_{n}$, i.e. $V_{i \Delta_{n}}$. It is convenient to extend this definition to $t \in [0, T]$. We do this by setting $\tilde{V}_{t} = \tilde{V}_{i \Delta_{n}}$, for $t \in [i \Delta_{n}, (i+1) \Delta_{n})$, while for $t > (Tn - h_{n} - k_{n} + 1) \Delta_{n}$: $\tilde{V}_{t} = \tilde{V}_{(Tn - h_{n} - k_{n} + 1) \Delta_{n}}$, thus holding the final spot variance estimate fixed up to time $T$.

Finally, we introduce the following constants, which depend on the weight function:
\begin{equation}
\psi_{1} = \int_{0}^{1} g'(s)^{2}ds \quad \text{and} \quad  \psi_{2} = \int_{0}^{1}g(s)^{2}ds.\footnote{In the proofs, after freezing volatility locally, $\psi_{1}^{n} = k_{n} \sum_{j=1}^{k_{n}} \big(g ( \frac{j}{k_{n}}) - g( \frac{j-1}{k_{n}}) \big)^{2}$ and $\psi_{2}^{n} = \frac{1}{k_{n}} \sum_{j=1}^{k_{n}-1} g^{2} ( \frac{j}{k_{n}} )$ appear in the conditional expectation of $\hat{V}_{i\Delta_{n}}$. As $\Delta_{n} \to 0$, $\psi_{i}^{n} \rightarrow \psi_{i}$ and the error has order $\psi_{i}^{n} - \psi_{i} = O \big( \Delta_{n}^{1/2} \big)$, for $i = 1,2$. This means we can work with $\psi_{i}$ in the expressions below, and this substitution has no impact on the asymptotic analysis. In contrast, $\psi_{i}^{n}$ can differ materially from $\psi_{i}$ if $k_{n}$ is small. As a practical recommendation, it is therefore better to work with $\psi_{i}^{n}$.}
\end{equation}
With this notation in hand, we have the following result for our preliminary estimator of the spot variance.
\begin{lemma} \label{lemma:spot-variance}
Suppose that Assumption \ref{assumption:noise} holds and $r \in[0,2]$. For each $t \in [0,T]$, as $\Delta_{n} \rightarrow 0$,
\begin{equation} \label{equation:spot-variance-estimator}
\tilde{V}_{t} \xrightarrow{ \mathbb{P}} \theta \psi_{2} V_{t} + \frac{1}{ \theta} \psi_{1} \omega_t^{2}.
\end{equation}
\end{lemma}
As an aside, we should point out it is also possible to show that $\tilde{V}_{i \Delta_{n}}$ retains its consistency even in absence of truncation. However, in order to derive rates of convergence, which play an important role in the subsequent theory, truncation is required when $X$ is discontinuous. We therefore do not pursue this idea further.\footnote{The corresponding analysis of rates of convergence with the non-truncated spot variance estimator and continuous $X$ appeared in an earlier working paper version of this article (available at request).}

To strip out the residual noise variation, we need an estimator of $\omega_{t}^{2}$. There are several options around \citep*[e.g.,][]{bandi-russell:06a, hansen-lunde:06b, oomen:06a} and each one has its own merits and disadvantages \citep*[see, e.g.,][for a comparison]{gatheral-oomen:10a}. In this paper, we adopt the following:
\begin{equation} \label{equation:noise-variance-estimator}
\hat{\omega}_{i \Delta_n}^{2} = - \frac{1}{h_{n} - 1} \sum_{m = 1}^{h_{n}-1} \Delta_{i+m}^{n}Z \Delta_{i+m+1}^{n}Z,
\end{equation}
which is a consistent estimator of $\omega_{i \Delta_n}^{2}$ (as demonstrated in the proof of Theorem \ref{theorem:uniform-convergence-fixedT}). If we define $\hat{ \omega}_{t}^{2} = \hat{ \omega}_{i \Delta_{n}}^2$ for $t \in [i \Delta_{n}, (i+1) \Delta_{n})$, it then follows that
\begin{equation} \label{equation:feasible-spot-variance-estimator}
\hat{V}_{t} \equiv \frac{1}{ \theta \psi_{2}} \tilde{V}_{t} - \frac{ \psi_{1}}{ \theta^{2} \psi_{2}} \hat{ \omega}_t^{2} \xrightarrow{ \mathbb{P}} V_{t}.
\end{equation}
Subtracting the bias implies $\hat{V}_{t}$ can be negative in finite samples. Nevertheless, not a single point estimate fell below zero neither in our simulations nor empirical work.

We then construct the REDF:
\begin{equation} \label{equation:redf}
F_{n,T}(x) = \frac{1}{T} \int_{0}^{T} 1_{ \{ \hat{V}_{t} \leq x \}}dt,
\end{equation}

\section{Asymptotic properties of the REDF} \label{section:redf}

In this section, we cover the asymptotic theory of the REDF. In Section \ref{section:infill}, we deal first with consistency for the EDF in the infill setting, where we assume that the noisy log-price is recorded over ever shorter intervals but the total time elapsed is constant, as formalized by $\Delta_{n} \rightarrow 0$ with $T$ fixed. In the following Section \ref{section:joint}, we then further assume high-frequency data are collected on an expanding time window by also letting $T \rightarrow \infty$ and deduce convergence toward the marginal distribution function. We exploit these theoretical insights to develop some new goodness-of-fit tests for stochastic volatility models in Section \ref{section:goodness-of-fit}.

\subsection{Infill setting} \label{section:infill}

We start this section with showing that $F_{n,T}$ is a consistent estimator of the EDF.
\begin{theorem} \label{theorem:uniform-convergence-fixedT}
Suppose that Assumptions \ref{assumption:pathwise} -- \ref{assumption:noise} hold and $r \in [0,2]$. Then, for each $T > 0$, as $\Delta_{n} \rightarrow 0$:
\begin{equation}
\sup_{x \in \mathbb{R}_{+}} \big|F_{n,T}(x)-F_{T}(x) \big| \xrightarrow{ \mathbb{P}} 0.
\end{equation}
\end{theorem}
The consistency of $F_{n,T}$ is derived directly from that of the spot volatility estimator and the continuity of the EDF. It thus inherits the jump robustness of the estimator in \eqref{equation:variance-estimator}, regardless of the activity level of jumps in $X$.

%

We define $Q_{n,T}( \alpha) = \inf \big\{x:F_{n,T}(x) \geq \alpha \big\}$ and $Q_{T}( \alpha) = \inf \big\{x:F_{T}(x) \geq \alpha \big\}$ as the uniquely determined $\alpha$-quantile of $F_{n,T}$ and $F_{T}$, for $\alpha \in(0,1)$. Then, using the uniform convergence in probability of $F_{n,T}$, we deduce $Q_{n,T}( \alpha)$ is a consistent estimator of $Q_{T}( \alpha)$. As it is an implication of Lemma 21.2 in \citet*{vaart:98a}, we omit a proof.
\begin{corollary}
Suppose that Assumptions \ref{assumption:pathwise} -- \ref{assumption:noise} hold and $r\in[0,2]$. Then, for each $T > 0$, if $Q_{T}( \alpha)$ is continuous at $\alpha$ and as $\Delta_{n} \rightarrow 0$:
\begin{equation}
Q_{n,T}( \alpha) \xrightarrow{ \mathbb{P}} Q_{T}( \alpha).
\end{equation}
\end{corollary}

\subsection{Joint infill and long-span setting} \label{section:joint}

The goal here is to extend the above analysis (with $T$ fixed) to estimation of the stationary distribution function, $F$. To do so, we study an asymptotic framework, where the time span $T \rightarrow \infty$ jointly with $\Delta_{n} \rightarrow 0$. The key result is a feasible CLT for the REDF, which enables us to compute the accuracy with which the marginal distribution can be recovered from the data. To facilitate the derivation of this theory we require some additional assumptions on the volatility process.

\begin{assumption} \label{assumption:mixing-condition}
$(V_{t})_{t \geq 0}$ is stationary and strongly mixing with mixing coefficient function $\alpha(t) = O(t^{- \gamma})$, as $t \rightarrow \infty$ and for some $\gamma > 1$.
\end{assumption}
The stationarity condition in Assumption \ref{assumption:mixing-condition} is required to make the target of inference well-defined. We then restrict the memory of the volatility, so that $(V_{t})_{t \geq 0}$ is not ``too'' strongly dependent. This implies that $F_{T}$ (as an estimator of the marginal distribution, $F$) is consistent, as $T \rightarrow \infty$. We exploit this to deduce that $F_{n,T}$ also converges in probability to $F$, once we show that the discretization error embedded in the recovery of the volatility path is asymptotically negligible (as $\Delta_{n} \to 0$ fast enough, because the errors accumulate with $T$). Moreover, it is an essential part in showing the weak convergence of the empirical processes that we construct in the derivation of our goodness-of-fit test statistics in Section \ref{section:goodness-of-fit}.

The mixing condition is slightly weaker than the comparable assumptions made in related work with joint infill and long-span asymptotics \citep*[see, e.g.,][]{todorov-tauchen:12a, andersen-thyrsgaard-todorov:19a}. It encompasses several stochastic volatility models, for instance a large class of processes driven by Brownian motion \citep*[e.g.,][]{heston:93a} or the L\'{e}vy-driven Ornstein-Uhlenbeck model of \citet*{barndorff-nielsen-shephard:01a}, where volatility is governed by general (positive) processes.\footnote{Assume that $V$ is a solution of the stochastic differential equation: $dV_{t} = \tilde{b}(t, V_{t})dt + \tilde{ \sigma}(t, V_{t})dW_{t}$,
where $\tilde{ \sigma}(t, V_{t}) \leq K \big(1 + | V_{t}|^{1/2} \big)$, for some $K > 0$. If there exists an $S$ such that for all $|s|\geq S$ and $t\geq 0$, $b(t,s) \leq -\gamma$, for some $\gamma>0$, Theorem 2 in \citet*{veretennikov:88a} implies that $\alpha(t) = O(t^{-\gamma})$.}\footnote{If $V$ is of the form $d V_{t} = -\kappa V_{t} dt + dZ_{ \kappa t}$, where $\kappa > 0$ and $Z$ is a positive L\'{e}vy process (e.g., a subordinator) with L\'{e}vy measure $\nu$, then it follows from \citet*{jongbloed-meulen-vaart:05a} that $V$ is stationary if $\int_{2}^{\infty} \ln(x) \nu(dx) < \infty$. If further $\mathbb{E} \big[| V_{1} |^{p} \big] < \infty$, for some $p>0$, there exists $a > 0$ such that $\alpha(t) = O (e^{-at})$. The decay of the mixing coefficient can even be deduced if the driving process in volatility is a fractional Brownian motion \citep*[see, e.g.,][]{magdziarz-weron:11a}.} As a result, the setup is not restrictive in practice as it can capture a wide variety of marginal distributions.

Next, we replace the pathwise smoothness imposed on the EDF of volatility from Assumption \ref{assumption:pathwise} with the following condition.

\begin{assumption} \label{assumption:marginal-distribution}
$F$ is differentiable with bounded derivative $f$.
\end{assumption}

\begin{theorem} \label{theorem:weak-convergence}
Suppose that Assumptions \ref{assumption:noise-strong} -- \ref{assumption:marginal-distribution} hold true and $r\in[0,2)$. If either of the following conditions is fulfilled: \\[0.25cm]
\textbf{\textnormal{(i)}} Assumption \ref{assumption:volatility}\textnormal{(i)} with $H \in(0,1)$, $h_{n} \asymp \Delta_{n}^{- \frac{4H+1}{4H+2}}$ and $T^{1/2+ \iota} \Big( \Delta_{n}^{ \frac{H}{4H+2}- \iota} \vee \Delta_{n}^{(2-r) \bar{ \omega}- \iota} \Big) \rightarrow 0$ as $\Delta_{n} \rightarrow 0$ and $T \rightarrow \infty$ for some $\iota > 0$, $\bar{ \omega} \in \Big( \frac{H+1}{8 H+4}, \frac{1}{4} \Big)$, or \\[0.25cm]
\textbf{\textnormal{(ii)}} Assumption \ref{assumption:volatility}\textnormal{(ii)}, $h_{n} \asymp \Delta_n^{-3/4}$ and $T^{1/2 + \iota} \left( \Delta_{n}^{1/8- \iota} \vee \Delta_{n}^{(2-r) \bar{ \omega}- \iota} \right) \rightarrow 0$ as $\Delta_{n} \rightarrow 0$ and $T \rightarrow \infty$ for some $\iota > 0$,
$\bar{ \omega} \in \Big( \frac{3}{16}, \frac{1}{4} \Big)$. \\[0.25cm]
Then, for fixed $x \in \mathbb{R}_{+}$, it holds that
\begin{equation} \label{equation:large-T-clt}
\sqrt{T} \big( F_{n,T}(x) - F(x) \big) \xrightarrow{d} N(0, \Sigma(x)),
\end{equation}
where $\Sigma(x) = 2 \int_{0}^{ \infty} \big( F_{t}(x,x) - F(x)^{2} \big) dt$ with $F_{t}(x,y) = \mathbb{P} \left( V_0 \leq x, V_{t} \leq y \right)$.
\end{theorem}

In Theorem \ref{theorem:weak-convergence}, the block length $h_{n}$ is selected so that the discrepancy between the empirical process associated with the REDF and EDF vanishes at the fastest rate. To control this error, we need to restrict the relative speed at which $T \rightarrow \infty$ and $\Delta_{n} \rightarrow0$.

The first restriction is due to the untangling of the spot variance and Brownian component. The least binding is to take $h_{n}$ as indicated. Note that $h_{n}$ is an increasing function of $H$. In particular, if $\sigma$ is a continuous semimartingale (i.e., $H = 1/2$), the contribution of this error to $\sqrt{T} \big(F_{n,T}(x) - F_T(x) \big)$ has order $T^{1/2+ \iota} \Delta_{n}^{1/8- \iota}$. With a smoother volatility ($H \rightarrow 1$), the rate of convergence improves to $T^{1/2 + \iota} \Delta_{n}^{1/6- \iota}$, whereas it gets arbitrarily slow as $H \rightarrow 0$. The intuition is that it is virtually impossible to retrieve spot variance from a ``rough'' path. On the other hand, if $H = 1/2$ the optimal choice of $h_{n}$ and the associated speed condition is unchanged, irrespective of whether volatility jumps or not.

The second is due to the presence of jumps in $X$. It gets more restrictive, the higher $r$ is. Indeed, if $r > \frac{2}{H+1}$ this is the leading term, but for that to happen, the price jumps must not only be of infinite activity, but also infinite variation (i.e., $r>1$).

The noise-robustness of $F_{n,T}$ plays a critical role in determining the above condition. To compare, if there was no error in our measurement of $X$, a standard non-noise-robust realized measure suffices for estimation of spot volatility \citep*[see, e.g.,][]{jacod-protter:12a}. Here, the rate condition improves to $T^{1/2 + \iota} \Delta_{n}^{1/4-\iota} \rightarrow 0$ with $H = 1/2$, as is consistent with the fixed $T$ asymptotic theory in \citet*{li-todorov-tauchen:13a}. The deterioration of the speed condition is due to the added complexity from retrieving volatility robustly, which cuts the rate of convergence in half. This problem is well-known in the literature on estimation of integrated variance \citep*[see, e.g.,][]{zhang-mykland-ait-sahalia:05a, barndorff-nielsen-hansen-lunde-shephard:08a, jacod-li-mykland-podolskij-vetter:09a}. Moreover, while a convergence rate of (almost) $\Delta_{n}^{1/8}$ may appear slow, it is optimal for extraction of spot variance in noisy diffusion models \citep*[e.g.,][p. 296]{ait-sahalia-jacod:14a}.

In addition, note that the error of the spot noise variance estimator in \eqref{equation:noise-variance-estimator} is always negligible. For example, in Theorem \ref{theorem:weak-convergence}\textbf{\textnormal{(i)}} $| \hat{ \omega}_{t}^{2} - \omega_{t}^{2}| = O_{p} \big( h_{n}^{-1/2} \vee h_{n} \Delta_{n} \big)$, where $h_{n} \asymp \Delta_{n}^{- \frac{4H+1}{4H+2}}$, which is smaller than the spot variance estimation error $| \hat{V}_{t} - V_{t}|$ of at least order $\Delta_{n}^{ \frac{H}{4H+2}}$ (the latter is bigger than $\Delta_{n}^{1/6}$, as seen by letting $H \rightarrow 1$).

Notice that the asymptotic variance in Theorem \ref{theorem:weak-convergence}, $\Sigma(x)$, is non-stochastic, so that the usual machinery of stable convergence from high-frequency analysis is not required to conclude that $\sqrt{T} \big( F_{n,T}(x) - F(x) \big) / \sqrt{ \Sigma(x)} \xrightarrow{d} N(0,1)$.

$\Sigma(x)$ depends crucially on the memory of the volatility process. If a series is highly dependent, which is true empirically for volatility, it becomes harder to recover its marginal distribution, which results in larger values of $\Sigma(x)$ and wider confidence intervals for $F(x)$. The only way to compensate for this effect is to collect a larger sample.

$\Sigma(x)$ can be computed either analytically or, for example, by numerical integration. As shown by \citet*{dehay:05a}, however, for finite $T$:
\begin{equation}
\label{equation:actual_variance}
\text{var} \left( \sqrt{T} \big(F_{T}(x) - F(x) \big) \right) = 2 \int_{0}^{T} \left( 1 - \frac{t}{T} \right) \left(F_{t}(x,x) - F(x)^{2} \right) dt.
\end{equation}
The reduction by $(1-t/T)$ in the true variance of $F_{T}$ is an edge effect. We compute \eqref{equation:actual_variance} in the simulation section, as it provides an infeasible assessment for the accuracy of the approximation in \eqref{equation:large-T-clt} for finite $T$. On the other hand, if $\Delta_{n}$ is not small enough to ignore the pathwise discretization error, there is an extra source of randomness in $F_{n,T}$ induced by the estimation of spot volatility. As such, \eqref{equation:actual_variance} may actually understate the variation of $F_{n,T}$.

To estimate $\Sigma(x)$, we fix $\xi \in (0, 1/3)$ and set:
\begin{equation}
\Sigma_{n,T}(x) = 2 \int_0^{T^{ \xi}} \left( 1 - \frac{t}{T^{\xi}} \right) \left(F_{t,n,T}(x) - F_{n,T}(x)^{2} \right)dt,
\end{equation}
where
\begin{equation}
F_{t,n,T}(x) = \frac{1}{T - T^{\xi}} \int_{0}^{T - T^{\xi}} 1_{\{\hat{V}_{s + t} \leq x, \hat{V}_{s} \leq x\}} ds.
\end{equation}
Then, as shown in the proof of Theorem \ref{theorem:weak-convergence}, $\Sigma_{n,T}(x) \xrightarrow{ \mathbb{P}} \Sigma(x)$.

The estimation of $\Sigma(x)$ is related to the problem of estimating the long-run variance in a stationary time series. In that context, the lag length has to grow slower than $\sqrt{T}$ to ensure consistency, and the optimal bandwidth is usually $O(T^{1/3})$, see, e.g., \citet*{newey-west:94a}. The condition imposed on $\xi$ is identical to that.

In practice, we may also be interested in determining how accurate the associated quantile(s) of $F$ are estimated (e.g., for risk management reporting). These questions are, of course, intertwined. Indeed, from the assumed existence of the density $f$ and Corollary 21.5 in \cite{vaart:98a}, we deduce the following result.
\begin{corollary}
Suppose that the conditions of Theorem \ref{theorem:weak-convergence} are fulfilled. Then, for any $\alpha \in (0,1)$ with $f \big(Q( \alpha) \big) > 0$, it holds that
\begin{equation}
\sqrt{T} \big( Q_{n,T}( \alpha) - Q( \alpha) \big) \xrightarrow{d} N \Bigg(0, \frac{\Sigma \big( Q(\alpha) \big)}{f \big(Q( \alpha) \big)^{2}} \Bigg).
\end{equation}
\end{corollary}
Thus, inference about any given quantile is feasible, so long as an estimator of the density function of the marginal distribution evaluated at that quantile is available.

\subsection{Goodness-of-fit of the volatility distribution} \label{section:goodness-of-fit}

The preceding theory shows that $F_{n,T}$ converges to the stationary distribution function of volatility, as $\Delta_{n} \to 0$ and $T \rightarrow \infty$. In this section, we build upon this analysis to construct realized extensions of some classical goodness-of-fit tests for the volatility. To this end, we first derive a functional central limit theorem for the process $G_{n,T} = \Big\{ \sqrt{T} \big (F_{n,T}(x) - F(x) \big), x \in \mathbb{R}_+ \Big\}$.
\begin{theorem} \label{theorem:functional-clt}
Suppose that the conditions of Theorem \ref{theorem:weak-convergence} are fulfilled. Then, $G_{n,T}$ converges weakly in the space $\mathbb{D}(\mathbb{R}_+)$ of c\`{a}dl\`{a}g functions equipped with the uniform topology to a Gaussian process, $G_{F}$, with mean zero and covariance function $\Sigma(x,y) = \text{\upshape{cov}}\big(G_F(x),G_F(y) \big)$.
\end{theorem}
Note that Theorem \ref{theorem:functional-clt} can also be extended to the empirical process associated with the quantile estimator: $G_{n,T}^{Q} = \Big\{ \sqrt{T} \big(Q_{n,T}( \alpha) - Q( \alpha) \big), \alpha \in (0,1) \Big\}$.

As shown in Lemma A.3 in Appendix \ref{appendix:proofs}, $\Sigma(x,y)$ can be consistently estimated by
\begin{equation}
\Sigma_{n,T}(x,y) = \int_{0}^{T^{\xi}} \big(F_{t,n,T}(x,y)+F_{t,n,T}(y,x)-2F_{n,T}(x)F_{n,T}(y) \big)dt,
\end{equation}
for $\xi \in (0,1/3)$, where
\begin{equation}
F_{t,n,T}(x,y) = \frac{1}{T-T^{ \xi}} \int_{0}^{T-T^{ \xi}} 1_{ \big\{ \hat{V}_{s+t} \leq x, \hat{V}_{s} \leq y \big\}} ds.
\end{equation}
Appealing to Theorem \ref{theorem:functional-clt}, we can now construct the test statistics for our goodness-of-fit tests, which evaluate the fit of an assumed stochastic volatility model:
\begin{equation}\label{equation:test-statistic}
rKS(F) = \sup_{x \in \mathbb{R}_{+}} \big| G_{n,T}(x) \big| \qquad \text{and} \qquad
rL^{2}(F) = \int_{\mathbb{R}_{+}} G_{n,T}(x)^{2}w(x)dx,
\end{equation}
where $w: \mathbb{R}_{+} \rightarrow \mathbb{R}_{+}$ is a continuous weight function.

$rKS$ resembles a classical Kolmogorov-Smirnov test for an observed process, whereas $rL^{2}$ is a realized version of the goodness-of-fit test based on a weighted $L^{2}$ norm. The latter nests several existing tests, such as the Cramer-von-Mises (i.e., $w(x) = f(x)$) or Anderson-Darling (i.e., $w(x) = f(x)/[F(x)(1-F(x))]$). $rKS$ has the advantage that it is rather trivial to implement, once the REDF is constructed. On the other hand, the literature suggests that an $L^{2}$ statistic, at least in the i.i.d. setting, may be more powerful in finite samples, but it can also be tedious to compute, if a closed-form expression for $w$ is unavailable. This is, for example, the case for a Cramer-von-Mises test if the marginal density is unknown or hard to evaluate.

\begin{corollary} \label{corollary:weak-convergence}
Suppose that the conditions of Theorem \ref{theorem:weak-convergence} are fulfilled. Then, it holds that
\begin{equation}
rKS(F) \xrightarrow{d} \sup_{x \in \mathbb{R}_+} |G_{F}(x)| \qquad \text{\upshape{and}} \qquad rL^{2}(F) \xrightarrow{d} \int_{ \mathbb{R}_{+}} |G_{F}(x)|^{2}w(x)dx.
\end{equation}
\end{corollary}

Now, suppose we are interested in testing the fit of a stochastic volatility model, indexed by a known parameter vector $\upsilon$. Later in this section, we return to the question of how parameter estimation affects the proposed tests. We let $F_{ \upsilon}(x) = \mathbb{P}( V_{t} \leq x; \upsilon)$ describe the marginal distribution of the model.\footnote{As $\upsilon \mapsto F_{ \upsilon}$  is not one-to-one, in general, the model itself is not fully identified by $F_{ \upsilon}$, unless further constraints are imposed on the drift and volatility \citep*[e.g.,][]{ait-sahalia:96a,bibby-skovgaard-sorensen:05a}.} The null---and associated alternative---hypothesis is then written as:
\begin{equation}
\mathcal{H}_{0}: F = F_{ \upsilon} \qquad \text{and} \qquad \mathcal{H}_{a}: F \neq F_{ \upsilon}.
\end{equation}
It follows that---under $\mathcal{H}_{0}$---$rKS(F_{ \upsilon}) \xrightarrow{d} \sup_{x \in \mathbb{R}_+} |G_{F_{ \upsilon}}(x)|$, whereas $rKS(F_{ \upsilon}) \overset{ \mathbb{P}}{ \rightarrow} \infty$ under $\mathcal{H}_{a}$, because the alternative means there is at least one $x$, such that $F_{n,T}(x) \overset{ \mathbb{P}}{ \nrightarrow} F_{ \upsilon}(x)$. An equivalent result holds for $rL^{2}(F)$. As large values of the $t$-statistics discredit that $F = F_{ \upsilon}$, the tests are therefore one-sided.

A drawback of this approach is that the limiting distribution of $rKS(F)$ and $rL^{2}(F)$ depends on the model under $\mathcal{H}_{0}$. This differs from the classical theory, for example the Kolmogorov-Smirnov statistic in the i.i.d. case with $F$ continuous and known \citep*[see, e.g.,][]{vaart:98a}. Even there, however, the asymptotic distribution is not readily available if the parameters describing $F_{ \upsilon}$ are estimated, as we do below, although in that setting it often suffices to prepare a single table of family-wise critical values in advance---e.g., as in \citet*{lilliefors:67a,lilliefors:69a}---if $F_{ \upsilon}$ belongs to a location-scale family. This is due to the fact that the limiting distribution, in that case, does not depend on nuisance parameters of $F$, only the functional form of the probability distribution \citep*[e.g.,][]{david-johnson:48a}. This is not true here, and we are therefore forced to retrieve critical values through simulation on a case-by-case basis. In Appendix B, we offer a detailed recipe of how this procedure works both for Corollary \ref{corollary:weak-convergence} and \ref{corollary:function-clt}.

In principle, we can construct a $t$-statistic that can be evaluated independently of $F$. To explain how, define $||f||_w = \sqrt{ \int_{ \mathbb{R}_{+}}w(x)f(x)^{2}dx}$, such that $\mathbb{E} \big[ ||G_{F}||_{w}^{2} \big] < \infty$.

The other ingredient is a normalizing constant:
\begin{equation}
\tilde{ \Sigma} = \int_{ \mathbb{R}_{+}} \Sigma(x)w(x)dx,
\end{equation}
with $\tilde{ \Sigma} < \infty$. Then, we set:
\begin{equation}
T(F) = \frac{||G_{n,T}||_w^{2}}{ \tilde{ \Sigma}}.
\end{equation}

We can then define a test statistic that rejects $\mathcal{H}_{0}$, if we observe that $T(F) > \chi_{1 - \alpha}^{2}$, where $\chi_{1 - \alpha}^{2}$ is the $(1 - \alpha)$-quantile of a chi-squared distribution with one degree of freedom. Let $\alpha_{T(F)}$ be the significance level achieved by such a test. The following proposition summarizes the details.
\begin{proposition} \label{proposition:distribution-free}
Suppose that the conditions of Theorem \ref{theorem:weak-convergence} are fulfilled. Then, for any $\alpha \in (0,0.215)$, it holds that under $\mathcal{H}_{0}$:
\begin{equation}
\lim_{ \Delta_{n} \rightarrow 0, T \rightarrow \infty} \alpha_{T(F)} \leq \alpha,
\end{equation}
while under $\mathcal{H}_{a}$:
\begin{equation}
\mathbb{P} \big( T(F) > c \big) \rightarrow 1,
\end{equation}
for any $c > 0$.
\end{proposition}
We can thus construct a goodness-of-fit test for which critical values are readily available, however it achieves a significance level of at most $\alpha$ and may therefore be overly conservative for some $F$. Moreover, while the constant $\tilde{ \Sigma}$ can be calculated under $\mathcal{H}_{0}$, so that no estimation is required, it is complicated to derive for many common stochastic volatility models. We therefore do not explore this test further.

In practice, the parameter vector $\upsilon$ is typically unknown and not fixed a priori. It therefore has to be estimated, which invalidates the previous analysis (assuming $F_{ \upsilon}$ known). The intuition is that the estimation moves the model-implied marginal distribution closer to the EDF, making inference based on the preceding theory too conservative, as is well-known from the traditional literature on goodness-of-fit testing.

The complication brought about by parameter estimation can be viewed as a different testing problem that applies if we are interested in testing an entire class of stochastic volatility processes, as opposed to a particular member of that class.

The null is thus a composite hypothesis:
\begin{equation}
\mathcal{H}_0 : \exists \upsilon \in \Upsilon : F = F_{ \upsilon} \qquad \text{and} \qquad \mathcal{H}_{a} : \forall \upsilon \in \Upsilon : F \neq F_{ \upsilon},
\end{equation}
where $\Upsilon$ is the admissible parameter space (i.e., $\Upsilon = \{ ( \kappa, v_{0}, \xi) \in \mathbb{R}^{3}_{+} : 2 \kappa v_{0} \geq \xi^{2} \}$ for the \citet*{heston:93a} model due to the Feller condition).

We assume a consistent estimator of $\upsilon$, say $\hat{ \upsilon}$, has been found and introduce a pseudo empirical process $\tilde{G}_{n,T} = \Big\{ \sqrt{T} \big( F_{n,T}(x) - F_{ \hat{ \upsilon}}(x) \big), x \in \mathbb{R}_{+} \Big\}$. At a theoretical level, the problem then arises because the parameters of the underlying distribution typically cannot be recovered at a convergence rate faster than $T^{-1/2}$, i.e $\hat{ \upsilon} - \upsilon = O_{p} (T^{-1/2})$. This cancels with the $\sqrt{T}$ scaling in the construction of $\tilde{G}_{n,T}$, so that the estimation error inevitably affects the asymptotic distribution.\footnote{A standard rate of convergence is required in the following. Although we are not aware about the presence of so-called ``super'' consistent estimators in the stochastic volatility literature---at least not in the stationary setting---we point out that if $\hat{ \upsilon} - \upsilon = o_{p}(T^{-1/2})$ the sampling error is asymptotically negligible and the problem reverts back to Corollary \ref{corollary:weak-convergence}. On the other hand, a slower rate of convergence implies the error is blown up by the $\sqrt{T}$ scaling and then our approach just goes out the window.}

\begin{corollary} \label{corollary:function-clt}
Suppose that the conditions of Theorem \ref{theorem:weak-convergence} are fulfilled. In addition, we assume $\hat{ \upsilon}$ is asymptotically linear with influence function $\psi$ and that $\displaystyle \frac{ \partial F_{ \upsilon}}{\partial \upsilon}(x)$ is bounded and continuous in both $\upsilon$ and $x$ over the set $\mathbb{R}_{+} \times \bar{ \Upsilon}$ for a neighbourhood $\bar{ \Upsilon}$ around $\upsilon$. Then, $\tilde{G}_{n,T}$ converges weakly in $\mathbb{D}( \mathbb{R}_{+})$ equipped with the uniform topology to a Gaussian process of the form $\displaystyle G_{F_{ \upsilon}} - \frac{ \partial F_{\upsilon}}{ \partial \upsilon}^{\top} \int \psi dG_{F_{ \upsilon}}$.
\end{corollary}
The requirement that $\hat{ \upsilon}$ should be asymptotically linear is fulfilled by most $\sqrt{T}$-consistent estimators proposed in the literature \citep*[e.g.,][]{bickel-klaassen-ritov-wellner:98a}.

\section{Simulation study} \label{section:simulation}

We now appraise, via Monte Carlo simulation, our estimator of the EDF of spot variance introduced in Section \ref{section:theory}. We also evaluate the size and power of the goodness-of-fit tests for the marginal distribution of volatility proposed in Section \ref{section:goodness-of-fit}.

Starting at $X_{0} = 0$, and recalling that $\sigma_{t} = \sqrt{V_{t}}$, the efficient log-price is a convolution of a \citet*{heston:93a}-type stochastic volatility model and a pure-jump process:
\begin{align} \label{equation:heston}
\begin{split}
dX_{t} &= \sqrt{V_{t}} dW_{t}  + dJ_{t},\\[0.25cm]
dV_{t} &= \kappa \big( v_{0} - V_{t} \big)dt + \xi \sqrt{V_{t}}dB_{t},
\end{split}
\end{align}
where $W$ and $B$ are Brownian motions with $\text{corr}[dW_{t}dB_{t}] = \rho = -\sqrt{0.5}$. The other parameters are $\kappa = 0.05$, $v_{0} = 1$, and $\xi = 0.2$, which is calibrated to our empirical data and broadly aligns with previous studies \citep*[e.g,][]{ait-sahalia-kimmel:07a}.

In this framework, the law of $V_{t}$ (conditional on $V_{s}$, for $s < t$) is---up to a scale factor---non-central chi-square with $d = 4 v_{0} \kappa \xi^{-2}$ degrees of freedom and non-centrality parameter $\displaystyle \lambda = \frac{4 \kappa e^{- \kappa(t-s)}}{\xi^{2}(1-e^{- \kappa(t-s)})} V_{s}$ \citep*[e.g.,][]{cox-ingersoll-ross:85a}. As $t \to \infty$, the conditional distribution converges to a $\text{Gamma}(2 \kappa v_{0} \xi^{-2}, 2\kappa \xi^{-2})$, which is also the unconditional law of $V_{t}$ for any finite $t$, if $V_{0}$ is drawn from this distribution. We do that here. The variance of $F_{T}(x)$ in \eqref{equation:actual_variance} is then computed from these expressions by numerical integration of the joint density function.

$J_{t}$ is a symmetric tempered stable process with L\'{e}vy measure \citep*[e.g.,][]{rosinski:07a}:
\begin{equation}
\nu(dx) = c\frac{e^{- \lambda x}}{x^{1+r}} dx, \quad \text{for } x > 0,
\end{equation}
where $c > 0$, $\lambda > 0$ and $r \in [0,2)$. The degree of price jump activity is measured by $r$. We choose $r = 0.5$, which produces an infinite-activity, finite-variation process with many small jumps and a few large ones. We assume $\lambda = 3$ and calibrate $c$ such that $J_{t}$ accounts for 20\% of the quadratic variation. This complies with previous work in the field \citep*[e.g.,][]{ait-sahalia-jacod-li:12a, ait-sahalia-xiu:16a}.

\begin{figure}[t!]
\begin{center}
\caption{Estimation of the EDF. \label{figure:fixed-sigma}}
\begin{tabular}{cc}
\small{Panel A: Sample path of $V_{t}$.} & \small{Panel B: EDF and REDF.}\\
\includegraphics[height=0.4\textwidth,width=0.48\textwidth]{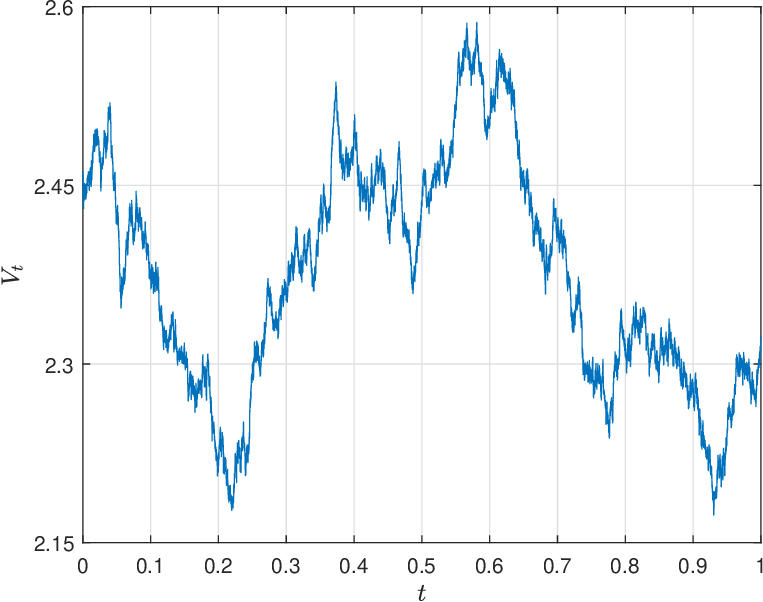} &
\includegraphics[height=0.4\textwidth,width=0.48\textwidth]{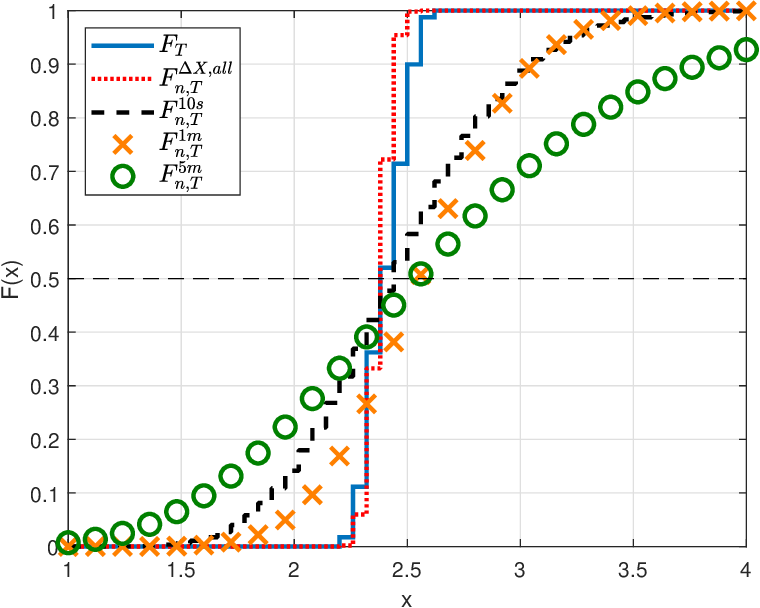}\\
\end{tabular}
\begin{scriptsize}
\parbox{\textwidth}{\emph{Note.} This figure plots a simulated path of $V_{t}$ in Panel A. The associated EDF of $V_{t}$, $F_{T}$, appears in Panel B (solid line). We estimate $F_{T}$ with $F_{n,T}$ over 10,000 simulations and plot the average (dashed line). The non-noise-robust version proposed in \citet*{li-todorov-tauchen:13a, li-todorov-tauchen:16a} based on 1- and 5-minute sparse sampling is shown as a comparison ('$\times$' and '$\circ$'). The dotted line is an infeasible estimate that has access to the whole record of $X$, $(X_{i \Delta_{n}})_{i = 0}^{n}$.}
\end{scriptsize}
\end{center}
\end{figure}

We generate 10,000 ``continuous-time'' paths with $23,400$ updates per unit of time $T$. As described in \citet*{todorov-tauchen-grynkiv:14a}, $J_{t}$ is simulated as the difference between two spectrally positive tempered stable processes using the acceptance-rejection algorithm of \citet*{baeumer-meerschaert:10a}. The rest of the system is discretized by an Euler approach. To gauge how close the REDF is to the population counterpart, we extract a coarser grid of $n = 2,340T$ log-price increments. This emulates a financial market, where new transactions arrive regularly every tenth second over a 6.5 hour trading day, which is aligned to the sample sizes in our empirical work.

We add heteroscedastic noise $Z_{i \Delta_{n}} = X_{i \Delta_{n}} + U_{i \Delta_{n}}$ with $U_{i \Delta_{n}} \sim N(0, \omega_{i \Delta_{n}}^{2})$, where $\omega_{i \Delta_{n}}^{2}$ is set by fixing the noise-to-signal ratio \citep*[e.g.,][]{oomen:06a}: $\gamma = \sqrt{ \displaystyle \frac{1}{\Delta_{n}} \frac{\omega_{i \Delta_{n}}^{2}}{ V_{i \Delta_{n}}}}$. As consistent with \citet*{christensen-oomen-podolskij:14a}, we take $\gamma = 0.5$.\footnote{We made a robustness check with $\gamma = 2.0$ as in \citet*{ait-sahalia-jacod-li:12a}. The pre-averaging estimator was hardly affected by this change, while the sparse estimator was substantially more biased, due to the amplification of the noise.} Thus, $(Z_{i \Delta_{n}})_{i=0}^{n}$ comprises our discretely observed sample used to estimate the spot variances, which we recover with $\tilde{V}_{i \Delta_{n}}$ from \eqref{equation:variance-estimator} combined with \eqref{equation:spot-variance-estimator}. We pre-average with $g(x) = \min(x,1-x)$, $k_{n} = [\theta / \sqrt{ \Delta_{n}}]$ and $\theta = 1/3$. We also truncate using a standard approach. In particular, a pre-averaged increment is reset to zero if $|\bar{Z}_{i}| \geq q_{1-\alpha} \sqrt{ \widehat{IV}} \Delta_{n}^{ \bar{ \omega}}$, where $q_{1- \alpha}$ is the $1- \alpha$ quantile of the standard normal distribution, $\widehat{IV}$ is the pre-averaged bipower variation, and $\bar{ \omega} \in (0,0.25)$ controls how fast the cutoff goes to zero with $\Delta_{n}$ (note $\bar{Z}_{i} = O_{p} \big( \Delta_{n}^{1/4} \big)$ if $X$ is continuous). Here, $\alpha = 0.001$ and $\bar{ \omega} = 0.20$. The debiasing in $\hat{V}_{i \Delta_{n}}$ is done with $\hat{ \omega}_{i \Delta_{n}}^{2}$ from \eqref{equation:noise-variance-estimator}, which we measure locally on a block that covers all the noisy high-frequency data that were included in the calculation of the pre-averaged increments that are part of the computation of $\tilde{V}_{i \Delta_{n}}$.

In Figure \ref{figure:fixed-sigma}, we examine how $F_{n,T}$ recovers $F_{T}$. We set $T = 1$ and freeze volatility across replica to keep the estimation target fixed. The sample path of $V_{t}$ is shown in Panel A, while the associated EDF appears in Panel B. We report our estimator and compare it to \citet*{li-todorov-tauchen:13a}. We construct the latter statistic from 1- and 5-minute low-frequency returns. As recommended in their work and to facilitate comparison, $h_{n}$ is set throughout so that it spans about 1.5 hours worth of data.

At the 1-minute resolution, the \citet*{li-todorov-tauchen:13a} estimator is severely distorted by market frictions, which leads to a systematic upward bias. This was to be expected, as it is not resistant to noise. On the other hand, while the 5-minute statistic does not accumulate any discernible bias, it delivers imprecise estimates, resulting in a highly overdispersed REDF. In contrast, as our noise-robust estimator is able to capitalize on tick-by-tick information, it is more accurate. This suggests the inferior rate of convergence embedded in pre-averaging is more than offset by being able to employ all the data.

\begin{figure}[t!]
\begin{center}
\caption{Mean relative bias and absolute error. \label{figure:bias-mae}}
\begin{tabular}{cc}
\small{Panel A: Relative bias.} & \small{Panel B: Absolute error.}\\
\includegraphics[height=0.4\textwidth,width=0.48\textwidth]{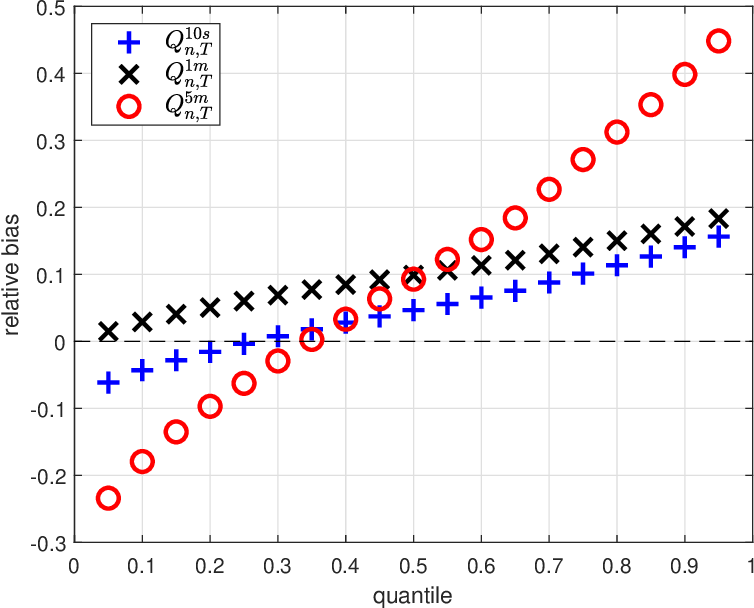} &
\includegraphics[height=0.4\textwidth,width=0.48\textwidth]{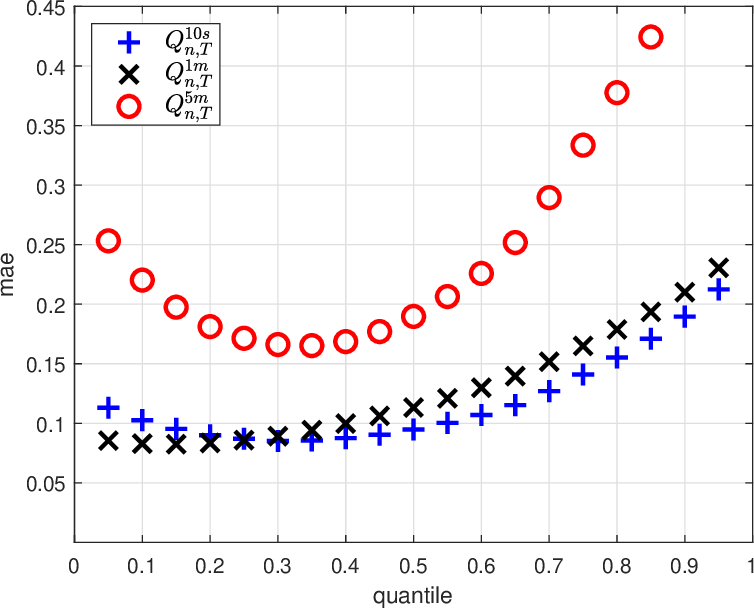}\\
\end{tabular}
\begin{scriptsize}
\parbox{\textwidth}{\emph{Note.} This figure plots the average relative bias in Panel A and absolute error in Panel B of our realized estimator of the EDF of spot variance ('$+$'). As a comparison, we also plot the \citet*{li-todorov-tauchen:13a} no-noise estimator based on 1- and 5-minute sparse sampling ('$\times$' and '$\circ$').}
\end{scriptsize}
\end{center}
\end{figure}

Next, we vary $V_{t}$ randomly in each iteration and estimate $F_{T}$ as above. In Figure \ref{figure:bias-mae}, we plot the relative error $Q_{n,T}( \alpha)/Q_{T}( \alpha) - 1$ in Panel A and absolute error $\big|Q_{n,T}( \alpha) - Q_{T}( \alpha) \big|$ in Panel B, which are averaged across Monte Carlo trials, for the vector of quantiles $\alpha = (0.05, 0.10, \ldots,0.95)'$. As evident---and consistent with Figure \ref{figure:fixed-sigma}---our estimator is less biased, especially in the tails. The shape of the upward sloping bias, common among estimators, is explained by the fact that the sampling distribution of $\hat{V}_{t}$, say, is spread out (not necessarily centered) around $V_{t}$. Hence, when volatility is low, $\hat{V}_{t}$ sometimes underestimates $V_{t}$, which translates into a downward bias in the low quantiles of the REDF vis-\'{a}-vis the EDF, and vice versa. The bias is more pronounced when $V_{t}$ is high, because it is harder to estimate spot variance in a volatile market. Overall, the REDF proposed in this paper is nevertheless roughly unbiased over a large spectrum of the distribution with a relative error within a few percent. This is further corroborated by Panel B. In general, while all estimators do a relatively poorer job at fitting the tails of the distribution of spot variance, there are again notable gains by applying our estimator.

\begin{figure}[t!]
\begin{center}
\caption{Inference about marginal distribution, $T = 250$. \label{figure:large-T}}
\begin{tabular}{cc}
\small{Panel A: Point estimate.} & \small{Panel B: Kernel density.}\\
\includegraphics[height=0.4\textwidth,width=0.48\textwidth]{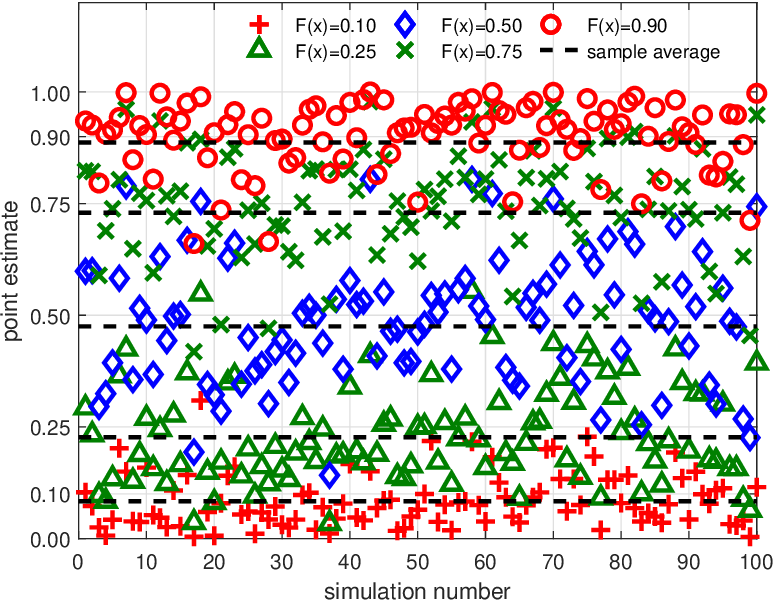} &
\includegraphics[height=0.4\textwidth,width=0.48\textwidth]{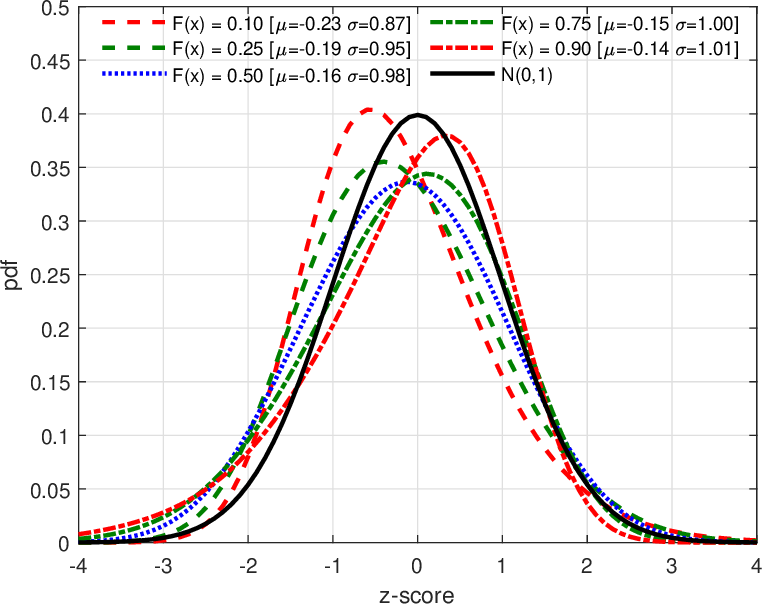}\\
\end{tabular}
\begin{scriptsize}
\parbox{\textwidth}{\emph{Note.} In Panel A, we report point estimates of $F_{n,T}(x)$ for the first one hundred Monte Carlo replica and sample averages across all simulations, for $F(x) = 0.10, 0.25, 0.50, 0.75$ and $0.90$. In Panel B, we plot kernel density estimates (with bandwidth selection via Silverman's rule of thumb) of the standardized statistic $\sqrt{T} \big( F_{n,T}(x) - F(x) \big) / \sqrt{ \Sigma(x)}$. The Gaussian curve is superimposed as a visual reference point. Throughout, $\Delta_{n} = 1/2,340$ and $T = 250$.}
\end{scriptsize}
\end{center}
\end{figure}

We now turn to estimation of the marginal distribution of volatility based on Theorem \ref{equation:large-T-clt}. We work with $T = 250$, corresponding to the number of trading days in about a year. This relatively low value is meant to be conservative and illustrate the potential of our estimator. As a comparison, $T = 2,517$ in our empirical work in Section \ref{section:empirical}. Note that $n$ is kept fixed relative to $T$, so that $\Delta_{n}$ is constant. In Figure \ref{figure:large-T}, we report point estimates---for the first 100 realizations---and sample averages across all 10,000 simulations of $F_{n,T}(x)$ and kernel densities of the asymptotic pivot $\sqrt{T} \big( F_{n,T}(x) - F(x) \big) / \sqrt{ \Sigma(x)} \xrightarrow{d} N(0,1)$ evaluated at five distinct points, which cover a variety of low-to-high volatility states, i.e. $F(x) = 0.10, 0.25, 0.50, 0.75$ and $0.90$. We see that in general the REDF is unbiased and correctly scaled. If we look far in the tails, in particular the right-hand one, the density estimate is slightly skewed toward the center, which is intuitive due to the natural bounds on $F_{n,T}$. Still, the limiting distribution provides a good description of the finite sample variation of $F_{n,T}$.

Last but not least, we explore the properties of the realized Kolmogorov-Smirnov and Cramer-von-Mises (i.e., a weighted $L^{2}$ norm with $w(x) = f(x)$) goodness-of-fit test. This piece of analysis is based on a shorter run of 1,000 Monte Carlo replica due to the increased computational cost of the resampling scheme. We draw 500 bootstrap samples in each simulation trial and inspect both the setting where $v$ is known and estimated. We follow \citet*{corradi-distaso:06a} and recover the parameters via $\hat{v}$ using GMM to match the sample mean, variance, plus the first and second autocovariance of the bias-corrected pre-averaged bipower variation of \citet*{podolskij-vetter:09a} to the corresponding model-based moments of integrated variance (the former is consistent for the latter).

To be able to define a measure of the testing power we simulate from another model, i.e. the log-normal---or exponential Ornstein-Uhlenbeck---process, which is the solution of the stochastic differential equation:
\begin{equation}
d\ln V_{t} = \kappa (v_{0} - \ln V_{t}) dt + \xi dB_{t},
\end{equation}
where $v = ( \kappa, v_{0}, \xi) = (0.08, -0.3, 0.45)$ is comparable to what we observe in the real data. We leave the leverage correlation unchanged at $\rho = -\sqrt{0.5}$.

In Table \ref{table:mc-size-and-power}, we report the size and power of the test for a nominal significance level of $\alpha = 5\%$ and $T = 250$, $500$, $1,000$, and $2,000$. As seen, the test is roughly unbiased and has a rejection rate close to $\alpha$. We next comment on the power, as reported in the second half of Table \ref{table:mc-size-and-power}. It is calculated as the rejection rate achieved by simulating log-normal volatility and testing the square-root process. $\mathcal{H}_{0}$ without estimation is here based on the configuration of the \citet*{heston:93a} model with the true parameters used to simulate volatility for the size analysis. As shown, the test has moderate power if the sample size is small, but the rejection rate increases steadily as $T$ grows larger. At $T = 2,000$, the power is in the 60--90\% range. The largest choice of $T$ corresponds to about eight years worth of high-frequency data, which is widely available for many assets in practice. The impact of parameter estimation is to ``flatten'' the power curve adding an upward bias for small $T$ and dampening it for large $T$. We conjecture this effect is caused by the inherent sampling error in the pre-averaged bipower variation, which leads to systematic upward biases in the estimate of both the mean reversion and volatility-of-volatility coefficient $\kappa$ and $\xi$.\footnote{The \citet*{heston:93a} model has an ARMA(1,1) representation for the integrated variance. We could exploit this structure by casting the whole system into state-space representation and then use a Kalman filter to extract a filtered (or smoothed) volatility series before estimating the model, as discussed in \citet*{barndorff-nielsen-shephard:02a} \citep*[see also, e.g.,][for an alternative approach]{todorov-tauchen-grynkiv:11a}. This tends to deliver less biased parameter estimates. However, we do not pursue this idea here.} The misspecification of the dynamic properties of the system invariably yields simulated critical values that are slightly off target and this effect appears more pronounced under the alternative. Finally, we notice the realized Kolmogorov-Smirnov test has slightly higher power than our weighted $L^{2}$ statistic, which is a somewhat surprising finding compared to the classical results for i.i.d. data.

\begin{table}[t!]
\setlength{\tabcolsep}{0.33cm}
\begin{center}
\caption{Properties of the realized goodness-of-fit tests. \label{table:mc-size-and-power}}
\smallskip
\begin{tabular}{lrrccccccccccc}
\hline\hline
      & & \multicolumn{5}{c}{Size} & & \multicolumn{5}{c}{Power} \\
\cline{3-7} \cline{9-13}
      & & \multicolumn{2}{c}{$rKS(F)$} & & \multicolumn{2}{c}{$rL^{2}(F)$}
      & & \multicolumn{2}{c}{$rKS(F)$} & & \multicolumn{2}{c}{$rL^{2}(F)$} \\
\cline{3-4} \cline{6-7} \cline{9-10} \cline{12-13}
& & $v$ & $\hat{v}$ & & $v$ & $\hat{v}$ & & $v$ & $\hat{v}$ & & $v$ & $\hat{v}$ \\
$T = $&  250& 0.042 & 0.086 & & 0.048 & 0.086 & & 0.091 & 0.422 & & 0.074 & 0.314\\
      &  500& 0.044 & 0.060 & & 0.047 & 0.082 & & 0.192 & 0.502 & & 0.115 & 0.389\\
      & 1000& 0.043 & 0.034 & & 0.043 & 0.054 & & 0.471 & 0.596 & & 0.266 & 0.451\\
&       2000& 0.035 & 0.026 & & 0.049 & 0.022 & & 0.951 & 0.793 & & 0.893 & 0.637\\
\hline \hline
\end{tabular}
\smallskip
\begin{scriptsize}
\parbox{\textwidth}{\emph{Note.} The table reports rejection rates of the goodness-of-fit tests for stochastic volatility models proposed in Section \ref{section:goodness-of-fit}. $rKS(F)$ ($rL^{2}(F)$) is the realized Kolmogorov-Smirnov (Cramer-von-Mises) test. The number of Monte Carlo simulations is 1,000 with 500 bootstrap samples used to compute critical values for the $t$-statistic in each trial. $v$ ($\hat{v}$) is based on the true (estimated) parameter vector. The power is calculated as the rejection rate achieved by simulating log-normal volatility and testing whether the \citet*{heston:93a} process is the true model. Further details are available in the main text.}
\end{scriptsize}
\end{center}
\end{table}

\section{Empirical application} \label{section:empirical}

We here illustrate the application of our new nonparametric estimator of the EDF of spot variance and the associated goodness-of-fit tests to real high-frequency data. We construct the $F_{n,T}$ measure from tick data based on selected exchange-traded funds (ETFs) that cover different sectors of the U.S. stock market. In addition to a proxy for the market index (ticker symbol SPY), we add the nine industry portfolios of the S\&P 500, yielding a total of ten securities.\footnote{There is also a 10th sector ETF (XLRE), which is a diversified portfolio of companies exposed to real estate. However, as the inception date of this fund is rather recent, we exclude it from consideration due to the limited availability of high-frequency data. Further information about the ETFs is available at http://www.sectorspdr.com/.} High-frequency data was acquired from the TAQ database and comprise a complete series of trades and quotes for each stock for the sample period July 2008 -- June 2018. The analysis here is based on transaction data, which were filtered for outliers as in \citet*{christensen-oomen-podolskij:14a} and subsequently pre-ticked to an equidistant 10-second grid from 9:30am to 4:00pm EST, resulting in $T = 2,517$ days of $n = 2,340$ intraday returns.\footnote{Trading at NYSE terminates early at 1:00pm on a few regularly scheduled business days each year in observance of upcoming holidays. We do not collect data after the exchange has officially closed, as there is typically very little liquidity, and therefore use a shorter sample of $n = 1,260$ log-price increments on these days.} A list of ticker symbols and descriptive statistics of the sample are presented in Table \ref{table:descriptive}.

\begin{table}[t!]
\setlength{\tabcolsep}{0.20cm}
\begin{center}
\caption{Descriptive statistics of high-frequency data. \label{table:descriptive}}
\smallskip
\begin{tabular}{lcccccccccccccc}
\hline\hline
     & & & && \multicolumn{8}{c}{goodness-of-fit test (p-value)}\\ \cline{6-13}
code & effective & $\sigma$ & $\gamma$ && \multicolumn{2}{c}{gamma} & & \multicolumn{2}{c}{log-normal} & & \multicolumn{2}{c}{inverse gaussian}\\ \cline{6-7} \cline{9-10} \cline{12-13}
& sample & & && $rKS(F)$ & $rL^{2}(F)$ && $rKS(F)$ & $rL^{2}(F)$ && $rKS(F)$ & $rL^{2}(F)$ \\
\hline
XLB & 0.756 & 0.204 & 0.253 && 0.001 & 0.000 && 0.000 & 0.000 && 0.080 & 0.039 \\
XLE & 0.938 & 0.234 & 0.136 && 0.000 & 0.000 && 0.007 & 0.005 && 0.219 & 0.161 \\
XLF & 0.930 & 0.246 & 0.895 && 0.000 & 0.000 && 0.000 & 0.000 && 0.009 & 0.004 \\
XLI & 0.809 & 0.176 & 0.334 && 0.000 & 0.000 && 0.006 & 0.004 && 0.130 & 0.109 \\
XLK & 0.761 & 0.165 & 0.506 && 0.000 & 0.000 && 0.052 & 0.025 && 0.035 & 0.045 \\
XLP & 0.723 & 0.124 & 0.558 && 0.003 & 0.001 && 0.021 & 0.009 && 0.015 & 0.022 \\
XLU & 0.778 & 0.167 & 0.381 && 0.000 & 0.000 && 0.083 & 0.060 && 0.041 & 0.071 \\
XLV & 0.768 & 0.144 & 0.342 && 0.000 & 0.000 && 0.000 & 0.000 && 0.046 & 0.035 \\
XLY & 0.768 & 0.174 & 0.245 && 0.003 & 0.000 && 0.000 & 0.000 && 0.054 & 0.021 \\
SPY & 0.999 & 0.153 & 0.145 && 0.003 & 0.000 && 0.029 & 0.007 && 0.300 & 0.360 \\
\hline \hline
\end{tabular}
\smallskip
\begin{scriptsize}
\parbox{\textwidth}{\emph{Note.} We report descriptive statistics of the equity high-frequency data. ``code'' is the ticker symbol, ``effective sample'' (a measure of liquidity) is the fraction of previous-tick interpolated 10-second prices originating from a new transaction rather than a repetition, $\sigma$ is annualized volatility based on a pre-averaged bipower variation of \citet*{podolskij-vetter:09a}, while $\gamma$ is the noise-to-signal ratio defined in the main text. The numbers are computed daily and averaged over the sample. The p-value is for the realized Kolmogorov-Smirnov and Cramer-von-Mises goodness-of-fit test with marginal distribution under $\mathcal{H}_{0}$ indicated by the designated column label.}
\end{scriptsize}
\end{center}
\end{table}

\begin{figure}[ht!]
\begin{center}
\caption{Properties of stochastic variance in SPY. \label{figure:SPY}}
\begin{tabular}{cc}
\small{Panel A: PDF of spot variance.} & \small{Panel B: QQ plot.}\\
\includegraphics[height=0.4\textwidth,width=0.48\textwidth]{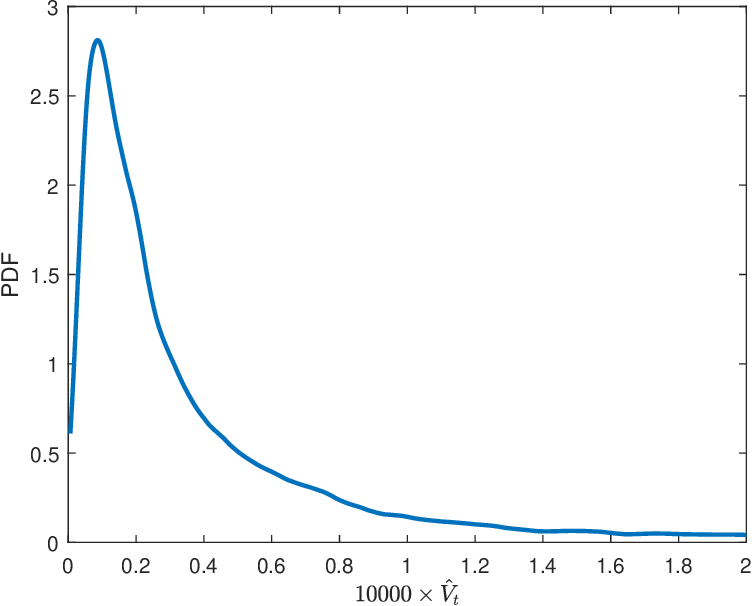} &
\includegraphics[height=0.4\textwidth,width=0.48\textwidth]{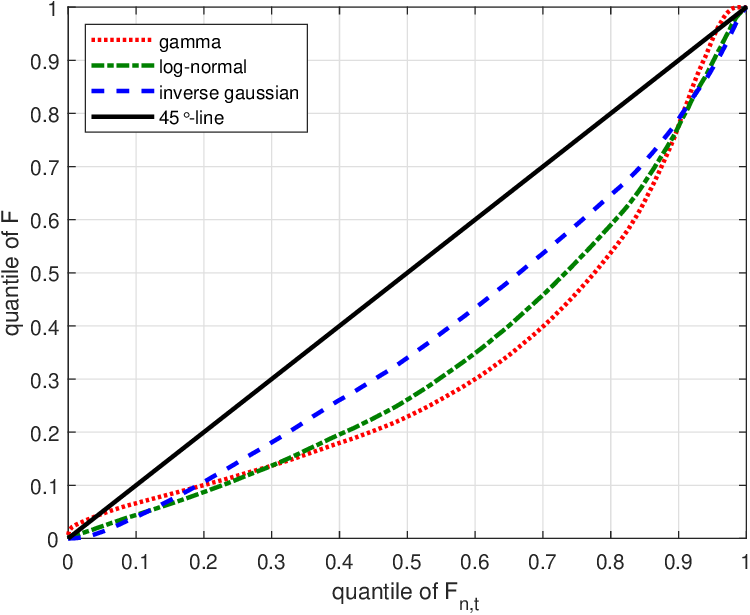}
\end{tabular}
\begin{scriptsize}
\parbox{\textwidth}{\emph{Note.} In Panel A, we plot a kernel-based estimate of the probability density function of spot variance for SPY. In Panel B, we create a QQ plot, where the quantiles of $F_{n,T}$ (on the $x$-axis) are mapped against the quantiles of a fitted distribution function $F$ (on the $y$-axis). The gamma, log-normal and inverse Gaussian distribution are included for the latter comparison.}
\end{scriptsize}
\end{center}
\end{figure}

The calculation of $\hat{V}_{t}$ proceeds as in Section \ref{section:theory} -- \ref{section:simulation} in terms of tuning parameters. We further account for the diurnal pattern in intraday spot volatility. We follow the standard approach in the literature, which is to compute the average value of $\hat{V}_{t}$ at each time point of the day over the sample. The diurnal factor is then estimated by normalizing the sum of these averages to one. We correct $\hat{V}_{t}$ by this quantity and use the adjusted time series to construct $F_{n,T}$. In Panel A of Figure \ref{figure:SPY}, we show a kernel-based estimate of the probability density function implied by the time series of variance estimates of SPY, as representative of our data.

We compute $rKS(F)$ and $rL^{2}(F)$ with three choices of $F$ under $\mathcal{H}_{0}$. The first are the gamma and log-normal distribution inherent in the square-root and log-normal diffusion models that were presented in the simulation section. Furthermore, we inspect the goodness-of-fit of a non-Gaussian Ornstein-Uhlenbeck process \citep*[e.g.,][]{barndorff-nielsen-shephard:01a}. The latter is a pure-jump specification with SDE:
\begin{equation}
\label{equation:levy}
dV_{t} = -\kappa V_{t} dt + dL_{t},
\end{equation}
where $L_{t}$ is a subordinator with L\'{e}vy measure $\nu_{L}(dx)$.

To maintain a tractable setting, we proceed as in \citet*{todorov-tauchen-grynkiv:11a} by assuming that the distribution of the increments of the volatility process corresponds to that of a tempered stable process. This means it has a L\'{e}vy density:
\begin{equation}
\nu_{V}(dx) = c \frac{e^{- \lambda x}}{x^{1 + \beta}}dx, \quad \text{for } x > 0,
\end{equation}
where $c > 0$, $\lambda > 0$ and $\beta < 1$. The L\'{e}vy measure of the background driving process $\nu_{L}(dx)$ can then be backed out from $\nu_{V}(dx)$, see e.g., \citet*{barndorff-nielsen-shephard:01a}. We fix the activity level of the volatility jumps at $\beta = 0.5$, so that the resulting marginal distribution of volatility is inverse Gaussian, i.e. $V_{t} \sim IG(\mu, \nu)$ with:
\begin{equation}
\mu = c \sqrt{ \frac{\pi}{ \lambda}}, \qquad \nu = 2 \pi c^{2},
\end{equation}
and
\begin{equation}
\displaystyle f(x; \mu, \nu) =  \sqrt{ \frac{ \nu}{2 \pi x^{3}}} \exp \bigg\{-\frac{ \nu}{2 \mu^{2}x}(x - \mu)^{2} \bigg\}, \quad \text{for } x > 0,
\end{equation}
is the density function of $V_{t}$.

The procedure in Section \ref{section:simulation} is applied to report a p-value for the $t$-statistic. The exception is that we boost the number of bootstrap samples to 1,000 for higher precision. The non-Gaussian Ornstein-Uhlenbeck process is discretized and simulated as explained in the appendix of \citet*{todorov-tauchen-grynkiv:11a}.\footnote{As in \cite{meddahi:03a}, we exploit the ARMA(1,1) structure for the integrated variance of this model to compute the relevant moments: $\displaystyle \mathbb{E}[IV_{t}] = c \sqrt{ \frac{ \pi}{ \lambda}}$, $\displaystyle \text{var}(IV_{t}) = \frac{c \sqrt{ \pi}}{ \lambda^{3/2} \kappa^{2}} ( e^{- \kappa}+\kappa-1)$, $\displaystyle \text{cov}(IV_{t}, IV_{t-1}) = \sqrt{ \frac{ \pi c^{2}}{4 \lambda^{3}}} \frac{(1 - e^{-\kappa})^{2}}{ \kappa^{2}}$, and $\text{cov}(IV_t,IV_{t+j}) = e^{-\kappa} \text{cov}(IV_{t},IV_{t+j-1})$ for $j \geq 2$. We also calculated these expressions based on the ARMA coefficients reported in the appendix of \citet*{todorov-tauchen-grynkiv:11a}, but the results differ and we suspect there are some typos in that paper.}

In Table \ref{table:descriptive}, we present the outcome of these efforts, while Panel B of Figure \ref{figure:SPY} shows a QQ-style plot, where the quantiles of $F_{n,T}$ (on the $x$-axis) are compared to the quantiles of the best fitting distribution function $F$ in each class (on the $y$-axis). The graph should align with the 45$^\circ$-line if the model is correct. As readily seen, the gamma distribution does not capture stochastic equity variance at all. The main problem with the square-root process is that it cannot generate a sufficient level of volatility-of-volatility (via $\xi$) without violating the Feller restriction, which is highly binding in practice. The p-values of the log-normal model are higher on average, but it also does not deliver an acceptable fit of the data across sectors. Although there are a few close calls, for example XLK, XLU and SPY, the model is generally rejected at standard levels of significance. In contrast, the p-values of the inverse Gaussian are typically much higher. This is in line with \citet*{todorov-tauchen:12a}, and it appears relatively consistent across the ETF space. Nevertheless, the latter also struggles (depending a bit on which $t$-statistic we ask), for instance it does not capture return variation exhibited by the highly volatile financial sector ETF (XLF), but it is generally the best fitting model of the ones estimated in this paper.

\section{Conclusion} \label{section:conclusion}

We construct a nonparametric jump- and noise-robust realized measure of the EDF of the latent volatility of a general It\^{o} semimartingale sampled at high-frequency on a fixed time interval. We extend previous work of \citet*{li-todorov-tauchen:13a, li-todorov-tauchen:16a} in the noise-free setting and prove that in the presence of microstructure noise the pre-averaged version of their estimator is consistent in the infill asymptotic limit. The resistance to market frictions enables our statistic to fully exploit information about volatility available in tick-by-tick high-frequency data, which improves its accuracy in a simulation study.

In a subsequent analysis, we let the time span tend to infinity. We show that our estimator then converges to the marginal distribution function of volatility. This result is subject to some rate conditions, which are more restrictive in the presence of jumps or roughness in volatility. We establish a functional CLT, from which we prove the validity of a family of goodness-of-fit tests for stochastic volatility models. The limiting distribution of our $t$-statistic depends on the true model under the null hypothesis, but critical values can be found via simulation. We show how the procedure can be adapted to the case with parameter estimation.

We apply the theory to high-frequency data from several ETFs that track different sectors of the U.S. stock market. The marginal distribution embedded in several classes of popular stochastic volatility models is tested against our nonparametric measure. There is strong evidence against the gamma or log-normal distribution, while the inverse Gaussian---implied by the class of non-Gaussian Ornstein-Uhlenbeck processes of \citet*{barndorff-nielsen-shephard:01a} with tempered stable increments---has more support. These findings are consistent with \citet*{todorov-tauchen-grynkiv:11a}, but in disagreement with \citet*{corradi-distaso:06a}.

Our paper can serve as a starting point for building more refined models of stochastic volatility. It appears natural, for example, to inspect the properties of the generalized inverse Gaussian (GIG) as the marginal distribution of volatility. It has the PDF:
\begin{equation} \label{equation:gig}
f(x;a,b,p) = \frac{(a/b)^{p/2}}{2K_{p}(\sqrt{ab})} x^{(p-1)} e^{-(ax+b/x)/2}, \quad \text{for } x > 0,
\end{equation}
where $K_{p}$ is the modified Bessel function of the second kind, while $a > 0, b > 0$ and $p$ a real number are the parameters. The GIG is self-decomposable (necessary and sufficient for stationary solutions of \eqref{equation:levy} to exist) and nests the gamma distribution as $b \to 0$ (with $a > 0$) and the inverse Gaussian if $p = -1/2$. The extra degree of freedom may just be enough for a complete description of equity variance. We leave this assessment for future research.

\pagebreak

\appendix

\section{Appendix} \label{appendix:proofs}

We here prove the theoretical results presented in the main text. As usual, the localization procedure in \citet*[][Section 4.4.1]{jacod-protter:12a} implies we can assume that $b_{t}$, $\sigma_{t}$, $\tilde{b}_{t}$, $\tilde{ \sigma}_{t}$, $\tilde{ \sigma}'_{t}$, $\omega_t$ and the jump component of $X$ are bounded for the infill setting with $T$ fixed, i.e. Lemma 2.1, Theorem 3.1 and Corollary 3.2. Without loss of generality, the proofs are based on $r = 2$. Also, to preserve notation $C$ denotes a generic constant that changes value from one line to the next.

\subsection*{Proof of Lemma 2.1}

Fix a $t > 0$. If $t \in [i \Delta_{n}, (i+1) \Delta_{n})$, we set $t_{n} = i \Delta_{n}$ to highlight the dependence on $n$ and note that $V_{t_{n}} \rightarrow V_{t}$ and $\omega_{t_{n}}^2 \rightarrow \omega_{t}^2 $ due to their c\`{a}dl\`{a}g properties.
Thus, our job is to establish:
\begin{equation*}
\tilde{V}_{i \Delta_{n}} - \theta \psi_{2} V_{t_{n}} - \frac{1}{ \theta} \psi_{1} \omega_{t_n}^{2} \xrightarrow{ \mathbb{P}} 0.
\end{equation*}
We start by introducing an approximation of the pre-averaged return series $\bar{Z}_{i}$, in which the stochastic volatility and noise processes are fixed locally in each computation:
\begin{equation*}
\bar{Y}_{i,i+m} = \sum_{j=1}^{k_{n}-1} g \bigg( \frac{j}{k_{n}} \bigg) \big( \sigma_{i \Delta_{n}} \Delta_{i+m+j}^{n}W +
\omega_{i \Delta_{n}}\Delta_{i+m+j}^{n} \varepsilon \big),
\end{equation*}
and then we deduce the claim based on $\bar{Y}_{i, i+m}$, i.e.
\begin{equation} \label{A3}
\check{V}_{i \Delta_{n}} \equiv \frac{1}{h_{n} \sqrt{ \Delta_{n}}} \sum_{m=0}^{h_{n}-1} \bar{Y}_{i, i+m}^{2} - \theta \psi_{2} V_{t_{n}} - \frac{1}{ \theta} \psi_{1} \omega_{t_{n}}^{2} \xrightarrow{ \mathbb{P}} 0.
\end{equation}
Now, it is readily seen that for each $m \geq 0$, $\Delta^{-1/4}_{n} \bar{Y}_{i, i+m}$ has (conditionally on $\mathcal{F}_{(i+m) \Delta_{n}}$) mean zero and variance equal to $\displaystyle k_{n}\sqrt{ \Delta_{n}} \psi_{2}^{n} V_{t_{n}}+\frac{1}{k_{n}\sqrt{ \Delta_{n}}}\psi_{1}^{n} \omega_{t_n}^{2}$. Furthermore,
\begin{equation*}
k_{n} \sqrt{ \Delta_{n}} \psi_{2}^{n} V_{t_{n}} + \frac{1}{k_{n} \sqrt{\Delta_{n}}} \psi_{1}^{n} \omega_{t_n}^{2} - \theta \psi_{2} V_{t_{n}} - \frac{1}{\theta} \psi_{1} \omega_{t_n}^{2} \xrightarrow{ \mathbb{P}} 0,
\end{equation*}
as $n \rightarrow \infty$. It follows that
\begin{equation*}
\frac{1}{h_{n} \sqrt{ \Delta_{n}}} \sum_{m=0}^{h_{n}-1} \mathbb{E} \big[ \bar{Y}_{i, i+m}^{2} \mid \mathcal{F}_{(i+m) \Delta_{n}} \big] -
\theta \psi_{2} V_{t_{n}} - \frac{1}{ \theta} \psi_{1} \omega_{t_{n}}^{2} \xrightarrow{ \mathbb{P}} 0,
\end{equation*}
and therefore it suffices to prove that
\begin{align*}
\frac{1}{h_{n} \sqrt{ \Delta_{n}}} \sum_{m=0}^{h_{n}-1} \Big( \bar{Y}_{i, i+m}^{2} - \mathbb{E} \big[ \bar{Y}_{i, i+m}^{2} \mid \mathcal{F}_{(i+m) \Delta_{n}} \big] \Big) \xrightarrow{ \mathbb{P}} 0.
\end{align*}
To this end, we note that
\begin{align*}
\Delta^{-1}_{n} \mathbb{E} \Big[ \big( \bar{Y}_{i, i+m}^{2} - \mathbb{E} \big[ \bar{Y}_{i, i+m}^{2} \mid \mathcal{F}_{(i+m) \Delta_{n}} \big] \big) \big( \bar{Y}_{i, i+l}^{2} - \mathbb{E} \big[ \bar{Y}_{i, i+l}^{2} \mid \mathcal{F}_{(i+l) \Delta_{n}} \big] \big) \Big] &\leq C, \quad & \text{for } |m-l| \leq k_{n}, \\[0.10cm]
\Delta^{-1}_{n} \mathbb{E} \Big[ \big( \bar{Y}_{i, i+m}^{2} - \mathbb{E} \big[ \bar{Y}_{i, i+m}^{2} \mid \mathcal{F}_{(i+m) \Delta_{n}} \big] \big) \big( \bar{Y}_{i, i+l}^{2} - \mathbb{E} \big[ \bar{Y}_{i, i+l}^{2} \mid \mathcal{F}_{(i+l) \Delta_{n}} \big] \big) \Big] &= 0, \quad & \text{for } |m-l| > k_{n}.
\end{align*}
Thus,
\begin{align*}
\mathbb{E} \Bigg[ \bigg( \frac{1}{h_{n} \sqrt{ \Delta_{n}}} \sum_{m=0}^{h_{n}-1} \Big( \bar{Y}_{i, i+m}^{2} - \mathbb{E} \big[ \bar{Y}_{i,i+m}^{2} \mid \mathcal{F}_{(i+m) \Delta_{n}} \big] \Big) \bigg)^{2} \Bigg] \leq C \frac{k_{n}}{h_{n}} \rightarrow 0,
\end{align*}
and \eqref{A3} follows.

We are thus left with the assertion:
\begin{equation} \label{equation:A4}
S_{n} = \tilde{V}_{i \Delta_{n}} - \check{V}_{i \Delta_{n}} \xrightarrow{ \mathbb{P}} 0.
\end{equation}
To prove \eqref{equation:A4}, we define
\begin{align*}
X_t''&= \int_{0}^{t} \int_{ \mathbb{R}} \delta(s,z) 1_{ \{| \delta(s,z)| \leq 1 \}} (\mu-\nu)(d(s,z)), \quad
B_{t}'' = \int_{0}^{t} b_{s} ds + \int_{0}^{t} \int_{ \mathbb{R}} \delta(s,z) 1_{ \{| \delta(s,z)| > 1 \}} \mu(d(s,z)), \\[0.10cm]
Y_{t}' &= \int_{i \Delta_{n}}^{t}( \sigma_{s} - \sigma_{i \Delta_{n}})dW_{s} + ( \omega_{t}- \omega_{i \Delta_{n}}) \varepsilon_{t}.
\end{align*}
Note that $\Delta_{j}^{n} Z = \sigma_{i \Delta_{n}} \Delta_{j}^{n}W + \omega_{i \Delta_{n}} \Delta_{j}^{n} \varepsilon+ \Delta_{j}^{n}B'' + \Delta_{j}^{n}Y' + \Delta_{j}^{n}X''$, for $j = i+1, \ldots, i+h_{n}+k_{n}$. Now, from the inequality \citep*[][p. 258]{jacod-protter:12a}:
\begin{equation*}
\big|(x + y + z + w)^{2} 1_{ \{| x + y + z + w | \leq v \}} - x^{2} \big| \leq C \frac{x^{4}}{v^{2}} + \epsilon x^{2}+ \frac{C}{ \epsilon} \big((y^{2} \wedge v^{2}) + z^{2} + w^{2} \big)
\end{equation*}
with $v = v_{n} \Delta_{n}^{-1/4} = \alpha \Delta_{n}^{ \bar{ \omega} - 1/4}$, $\epsilon \in (0,1]$, and
\begin{equation*}
x = \frac{ \bar{Y}_{i,i+m}}{ \Delta_{n}^{1/4}}, \quad
y = \frac{ \bar{X}_{i+m}''}{ \Delta_{n}^{1/4}}, \quad
z = \frac{ \bar{Y}_{i+m}'}{  \Delta_{n}^{1/4}}, \quad
w = \frac{ \bar{B}_{i+m}''}{ \Delta_{n}^{1/4}},
\end{equation*}
we find that
\begin{equation*}
|S_{n}| \leq \frac{1}{h_{n}} \sum_{m=0}^{h_{n}-1} \Bigg[ C \Delta_{n}^{1/2-2 \bar{ \omega}} \bigg| \frac{ \bar{Y}_{i,i+m}}{ \Delta_{n}^{1/4}} \bigg|^{4}
+ \epsilon \bigg| \frac{ \bar{Y}_{i,i+m}}{ \Delta_{n}^{1/4}} \bigg|^{2} + \frac{C}{ \epsilon} \Bigg( \Delta_{n}^{2 \bar{ \omega}-1/2}\bigg| \frac{ \bar{X}_{i+m}''}{ \Delta_{n}^{\bar{\omega}}} \wedge 1 \bigg|^{2} + \bigg| \frac{ \bar{Y}_{i+m}'}{ \Delta_{n}^{1/4}} \bigg|^{2} + \bigg| \frac{ \bar{B}_{i+m}''}{ \Delta_{n}^{1/4}} \bigg|^{2} \Bigg) \Bigg].
\end{equation*}
As $| \Delta_{i}^{n} B''| \leq C \Delta_{n}$, uniformly in $i$, we see that
\begin{equation*}
\bigg| \frac{ \bar{B}''}{ \Delta_{n}^{1/4}} \bigg| \leq C \Delta_{n}^{1/4}.
\end{equation*}
Then, by means of the Burkholder-Davis-Gundy inequality:
\begin{equation} \label{equation:A13}
\mathbb{E} \Big[| \bar{Y}_{i+m}'|^{2} \mid \mathcal{F}_{(i+m) \Delta_{n}} \Big] \leq C k_{n} \Delta_{n}
\mathbb{E}\bigg[\sup_{s \in [i \Delta_{n},(i+h_{n}+k_{n}) \Delta_{n}]} \big(| \sigma_{s}- \sigma_{i \Delta_{n}}|^{2}+| \omega_{s}- \omega_{i \Delta_{n}}|^{2} \big) \mid \mathcal{F}_{(i+m) \Delta_{n}} \bigg].
\end{equation}
As $\sigma$ and $\omega$ are c\`{a}dl\`{a}g and can be assumed to be bounded, the last term converges to zero. Next, for the components involving $\bar{Y}_{i, i+m}$, it holds that \citep*[][Lemma 1]{podolskij-vetter:09a}:
\begin{equation} \label{equation:A14}
\mathbb{E} \Big[ | \bar{Y}_{i, i+m} |^{4} \mid \mathcal{F}_{(i+m) \Delta_{n}} \Big] \leq C \Delta_{n}.
\end{equation}
In addition, Corollary 2.1.9 in \citet*{jacod-protter:12a} applied to the pre-averaging case yields:
\begin{equation*}
\mathbb{E}\bigg[ \Big| \frac{\bar{X}_{i+m}''}{ \Delta_{n}^{\bar{ \omega}}} \wedge 1 \Big|^{2} \mid \mathcal{F}_{(i+m) \Delta_{n}} \bigg] \leq
C \Delta_{n}^{1/2-2 \bar{ \omega}} \phi_{n},
\end{equation*}
where $(\phi_n)_{n \in \mathbb{N}}$ is a sequence of real numbers such that $\phi_n \to 0$ as $n \to \infty.$
This, along with \eqref{equation:A13} -- \eqref{equation:A14}, implies that:
\begin{equation*}
\mathbb{E} \big[|S_{n}| \big] \leq C\left( \Delta_n^{1/2-2 \bar{\omega}} + \epsilon + \frac{1}{\epsilon} \Bigg( \phi_n + \mathbb{E}
\bigg[ \sup_{s \in [i \Delta_{n},(i + h_{n} + k_{n}) \Delta_{n}]} \big(|\sigma_{s}-\sigma_{i \Delta_{n}}|^{2}+|\omega_{s}-\omega_{i \Delta_{n}}|^{2}\big) \bigg] + \Delta_{n}^{1/2} \Bigg)\right).
\end{equation*}
By letting first $n$ to infinity, and then sending $\epsilon$ to zero, we complete the proof of Lemma 2.1. \qed

\subsection*{Proof of Theorem 3.1}
First, we prove that $\hat{ \omega}_{t}^{2} \xrightarrow{ \mathbb{P}} \omega_{t}^{2}$ for each $t \in [0,T]$. Due to the definition of $\hat{ \omega}_{i \Delta_{n}}^{2}$ and the c\`{a}dl\`{a}g property of $\omega_{t}$, it suffices to show that $\hat{ \omega}_{i \Delta_{n}}^{2} \xrightarrow{ \mathbb{P}} \omega_{i \Delta_{n}}^{2}$ for each $i$, where
\begin{equation*}
\hat{\omega}_{i \Delta_{n}}^{2} = - \frac{1}{ h_n - 1} \sum_{m =1}^{h_n-1} \Delta_{i+m}^{n}Z \Delta_{i+m+1}^{n}Z.
\end{equation*}
Note that $\hat{ \omega}_{i \Delta_{n}}^{2} = A_{n}^{(1)} + A_{n}^{(2)}$, where
\begin{align*}
A_{n}^{(1)} &= - \frac{1}{h_{n}-1} \sum_{m =1}^{h_{n}-1} \Delta_{i+m}^{n}U \Delta_{i+m+1}^{n}U, \\[0.10cm]
A_{n}^{(2)} &= - \frac{1}{h_{n}-1} \sum_{m =1}^{h_{n}-1} \Delta_{i+m}^{n}U \Delta_{i+m+1}^{n}X + \Delta_{i+m}^{n}X \Delta_{i+m+1}^{n}U + \Delta_{i+m}^{n}X \Delta_{i+m+1}^{n}X.
\end{align*}
In view of (2.1.44) in \citet*{jacod-protter:12a}, we obtain that $\sup_{1 \leq m \leq h_{n}} \mathbb{E} \big[( \Delta_{i+m}^{n}X)^{2} \big] \leq C \Delta_{n}$.
This, together with $\sup_{1 \leq m \leq h_{n}} \mathbb{E} \big[( \Delta_{i+m}^{n}U)^{2} \big] \leq C$, yields that $\mathbb{E} \Big[ |A_{n}^{(2)} | \Big] \leq C \Delta_{n}^{1/2}$. For the main term $A_{n}^{(1)}$, we use a further decomposition:
\begin{align*}
A_{n}^{(1)}- \omega_{i \Delta_{n}}^{2} =& - \frac{1}{h_{n}-1} \sum_{m=1}^{h_{n}-1} \big( \Delta_{i+m}^{n}U \Delta_{i+m+1}^{n}U- \omega_{i \Delta_{n}}^{2} \Delta_{i+m}^{n} \varepsilon \Delta_{i+m+1}^{n} \varepsilon \big) \\[0.10cm]
&- \omega_{i \Delta_{n}}^{2} \frac{1}{h_{n}-1} \sum_{m=1}^{h_{n}-1} \big( \Delta_{i+m}^{n} \varepsilon \Delta_{i+m+1}^{n} \varepsilon + 1 \big) \equiv B_{n}^{(1)} + B_{n}^{(2)}.
\end{align*}
We immediately notice that $B_n^{(2)} \xrightarrow{ \mathbb{P}} 0$. Moreover,
\begin{equation*}
\mathbb{E} \Big[|B_{n}^{(1)}| \Big] \leq C \bigg( \mathbb{E} \bigg[ \sup_{s \in [i \Delta_{n},(i+h_{n}) \Delta_{n}]} | \omega_{s}-\omega_{i \Delta_{n}}|^{2} \bigg] \bigg)^{1/2} \rightarrow 0,
\end{equation*}
as $\omega$ is c\`{a}dl\`{a}g. This verifies that $\hat{ \omega}_{i \Delta_{n}}^{2} \xrightarrow{ \mathbb{P}} \omega_{i \Delta_{n}}^{2}$.

The rest of the proof follows the structure of Lemma 1 and Theorem 1 in \citet*{li-todorov-tauchen:13a}. We start by showing that $\mathbb{P} \big( V_{t} = x \big) = 0$. By assumption, $F_{T}$ is continuous almost surely, so letting $F_{T}(x-) = \lim_{y \uparrow x}F_{T}(y)$, it holds that $\mathbb{P} \big(F_{T}(x)\neq F_{T}(x-) \big) = 0$. Hence,
\begin{equation*}
\int_{0}^{T} \mathbb{P} \big( V_{t} = x \big) dt = \mathbb{E} \bigg[ \int_{0}^{T}(1_{ \{ V_{t} \leq x \}} - 1_{\{ V_{t} < x \}})dt \bigg] = T\mathbb{E} \big(F_{T}(x)-F_{T}(x-) \big) = 0.
\end{equation*}
Thus, $\mathbb{P}( V_{t} = x) = 0$ for a.e. $t \in [0,T]$. Next, we show the pointwise consistency of $F_{n,T}(x)$. We therefore fix $x \in \mathbb{R}_{+}$ and notice that
\begin{align*}
\mathbb{E} \Big[ \big| F_{n,T}(x) - F_{T}(x) \big| \Big] &\leq \frac{1}{T} \int_{0}^{T} \mathbb{E} \Big[ \big| 1_{ \{ \hat{V}_{t} \leq x \}} - 1_{ \{ V_{t} \leq x \}} \big| \Big]dt \\[0.10cm]
&= \frac{1}{T} \int_{0}^{T} \mathbb{E} \Big[ \big| \big( 1_{ \{ \hat{V}_{t} \leq x \}} - 1_{ \{ V_{t} \leq x \}} \big) 1_{ \{ V_{t} \neq x \}} \big| \Big] dt.
\end{align*}
Applying Lemma 2.1 and Lebesgue's dominated convergence theorem (twice), we deduce that $\mathbb{E} \Big[ \big| F_{n,T}(x) - F_{T}(x) \big| \Big] \rightarrow 0$, which implies pointwise consistency, $F_{n,T}(x) \xrightarrow{ \mathbb{P}} F_{T}(x)$. As $F_{n,T}$ and $F_{T}$ are non-decreasing functions and $F_{T}$ is continuous a.s., this convergence is locally uniform \citep*[see, e.g.,][]{resnick:98a}. Furthermore, as the paths of $\sigma_{t}$ are c\`{a}dl\`{a}g, for any $\eta > 0$ there exists a constant $M$ such that $\mathbb{P}( \sup_{t \in[0,T]} V_{t} > M) < \eta$, which means that $\mathbb{P}(F_{T}(M) \neq 1) < \eta$. In turn,
\begin{align*}
\mathbb{P} \bigg( \sup_{x \in \mathbb{R}_{+}} \big| F_{n,T}(x) - F_{T}(x) \big| > \epsilon \bigg) &\leq \mathbb{P} \bigg( \sup_{x \in[0,M]} \big| F_{n,T}(x) - F_{T}(x) \big| > \epsilon \bigg) + \mathbb{P} \bigg( \sup_{x \geq M} \big| F_{n,T}(x)-F_{T}(x) \big| > \epsilon \bigg) \\[0.10cm]
&\leq \mathbb{P} \bigg( \sup_{x \in[0,M]} \big| F_{n,T}(x) - F_{T}(x) \big| > \epsilon \bigg) + \mathbb{P} \bigg( \sup_{x \geq M} \big| F_{n,T}(x) - F_{n,T}(M) \big| > \frac{ \epsilon}{3} \bigg) \\[0.10cm]
&+ \mathbb{P} \bigg( \sup_{x \geq M} \big|F_{n,T}(M) - F_{T}(M) \big| > \frac{ \epsilon}{3} \bigg) + \mathbb{P} \bigg( \sup_{x \geq M} \big| F_{T}(M) - F_{T}(x) \big| > \frac{ \epsilon}{3} \bigg) \\[0.10cm]
&\leq \mathbb{P} \bigg( \sup_{x \in[0,M]} \big| F_{n,T}(x) - F_{T}(x) \big| > \epsilon \bigg) + \mathbb{P} \bigg(1 - F_{n,T}(M) > \frac{ \epsilon}{3} \bigg) \\[0.10cm]
&+ \mathbb{P} \bigg( \big| F_{n,T}(M) - F_{T}(M) \big| > \frac{ \epsilon}{3} \bigg) + \mathbb{P} \bigg(1 - F_{T}(M) > \frac{ \epsilon}{3} \bigg).
\end{align*}
Taking $\limsup$ and using the pointwise and locally uniform convergence:
\begin{equation*}
\limsup_{n \rightarrow \infty} \mathbb{P} \bigg( \sup_{x \in \mathbb{R}_+} \big| F_{n,T}(x) - F_{T}(x) \big| > \epsilon \bigg) \leq 2 \eta,
\end{equation*}
and since $\eta$ is arbitrary, this shows that
\begin{equation*}
\mathbb{P} \bigg( \sup_{x \in \mathbb{R}_{+}} \big| F_{n,T}(x) - F_{T}(x) \big| > \epsilon \bigg) \rightarrow 0,
\end{equation*}
as $\Delta_{n} \rightarrow 0$. \qed

\subsection*{Three auxiliary results}
In this section, we prove three auxiliary lemmas that are used throughout the subsequent proofs. Before stating the first, we need some additional notation.

Let $X'$ denote the continuous part of $X$. Next, define $X'' = X-X'$ and $Z'_{i \Delta_n} = X'_{i \Delta_n} + U_{i \Delta_n}$, and let $\check{V}_{i \Delta_n}$ be a version of $\hat{V}_{i \Delta_{n}}$ that uses $Z'$ instead of $Z$, i.e.
\begin{equation*}
\check{V}_{i,n} = \frac{1}{ \theta \psi_{2}} \frac{1}{ h_{n} \sqrt{ \Delta_{n}}} \sum_{m=0}^{h_{n}-1} \big ( \bar{Z}_{i+m}')^{2} 1_{\{ | \bar{Z}_{i+m}'| \leq v_{n} \}} + \frac{ \psi_{1}}{\theta^{2} \psi_{2}} \frac{1}{h_{n} - 1} \sum_{m = 1}^{h_{n}-1} \Delta_{i+m}^{n}Z' \Delta_{i+m+1}^{n}Z'.
\end{equation*}

\begin{lemma} \label{lemma:jumps}
Suppose that Assumption \ref{assumption:volatility} holds, and let $h_{n}/k_{n} \rightarrow \infty$ as $\Delta_{n} \rightarrow 0$. Then, for any $r \in(0,2)$ and some $\iota>0$,
\begin{equation*}
\mathbb{E} \Big[ | \hat{V}_{i \Delta_{n}}- \check{V}_{i \Delta_{n}}|^{p} \Big] \leq C \big((h_{n}/k_{n})^{1-p} \Delta_{n}^{1/2-r \bar \omega- 2 p(1/4- \bar \omega) - \iota} + \Delta_{n}^{((2-r) \bar \omega - \iota)p} + h_{n}^{-p/2} \Delta_{n} \big),
\end{equation*}
uniformly in $i$ for $p \geq 2$.
\end{lemma}
\noindent \textbf{Proof.} First, we observe that:
\begin{align*}
\hat{V}_{i \Delta_{n}}- \check{V}_{i \Delta_{n}} = S_{i,n}^{(1)} + S_{i,n}^{(2)},
\end{align*}
where
\begin{align*}
S_{i,n}^{(1)} &= \frac{1}{ \theta \psi_{2}} \frac{1}{ h_{n} \sqrt{ \Delta_{n}}} \sum_{m=0}^{h_{n}-1} \left( \big( \bar{Z}_{i+m} \big)^{2} 1_{\{ | \bar{Z}_{i+m}| \leq v_{n} \}} - \big( \bar{Z}_{i+m}' \big)^{2} 1_{ \{ | \bar{Z}_{i+m}'| \leq v_{n} \}} \right), \\[0.10cm]
S_{i,n}^{(2)} &= \frac{ \psi_{1}}{ \theta^{2} \psi_{2}} \frac{1}{h_{n} - 1} \sum_{m = 1}^{h_{n}-1} \left( \Delta_{i+m}^{n}Z \Delta_{i+m+1}^{n}Z- \Delta_{i+m}^{n}Z' \Delta_{i+m+1}^{n}Z' \right).
\end{align*}
Next, for any $v \geq 1$ and $a \geq 0$, we note that:
\begin{equation*}
\big||x+y|^{2} 1_{ \{|x+y| \leq v \}}-|x|^{2} 1_{ \{|x| \leq v \}} \big| \leq C
\left(  \frac{|x|^{2+a}}{v^{a}}+ (1+|x|)(|y| \wedge v)^{2} \right),
\end{equation*}
which can be verified by looking at the cases:
\begin{equation*}
\mbox{(1)} |x+y| \leq v, |x| \leq v \mbox{ (2)} |x+y| \leq v, |x|>v, \mbox{ (3)} |x+y|>v, v/2<|x|\leq v, \mbox{ (4)} |x+y|>v, |x| \leq v/2.
\end{equation*}
If we apply this inequality with $x = \Delta_{n}^{-1/4} \bar{Z}_{i+m}'$, $y = \Delta_{n}^{-1/4} \bar{X}_{i+m}''$, and $v = \Delta_{n}^{-1/4} v_{n}= \alpha \Delta_{n}^{ \bar \omega-1/4}$ (and $n$ large enough such that $v \geq 1$), it leads to:
\begin{equation*}
\big|S_{i,n}^{(1)} \big| \leq \frac{C}{h_{n}} \sum_{m=0}^{h_{n}-1} \left( \Delta_{n}^{a(1/4- \bar \omega)} \left| \Delta_{n}^{-1/4} \bar{Z}_{i+m}' \right|^{2+a} + \Delta_{n}^{2 ( \bar \omega-1/4)} \big(1+ \left| \Delta_{n}^{-1/4} \bar{Z}_{i+m}' \right| \big)
\left( \frac{ \bar{X}_{i+m}''}{ \Delta_{n}^{ \bar \omega}} \wedge 1 \right)^{2} \right).
\end{equation*}
Another application of Corollary 2.1.9 in \citet*{jacod-protter:12a} to the pre-averaging framework implies that, for any $q \geq 2$,
\begin{equation*}
\mathbb{E} \bigg[ \Big| \frac{ \bar{X}_{i+m}''}{ \Delta_{n}^{ \bar \omega}} \wedge 1 \Big|^{q} \mid \mathcal{F}_{(i+m) \Delta_{n}} \bigg]
\leq \Delta_n^{1/2-r \bar \omega} \phi_n,
\end{equation*}
where $\phi_{n} \rightarrow 0$. Let
\begin{equation*}
\chi_{i}^{n} = \Delta_{n}^{a(1/4- \bar \omega)} \big| \Delta_{n}^{-1/4 } \bar{Z}_{i}' \big|^{2+a} + \Delta_{n}^{2 (\bar \omega-1/4)} \Big(1+ \big| \Delta_{n}^{-1/4} \bar{Z}_{i}' \big| \Big) \left( \frac{ \bar{X}_{i}''}{ \Delta_{n}^{ \bar \omega}} \wedge 1 \right)^{2}.
\end{equation*}
Then, for any $p \geq 1$ and some arbitrarily small $\iota>0$, H\"{o}lder's inequality leads to:
\begin{equation*}
\mathbb{E} \big[| \chi_{i}^{n}|^{p} \mid \mathcal{F}_{i \Delta_{n}} \big] \leq C \Big( \Delta_{n}^{p a(1/4- \bar \omega)} + \Delta_{n}^{(1/2-r \bar \omega)b - p(1/2-2 \bar \omega)} \Big) \leq C \Delta_{n}^{(1/2-r \bar \omega)- p(1/2-2 \bar \omega)- \iota},
\end{equation*}
where the last part follows by taking large enough $a$ (recall $1/4- \bar \omega>0$). A successive conditioning argument combined with Burkholder-Gundis-Davis' inequality results in:
\begin{equation*}
\mathbb{E} \Big[ \big|S_{i,n}^{(1)} \big|^{p} \Big] \leq C \Big((h_{n}/k_{n})^{1-p} \Delta_{n}^{1/2-r \bar \omega-p(1/2-2 \bar \omega)} + \Delta_{n}^{((2-r) \bar \omega - \iota)p} \Big).
\end{equation*}
We proceed with the term $S_{i,n}^{(2)}$. Notice that:
\begin{equation*}
S_{i,n}^{(2)} = \frac{ \psi_{1}}{\theta^{2} \psi_{2}} \frac{1}{h_{n} - 1} \sum_{m = 1}^{h_{n}-1} \zeta_{i+m}^{n},
\end{equation*}
where $\zeta_{i}^{n} = \Delta_{i}^{n} Z' \Delta_{i+1}^{n} X'' + \Delta_{i}^{n} X'' \Delta_{i+1}^{n}Z' + \Delta_{i}^{n} X'' \Delta_{i+1}^{n} X''$. Section 2.1 in \citet*{jacod-protter:12a} then implies that, for any $p \geq 2$, $\mathbb{E} \big[| \Delta_{i}^{n} Z'|^{p} \mid \mathcal{F}_{(i-1) \Delta_{n}} \big] \leq C$
and $\mathbb{E} \big[| \Delta_{i}^{n} X''|^{p} \mid \mathcal{F}_{(i-1) \Delta_{n}} \big] \leq C \Delta_{n}^2$. This means that $\mathbb{E} \big[| \zeta_{i}^{n}|^{p} \mid \mathcal{F}_{(i-1) \Delta_{n}} \big] \leq C \Delta_{n}$, for any $p \geq 1$. After subtracting the conditional means, an application of the discrete Burkholder inequality then delivers that:
\begin{equation*}
\mathbb{E} \Big[ \big| S_{n}^{(2)} \big|^{p} \Big] \leq C \big( h_{n}^{-p/2} \Delta_{n} + \Delta_{n}^{p} \big) \leq C h_{n}^{-p/2} \Delta_{n},
\end{equation*}
and this concludes the proof. \qed

\begin{lemma} \label{lemma:rate-of-convergence}
Suppose that Assumption 1\textnormal{(i)} with $H \in(0,1)$ and 4 -- 6 hold true, as does the rate condition $T^{1/2+ \iota} \Big( \Delta_{n}^{ \frac{H}{4H+2}- \iota} \vee \Delta_{n}^{ \frac{H}{2H+1} - (1/2-2 \bar \omega)-\iota} \vee \Delta_n^{(2- r) \bar{ \omega} -\iota} \Big) \rightarrow 0$, as $\Delta_n\rightarrow 0$ and $T \rightarrow \infty$, for some $\iota > 0$ and $h_{n} \asymp \Delta_{n}^{- \frac{4H+1}{4H+2}}$. Finally,  let $r \in [0,2)$ and $\bar{ \omega} \in \left( \frac{1}{8H+4}, \frac{1}{4} \right)$. Then,
\begin{equation}
\sqrt{T} \sup_{x \in \mathbb{R}_{+}} \big|F_{n,T}(x) - F_T(x) \big| \xrightarrow{ \mathbb{P}} 0.
\end{equation}
\end{lemma}

\noindent \textbf{Proof.} The proof consists of three parts. First, we show that for each $i$ and $p \geq 2$:
\begin{equation} \label{RatesInd}
\mathbb{E} \Big[ | \check{V}_{i \Delta_{n}} - V_{i \Delta_{n}}|^{p} \Big] \leq C \Big[ \big( k_{n}/h_{n} \big)^{p/2} + \big(h_{n} \Delta_{n} \big)^{p H} + \Delta_{n}^{p/4} \Big].
\end{equation}
Note that
\begin{equation} \label{equation:decomposition}
\hat{V}_{i \Delta_{n}} - V_{i \Delta_{n}} = R_{n}^{(0)}+R_{n}^{(1)} + R_{n}^{(2)} + R_{n}^{(3)} + R_{n}^{(4)} + R_{n}^{(5)},
\end{equation}
where
\begin{align*}
R_{n}^{(0)} &= -\frac{1}{h_{n} \sqrt{ \Delta_{n}}} \sum_{m=0}^{h_{n}-1} \big( \bar{Z}_{i+m} \big)^{2} 1_{ \{ |\bar{Z}_{i+m}|> v_{n}  \}}, \\[0.10cm]
R_{n}^{(1)} &= \frac{1}{ \theta \psi_{2}} \frac{1}{h_{n} \sqrt{ \Delta_{n}}} \sum_{m = 0}^{h_{n}-1} \bar{X}_{i+m}^{2} - \sigma_{i \Delta_{n}}^{2} \bar{W}_{i+m}^{2}, \\[0.10cm]
R_{n}^{(2)} &= \frac{1}{ \theta \psi_{2}} \frac{2}{h_{n}} \sum_{m = 0}^{h_{n} - 1} \Delta_{n}^{-1/2} \bar{X}_{i+m} \bar{U}_{i+m}, \\[0.10cm]
R_{n}^{(3)} &= \frac{1}{ \theta \psi_{2}} \frac{1}{h_{n}} \sum_{m = 0}^{h_{n} - 1} \Big( \Delta_{n}^{-1/2} \bar{U}_{i+m}^{2} - \frac{ \psi_{1}}{ \theta} \omega_{i \Delta_{n}}^{2} \Big), \\[0.10cm]
R_{n}^{(4)} &= \frac{ \psi_{1}}{ \theta^{2} \psi_{2}}( \omega^{2}_{i \Delta_{n}}- \hat{ \omega}_{i \Delta_{n}}^{2}), \\[0.10cm]
R_{n}^{(5)} &= \sigma_{i \Delta_{n}}^{2} \bigg( \frac{k_{n} \Delta_{n}^{1/2} \psi_2^{n}}{ \theta \psi_{2}} - 1 \bigg) + \frac{1}{ \theta \psi_{2}} \frac{1}{h_{n}} \sum_{m = 0}^{h_{n} - 1} \sigma_{i \Delta_{n}}^{2} \big( \Delta_{n}^{-1/2} \bar{W}_{i+m}^{2} - k_{n} \Delta_{n}^{1/2} \psi_{2}^{n} \big).
\end{align*}
To deal with the term $R_{n}^{(0)}$, fix $p \geq 2$. We choose large $q$ satisfying $q(1/4- \bar{ \omega}) \geq p/4$. Then:
\begin{equation*}
\mathbb{E} \Big[| \big( \bar{Z}_{i+m} \big)^{2} 1_{ \{ |\bar{Z}_{i+m}| > v_{n}  \}}|^{p} \Big] \leq v_{n}^{-q} \mathbb{E} \Big[| \big( \bar{Z}_{i+m} \big)^{2p+q} \Big] \leq C \Delta_{n}^{p/4 + p/2}.
\end{equation*}
Invoking the discrete H\"{o}lder inequality yields $\mathbb{E} \Big[|R_{n}^{(0)}|^{p} \Big] \leq C \Delta_{n}^{p/4}$.

To deal with the remaining terms, consider the random variable:
\begin{equation*}
S_{n} = \sum_{m = 0}^{l_{n} k_{n} - 1} Y_{m} = \sum_{i=1}^{k_{n}} S_{n}^{(i)},
\end{equation*}
where $(Y_{m})_{m \geq 1}$ are mean zero and $(k_{n}-1)$-dependent, such that for each $i$ $S_{n}^{(i)} = \sum_{m = 0}^{l_{n}} Y_{i-1+m k_{n}}$ is a sum of martingale differences and $\sup_{m \in \mathbb N} \mathbb{E} \big[|Y_{m}|^{p} \big] \leq C$. Applying the discrete H\"older and Burkholder inequality, we obtain for each $p \geq 2$:
\begin{equation*}
\mathbb{E} \big[|S_{n}|^{p} \big] \leq k_{n}^{p-1} \sum_{i=1}^{k_{n}} \mathbb{E} \big[|S_{n}^{(i)}|^{p} \big] \leq  Ck_{n}^{p} l_{n}^{p/2}.
\end{equation*}
This observation leads to
\begin{align*}
\mathbb{E} \Big[ \big|R_{n}^{(2)} \big|^{p} \Big]+ \mathbb{E} \Big[ \big|R_{n}^{(3)} \big|^{p} \Big] + \mathbb{E} \Big[ \big|R_{n}^{(5)} \big|^{p} \Big] \leq C \left(\big(k_{n}/h_{n} \big)^{p/2} +  \Delta_{n}^{p/4}\right).
\end{align*}
In addition, using the notation and results derived in the proof of Theorem \ref{theorem:uniform-convergence-fixedT}, the discrete H\"older and Burkholder inequalities imply that:
\begin{align*}
\mathbb{E} \Big[ \big|R_{n}^{(4)} \big|^{p} \Big] \leq C \bigg( \mathbb{E} \Big[ \big|A_{n}^{(2)} \big|^{p} \Big] + \mathbb{E} \Big[ \big|B_{n}^{(1)} \big|^{p} \Big] + \mathbb{E} \Big[ \big|B_{n}^{(2)} \big|^{p} \Big] \bigg) \leq C \big( \Delta_{n}^{p/2}+ (h_{n} \Delta_{n})^{p} + h_{n}^{-p/2} \big).
\end{align*}
Turning next to $R_{n}^{(1)}$, we have for each $i$ and $m$ that
\begin{equation*}
\bar{X}_{i+m} - \sigma_{i \Delta_{n}} \bar{W}_{i+m} = \sum_{j=1}^{k_{n}-1} g(j/k_{n}) A_{i+m+j},
\end{equation*}
where
\begin{equation*}
A_{i+m+j} = \int_{(i+m+j-1) \Delta_{n}}^{(i+m+j) \Delta_{n}} b_{s} ds + \int_{(i+m+j-1) \Delta_{n}}^{(i+m+j) \Delta_{n}} ( \sigma_{s} - \sigma_{i \Delta_{n}}) dW_{s}.
\end{equation*}
The Burkholder inequality then implies that, for each $p \geq 2$,
\begin{equation*}
\mathbb{E} \Big[ \big| \bar{X}_{i+m} - \sigma_{i \Delta_{n}} \bar{W}_{i+m} \big|^{p} \Big] \leq C \Big( \big(k_{n} \Delta_{n} \big)^{p} + \big(k_{n} \Delta_{n} \big)^{p/2} \big(h_{n} \Delta_{n} \big)^{p H} \Big) \leq C \Big( \Delta_{n}^{p/2} + \Delta_{n}^{p/4} \big(h_{n} \Delta_{n} \big)^{p H} \Big),
\end{equation*}
uniformly in $i$ and $m$. Using the above together with the identity $|a^{2}-b^{2}|^{p}=|a-b|^{p}|a+b|^{p}$, plus the discrete H\"older and Cauchy-Schwarz inequality, we deduce that:
\begin{align*}
\mathbb{E} \Big[ \big|R_{n}^{(1)} \big|^{p} \Big] & \leq \frac{C}{h_{n}} \sum_{m = 0}^{h_{n} - 1} \mathbb{E} \Big[ \big| \Delta_{n}^{-1/4} \big( \bar{X}_{i+m} - \sigma_{i \Delta_{n}} \bar{W}_{i+m} \big) \big|^{p} \big| \Delta_{n}^{-1/4} \big( \bar{X}_{i+m} + \sigma_{i \Delta_{n}} \bar{W}_{i+m} \big) \big|^{p} \Big] \\[0.10cm]
&\leq C \Big( \Delta_{n}^{p/4} + \big(h_{n} \Delta_{n} \big)^{p H} \Big),
\end{align*}
which finishes the proof of \eqref{RatesInd}.


We continue with the second part. We denote $\eta_{n,T} = \sup_{t \in[0,T]} \big| \hat{V}_{t} - V_{t} \big|$ and notice that
\begin{equation*}
\eta_{n, T} \leq \sup_{i \in A_{n, T}} | \hat{V}_{i \Delta_{n}} - V_{i \Delta_{n}}| + \sup_{i \in A_{n, T}} \sup_{t \in (i \Delta_{n}, (i+1) \Delta_{n}]} |V_{t} - V_{i \Delta_{n}}|.
\end{equation*}
where $A_{n,T} = \{ i : 0 \leq i \leq [T / \Delta_{n}] - h_{n} - k_{n} \}$. Now, Assumption 1(i) implies:
\begin{equation*}
\mathbb{E} \bigg[ \Big| \sup_{ t \in (i \Delta_{n}, (i+1) \Delta_{n}]} |V_{t} - V_{i \Delta_{n}}| \Big|^{p} \bigg] \leq C \Delta_{n}^{p H}.
\end{equation*}
This result, combined with Lemma \ref{lemma:jumps}, inequality in means, the maximal inequality and \eqref{RatesInd}, implies that for some $\iota>0$,
\begin{align*}
\mathbb{E} \Big[ \big| \eta_{n, T} \big|^{p} \Big] &\leq C \frac{T}{ \Delta_{n}} \Big[ \big( k_{n} / h_{n} \big)^{p/2} + \big(h_{n} \Delta_{n} \big)^{p H} + (h_{n}/k_{n})^{1-p} \Delta_{n}^{1/2-r \bar{ \omega}- 2 p(1/4- \bar{ \omega})-\iota} + \Delta_{n}^{((2-r) \bar{ \omega} -\iota)p} \Big],
\end{align*}
which immediately delivers the bound:
\begin{equation*}
\mathbb{E} \Big[ \big| \eta_{n, T} \big|^{p} \Big]^{1/p} \leq C T^{1/p} \Delta_{n}^{-1/p} \Big[ \big( k_{n} / h_{n} \big)^{1/2} + \big(h_{n} \Delta_{n} \big)^{H} + (h_n/k_n)^{(1-p)/p} \Delta_n^{(1/2-r \bar{ \omega}- 2 p(1/4- \bar{ \omega})-\iota)/p} + \Delta_{n}^{(2-r) \bar{ \omega} -\iota} \Big].
\end{equation*}
As $p$ can be arbitrarily large and optimizing $h_{n} \asymp \Delta_{n}^{- \frac{4H+1}{4H+2}}$, we deduce that
\begin{equation*}
\eta_{n,T} = O_{p} \bigg(T^{ \iota} \Big( \Delta_{n}^{ \frac{H}{4H+2}- \iota} \vee \Delta_{n}^{ \frac{H}{2H+1}-(1/2-2 \bar{ \omega})- \iota} \vee \Delta_{n}^{(2-r) \bar{ \omega} - \iota} \Big) \bigg).
\end{equation*}
As for the final step in the proof, recall that by definition of $\eta_{n,T}$:
\begin{align*}
F_{T}(x - \eta_{n,T}) \leq F_{n,T}(x) \leq F_{T}(x + \eta_{n,T}).
\end{align*}
Then,
\begin{align} \label{thm3.4-2}
\begin{split}
\sqrt{T} \big| F_{n,T}(x) - F_{T}(x) \big| & \leq \sqrt{T} \big| F_{T}(x + \eta_{n,T}) - F_{T}(x - \eta_{n,T}) \big| \\[0.10cm]
 & \leq \big| G_{T}(x+ \eta_{n,T}) - G_{T}(x- \eta_{n,T}) \big| + \sqrt{T}\big|F(x+ \eta_{n,T}) - F(x- \eta_{n,T}) \big|,
\end{split}
\end{align}
where $G_{T}(x) = \sqrt{T}(F_{T}(x) - F(x))$.

As the density of $F$ exists and is bounded:
\begin{align*}
\sqrt{T}\sup_{x \in \mathbb{R}_{+}}|F(x+ \eta_{n,T})-F(x- \eta_{n,T})| =
\sqrt{T}\sup_{x \in \mathbb{R}_{+}} \int_{x- \eta_{n,T}}^{x+ \eta_{n,T}}f(z)dz \leq 2K \sqrt{T} \eta_{n,T},
\end{align*}
which---by the rate condition on $\eta_{n,T}$---implies that $\sqrt{T} \sup_{x \in \mathbb{R}_{+}}|F(x+ \eta_{n,T})-F(x- \eta_{n,T})| \xrightarrow{ \mathbb{P}} 0$. And since $\eta_{n,T} \xrightarrow{ \mathbb{P}}0$, it follows that for any $\epsilon, \delta>0$ there exists $(n',T')$ such that $\mathbb{P} ( \eta_{n,T}> \delta) \leq \epsilon/2$ for all pairs $(n,T)$ with $n \geq n'$ and $T\geq T'$, which adheres to the rate condition. Let $\gamma>0$ be given. Then,
\begin{align} \label{lemma A.1.2}
\begin{split}
\mathbb{P} \bigg( \sup_{x \in \mathbb{R}_{+}} \big|G_{T}(x+ \eta_{n,T}) - G_{T}(x- \eta_{n,T}) \big|> \gamma \bigg) &\leq \mathbb{P} \bigg( \{ \sup_{x \in \mathbb{R}_{+}} \big|G_{T}(x+ \eta_{n,T}) - G_{T}(x- \eta_{n,T}) \big|> \gamma \} \cap \{ \eta_{n,T}< \delta \} \bigg) \\[0.10cm]
&+ \mathbb{P} \Big( \eta_{n,T} \geq \delta \Big) \\[0.10cm]
&\leq \mathbb{P} \bigg( \{ \sup_{|x-y| \leq 2 \delta} \big|G_{T}(x) - G_{T}(y) \big|> \gamma \} \bigg) + \mathbb{P} \Big( \eta_{n,T} \geq \delta \Big).
\end{split}
\end{align}
It follows from Theorem 5.2 in \citet*{dehay:05a} that for any $\epsilon, \gamma>0$ there exists $\delta>0$ such that the first term in \eqref{lemma A.1.2} is smaller than $\epsilon/2$. Hence,
\begin{equation*}
\mathbb{P} \bigg( \sup_{x \in \mathbb{R}_{+}} \big|G_{T}(x+ \eta_{n,T})-G_{T}(x- \eta_{n,T}) \big|> \gamma \bigg) \leq \epsilon.
\end{equation*}
As $\epsilon, \gamma$ are arbitrary, this verifies that $\sup_{x \in \mathbb{R}_{+}}|G_{T}(x+ \eta_{n,T}) - G_{T}(x- \eta_{n,T})| \xrightarrow{ \mathbb{P}}0$, concluding the proof of Lemma \ref{lemma:rate-of-convergence}. \qed

\begin{lemma} \label{lemma:rate-of-convergence-jumps}
Suppose that Assumption 1\textnormal{(ii)} and 4 -- 6 hold true, as does the rate condition \\$T^{1/2+ \iota}\left(\Delta_{n}^{1/8- \iota}\vee \Delta_n^{2\bar\omega - \frac{1}{4}}\vee \Delta_n^{(2-r)\bar\omega - \iota}\right) \rightarrow 0$, as $\Delta_{n} \rightarrow 0$ and $T \rightarrow \infty$, for some $\iota>0$ and $h_{n} \asymp \Delta_{n}^{-3/4}$. Finally, let $r\in[0,2)$ and $\bar\omega\in\left(\frac{1}{8},\frac{1}{4}\right)$. Then,
\begin{equation}
\sqrt{T} \sup_{x \in \mathbb{R}_{+}} \big|F_{n,T}(x) - F_T(x) \big| \xrightarrow{ \mathbb{P}} 0.
\end{equation}
\end{lemma}

\noindent \textbf{Proof.} Recall that if $\sigma_{t}$ satisfies Assumption 1(ii), the form of $V_{t} = \sigma_{t}^{2}$ is representable via (the general version of) It\^{o}'s Lemma:
\begin{equation*}
V_{t} = V_{0} + \int_{0}^{t} \bar{b}_{s} ds + \int_{0}^{t} \bar{ \sigma}_{s} dW_{s} + \int_{0}^{t} \bar{ \sigma}_{s}' dB_{s} + \int_{0}^{t} \int_{ \mathbb{R}} \bar{ \delta}(s, z)( \mu - \nu) (d(s, z)).
\end{equation*}
We split the variance in two: $V_{t} = V_{t}'( \epsilon) + V_{t}''( \epsilon)$, for each $\epsilon > 0$, where
\begin{align*}
V_{t}^{c} &= V_{0} + \int_{0}^{t} \bar{b}_{s}ds + \int_{0}^{t} \bar{ \sigma}_{s} dW_{s} + \int_{0}^{t} \bar{ \sigma}_{s}' dB_{s}, \\[0.10cm]
V_{t}'( \epsilon) &= V_{t}^{c} + \int_{0}^{t} \int_{ \{z: \tilde{ \Gamma}(z) \leq \epsilon \}} \bar{ \delta}(s, z)( \mu - \nu)(d(s,z)) - \int_{0}^{t} \int_{ \{z: \tilde{ \Gamma}(z) > \epsilon \}} \bar{ \delta}(s,z) \nu(d(s,z)), \\[0.10cm]
V_{t}''( \epsilon) &= \int_{0}^{t} \int_{ \{z: \tilde{ \Gamma}(z) > \epsilon \}} \bar{ \delta}(s, z) \mu(d(s,z)).
\end{align*}
Moreover, we denote
\begin{align*}
J_{n}( \epsilon) &= \{ i \in \mathbb{Z}: 0 \leq i \leq [T / \Delta_{n}] - h_{n} - k_{n}, \mu( [i \Delta_{n}, (i+1) \Delta_{n} ) \times \{ z: \tilde{ \Gamma}(z) > \epsilon \})=0 \}, \\[0.10cm]
S_{n}( \epsilon) &= \cup_{i \in J_{n}( \epsilon)} [i \Delta_{n}, (i+1) \Delta_{n}), \\[0.10cm]
A_{n,T} &= \{ i : 0 \leq i \leq [T / \Delta_{n}] - h_{n} - k_{n} \}.
\end{align*}
In words, $J_{n}( \epsilon)$ is a set of indices that excludes the big jumps and $S_{n}( \epsilon)$ is the union of these. It leads to a decomposition $F_{n, T}(x) = F_{n, T}(x, \epsilon) + R_{n, T}(x, \epsilon)$ and $F_{T}(x) = F_{T}(x, \epsilon) + R_{T}(x, \epsilon)$ with
\begin{align*}
F_{n,T}(x, \epsilon) &= \frac{1}{T} \int_{S_{n}( \epsilon)} 1_{ \{ \hat{V}_{t} \leq x  \}} dt, \quad
R_{n, T}(x, \epsilon) = \frac{1}{T} \int_{[0, T] \setminus S_{n}( \epsilon)}1_{ \{ \hat{V}_{t} \leq x  \}} dt, \\[0.10cm]
F_{T}(x, \epsilon) &= \frac{1}{T} \int_{S_{n}( \epsilon)} 1_{ \{ V_{t} \leq x  \}} dt, \quad R_{T}(x, \epsilon) = \frac{1}{T} \int_{ [0, T] \setminus S_{n}( \epsilon)} 1_{ \{ V_{t} \leq x  \}} dt.
\end{align*}
Here, $F$ is the main term, while $R$ is a remainder. A final piece of notation:
\begin{align*}
\eta_{n, T}( \epsilon) = \sup_{i \in J_{n}( \epsilon)}| \hat{V}_{i \Delta_{n}} - V_{i \Delta_{n}}| + \sup_{i \in A_{n,T}} \sup_{t \in (i \Delta_{n}, (i+1) \Delta_{n}]} |V_{t}'( \epsilon) - V_{i \Delta_{n}}'( \epsilon)|.
\end{align*}
Let $\epsilon_{n} \asymp \Delta_{n}^{1/8}$. The aim is to prove that
\begin{equation} \label{SmallJumps}
\eta_{n, T}( \epsilon_{n}) = O_p(T^{ \iota} \Delta_{n}^{1/8 - \iota}).
\end{equation}
We deal first with the last term of $\eta_{n, T}( \epsilon_{n})$, which we call
\begin{equation*}
\tilde{ \eta}_{n, T}( \epsilon_{n}) = \sup_{i \in A_{n,T}} \sup_{t \in (i\Delta_{n}, (i+1) \Delta_{n}] } |V_{t}'( \epsilon_{n}) - V_{i \Delta_{n}}'( \epsilon_{n})|.
\end{equation*}
Observe that
\begin{equation*}
\sup_{t \in (i \Delta_{n}, (i+1) \Delta_{n}]} |V_{t}'( \epsilon_{n}) - V_{i \Delta_{n}}'( \epsilon_{n})| \leq \gamma_{i,n}^{(1)} + \gamma_{i, n}^{(2)} + \gamma_{i, n}^{(3)},
\end{equation*}
where
\begin{align*}
\gamma_{i,n}^{(1)} &= \sup_{t \in (i \Delta_{n}, (i+1) \Delta_{n}]} |V_{t}^{c} - V_{i \Delta_{n}}^{c}|, \\[0.10cm]
\gamma_{i,n}^{(2)} &= \sup_{t \in (i \Delta_{n}, (i+1) \Delta_{n}]} \big| \int_{i \Delta_{n}}^{t} \int_{ \{z: \tilde{ \Gamma}(z) \leq \epsilon_{n} \}} \bar{ \delta}(s, z)( \mu - \nu) (ds, dz) \big|, \\[0.10cm]
\gamma_{i, n}^{(3)} &= \sup_{t \in (i \Delta_{n}, (i+1) \Delta_{n}]} \big| \int_{i \Delta_{n}}^{t} \int_{ \{z: \tilde{ \Gamma}(z) > \epsilon_{n} \}} \bar{ \delta}(s, z) \nu(ds, dz) \big|.
\end{align*}
Application of the Burkholder and maximal inequality yields that, for any $p \geq 2$,
\begin{equation} \label{equation:gamma-1}
\mathbb{E} \bigg[ \big| \sup_{i \in A_{n,T}} \gamma_{i, n}^{(1)} \big|^{p} \bigg] \leq C T \Delta_{n}^{p/2-1}.
\end{equation}
Moreover, for each $p \geq 2$ and $i$, Lemma 2.1.5 in \citet*{jacod-protter:12a} implies that
\begin{align*}
\mathbb{E} \Big[ \big| \gamma_{i,n}^{(2)} \big|^{p} \Big] &\leq C \Delta_{n} \int_{ \{z: \tilde{ \Gamma}(z) \leq \epsilon_{n} \}} \tilde{ \Gamma}(z)^{p} \lambda(dz) + C \Delta_{n}^{p/2} \bigg( \int_{ \{z: \tilde{ \Gamma}(z) \leq \epsilon_{n} \}} \tilde{ \Gamma}(z)^{2} \lambda (dz) \bigg)^{p/2} \\[0.10cm]
&\leq C \big( \Delta_{n} \epsilon_{n}^{p-2} + \Delta_{n}^{p/2} \big).
\end{align*}
In view of the maximal inequality:
\begin{align} \label{equation:gamma-2}
\mathbb{E} \bigg[ \big| \sup_{i \in A_{n,T}} \gamma_{i,n}^{(2)} \big|^{p} \bigg] \leq C T \big( \epsilon_{n}^{p-2} + \Delta_{n}^{p/2 - 1} \big).
\end{align}
Finally,
\begin{equation} \label{equation:gamma-3}
| \sup_{i \in A_{n,T}} \gamma_{i, n}^{(3)}| \leq \sup_{i \in A_{n,T}} \int_{i \Delta_{n}}^{(i+1) \Delta_{n}} \int_{ \{z: \tilde{ \Gamma}(z) > \epsilon_{n} \}} \tilde{ \Gamma}(z) \nu(ds,dz) \leq C \Delta_{n} \epsilon_{n}^{-1}.
\end{equation}
As a result, for any $p \geq 2$, a combination of \eqref{equation:gamma-1} - \eqref{equation:gamma-3} shows that
\begin{align} \label{equation:tilde-eta}
\mathbb{E} \Big[ \big| \tilde{ \eta}_{n,T}(\epsilon_{n}) \big|^{p} \Big]^{1/p} \leq C \big( T^{1/p} \Delta_{n}^{1/2 - 1/p} + T^{1/p} \epsilon_{n}^{1 - 2/p}+\Delta_{n} \epsilon_{n}^{-1} \big).
\end{align}
Now, turn to the first term of $\eta_{n, T}( \epsilon_{n})$, which is $\hat{ \eta}_{n,T}( \epsilon_{n}) \equiv \sup_{i \in J_{n}( \epsilon_{n})}| \hat{V}_{i \Delta_{n}} - V_{i \Delta_{n}}|$. As in \eqref{equation:decomposition} in the proof of Lemma \ref{lemma:rate-of-convergence}, we disentangle, for each $i \in J_{n}( \epsilon_{n})$,
\begin{equation*}
\check{V}_{i \Delta_{n}} - V_{i \Delta_{n}} = R_{n}^{(1)} + R_{n}^{(2)} + R_{n}^{(3)} + R_{n}^{(4)} + R_{n}^{(5)}.
\end{equation*}
It suffices to look at $R_{n}^{(1)}$, as the rates of the other four terms are equal. As above, utilizing Burkholder's inequality, we see that for each $p \geq 2$, $i \in J_{n}( \epsilon_{n})$ and $0 \leq m \leq h_{n}-1$:
\begin{align*}
\mathbb{E} \Big[ \big| \bar{X}'_{i+m} - \sigma_{i \Delta_{n}} \bar{W}_{i+m} \big|^{p} \Big] &\leq C \bigg( \big(k_{n} \Delta_{n} \big)^{p} + \big(k_{n} \Delta_{n} \big)^{p/2} \mathbb{E} \Big[ \sup_{t \in [i \Delta_{n}, (i+h_{n}) \Delta_{n}]} | \sigma_{t} - \sigma_{i \Delta_{n}}|^{p} \Big] \bigg) \\[0.10cm]
&\leq C \Big( \big(k_{n} \Delta_{n} \big)^{p} + \big(k_{n} \Delta_{n} \big)^{p/2} \big[(h_{n} \Delta_{n})^{p/2} + h_{n} \Delta_{n} \epsilon_{n}^{p-2} + \big(h_{n} \Delta_{n} \big)^{p} \epsilon_{n}^{-p} \big] \Big) \\[0.10cm]
&\leq C \Delta_{n}^{p/4} \Big[ \big(h_{n} \Delta_{n} \big)^{p/2} + h_{n} \Delta_{n} \epsilon_{n}^{p-2} + \big(h_{n} \Delta_{n} \big)^{p} \epsilon_{n}^{-p} \Big],
\end{align*}
uniformly in $i$ and $m$. Proceeding as in the proof of Lemma \ref{lemma:rate-of-convergence}, it follows that
\begin{align*}
\mathbb{E} \Big[ \big| R_{n}^{(1)} \big|^{p} \Big] &\leq \frac{C}{h_{n}} \sum_{m = 0}^{h_{n} - 1} \mathbb{E} \Big[ \big| \Delta_{n}^{-1/4} (\bar{X}'_{i+m} - \sigma_{i \Delta_{n}} \bar{W}_{i+m}) \big|^{p} \big| \Delta_{n}^{-1/4} ( \bar{X}'_{i+m} + \sigma_{i \Delta_{n}} \bar{W}_{i+m}) \big|^{p} \Big]  \\[0.10cm]
& \leq C \Big[ \big(h_{n} \Delta_{n} \big)^{p/2} + h_{n} \Delta_{n} \epsilon_{n}^{p-2} + \big(h_{n} \Delta_{n} \big)^{p} \epsilon_{n}^{-p} \Big].
\end{align*}
In conclusion, for each $p \geq 2$ and $i \in J_{n}(\epsilon_{n})$, we get
\begin{equation*}
\mathbb{E} \Big[ \big| \check{V}_{i \Delta_{n}} - V_{i \Delta_{n}} \big|^{p} \Big]
\leq C \Big[ (k_{n}/h_{n})^{p/2} + (h_{n} \Delta_{n})^{p/2} + h_{n} \Delta_{n} \epsilon_{n}^{p-2} + (h_{n} \Delta_{n})^{p} \epsilon_{n}^{-p} \Big].
\end{equation*}
Hence, for any $p \geq 2$, the maximal inequality, Lemma \ref{lemma:jumps}, along with $k_{n} \asymp \Delta_{n}^{-1/2}$ and $h_{n} \asymp \Delta_{n}^{-3/4}$ yield that
\begin{align*}
\mathbb{E} \bigg[ \big| \sup_{i \in J_{n}( \epsilon)}| \hat{V}_{i \Delta_{n}} - V_{i \Delta_{n}}| \big|^{p} \bigg]^{1/p} \leq C T^{1/p} \Delta_{n}^{-1/p} \Big[ \Delta_{n}^{1/8} + \epsilon_{n}^{1-2/p} &+ \Delta_{n}^{1/4} \epsilon_{n}^{-1} \\[0.10cm]
&+ \Delta_{n}^{((p-1)/4 + 1/2-r \bar{ \omega} -p(1/2-2 \bar{ \omega}))/p} + \Delta_n^{(2-r) \bar{ \omega} -\iota} \Big].
\end{align*}
Merging this with \eqref{equation:tilde-eta}:
\begin{align*}
\mathbb{E} \Big[ \big| \eta_{n, T}( \epsilon_{n}) \big|^{p} \Big]^{1/p} & \leq C T^{1/p} \Delta_{n}^{-1/p} \Big[ \Delta_{n}^{1/8} + \epsilon_{n}^{1-2/p} + \Delta_{n}^{1/4} \epsilon_{n}^{-1} + \Delta_{n}^{((p-1)/4+1/2-r \bar{ \omega}-p(1/2-2 \bar{ \omega}))/p}+ \Delta_{n}^{(2-r) \bar{ \omega} - \iota} \Big] \\[0.10cm]
& \leq C T^{1/p} \Delta_{n}^{-1/p} \Big[ \Delta_{n}^{1/8} + \Delta_{n}^{1/8-1/4p} +\Delta_{n}^{((p-1)/4+1/2-r \bar{ \omega}-p(1/2-2 \bar{ \omega}))/p} + \Delta_{n}^{(2-r) \bar{ \omega} - \iota} \Big],
\end{align*}
where $\epsilon_{n} \asymp \Delta_{n}^{1/8}$ was used in the last line. Now, choosing large enough $p$ implies \eqref{SmallJumps}.

Passing back to the main proof, it holds that
\begin{equation*}
\mathbb{E} \bigg[ T^{-1} \int_{[0, T] \setminus S_{n}( \epsilon_{n})} dt \bigg] \leq C \Delta_{n} \epsilon_{n}^{-2}.
\end{equation*}
This delivers a bound for the remainder term(s):
\begin{align} \label{equation:rate-remainder}
R_{n,T}(x, \epsilon_{n}) + R_{T}(x, \epsilon_{n}) = O_{p} ( \Delta_{n} \epsilon_{n}^{-2}).
\end{align}
As $V_{t}''( \epsilon_{n})$ vanishes on $S_{n}( \epsilon_{n})$, $\sup_{t \in S_{n}( \epsilon_{n})} | \hat{V}_{t} - V_{t}| \leq \eta_{n, T}( \epsilon_{n})$. Then, proceeding as in \eqref{thm3.4-2} we deduce that
\begin{align*}
\sqrt{T}|F_{n, T}(x, \epsilon_{n}) - F_{T}(x, \epsilon_{n})| &\leq \sqrt{T} \big|F_{T}(x + \eta_{n, T}( \epsilon_{n}), \epsilon_{n}) - F_{T}(x - \eta_{n, T}( \epsilon_{n}), \epsilon_{n}) \big| \\[0.10cm]
& \leq \sqrt{T} \big| F_{T}(x + \eta_{n, T}( \epsilon_{n})) - F_{T}(x - \eta_{n, T}( \epsilon_{n})) \big| \\[0.10cm]
&\leq \big| G_{T}(x+ \eta_{n,T}( \epsilon_{n})) - G_{T}(x- \eta_{n,T}( \epsilon_{n})) \big| + \sqrt{T} \big|F(x+ \eta_{n,T}( \epsilon_{n}))-F(x- \eta_{n,T}( \epsilon_{n})) \big|,
\end{align*}
where $G_T(x)=\sqrt{T}(F_T(x) - F(x)).$ Here, the second inequality is due to integrating over a larger set.
Now, arguing as in the end of the proof of Lemma \ref{lemma:rate-of-convergence} and using \eqref{equation:rate-remainder}, we conclude that
\begin{equation*}
\sqrt{T} \sup_{x \in \mathbb{R}_{+} }|F_{n, T}(x) - F_{T}(x)| = o_{p}(1) + O_{p}(T^{1/2- \iota} \Delta_{n}^{1/8- \iota}) + O_{p}(T^{1/2} \Delta_{n} \epsilon_{n}^{-2}).
\end{equation*}
As $\epsilon_{n} \asymp \Delta_{n}^{1/8}$, our work is done. \qed

\subsection*{Proof of Theorem \ref{theorem:weak-convergence}}
Notice that
\begin{equation} \label{thm3.5-1}
\sqrt{T} \big(F_{n,T}(x)-F(x) \big) = \sqrt{T} \big(F_{n,T}(x)-F_{T}(x) \big) + \sqrt{T} \big(F_{T}(x)-F(x) \big).
\end{equation}
Appealing to Lemma \ref{lemma:rate-of-convergence} -- \ref{lemma:rate-of-convergence-jumps}, it holds that $\sup_{x \in \mathbb{R}_{+}} \sqrt{T} \big(F_{n,T}(x) - F_{T}(x) \big) \xrightarrow{ \mathbb{P}} 0$. Furthermore, by Theorem 4.1 in \cite{dehay:05a} it follows that $\sqrt{T} \big(F_{T}(x) - F(x) \big) \xrightarrow{d} N(0, \Sigma(x))$, where
\begin{equation*}
\Sigma(x)=2 \int_{0}^{\infty} \big(F_{t}(x,x)-F(x)^{2} \big) dt
\end{equation*}
and $F_{t}(x,y)=\mathbb{P}\big(\sigma_0^{2}\leq x, V_{t}\leq y\big)$. We next prove the proposed estimator of the asymptotic variance $\Sigma_{n,T}(x)\xrightarrow{\mathbb{P}} \Sigma(x)$, for all $x \in \mathbb{R}_{+}$. Thus, let $\xi \in (0,1/3)$ be fixed and observe that:
\begin{align} \label{SigmaSplit}
\Sigma_{n,T}(x)&=2\int_0^{T^{\xi}}\big(F_{t,n,T}(x)-F_{n,T}(x)^{2}\big)dt \nonumber\\[0.10cm]
&=2\int_0^{T^{\xi}}\big(F_{t,n,T}(x)-F_{t,T}(x)\big)dt+2\int_0^{T^{\xi}}\big(F_{t,T}(x)-F_{T}(x)^{2}\big)dt+2T^{\xi} \big(F_{T}(x)^{2}-F_{n,T}(x)^{2} \big),
\end{align}
where $\displaystyle F_{t,T}(x) = \frac{1}{T}\int_{0}^{T-T^{\xi}}1_{\{ V_{s+t} \leq x, V_{s} \leq x \}}ds$. The last term in the above display is asymptotically negligible by Lemma \ref{lemma:rate-of-convergence} -- \ref{lemma:rate-of-convergence-jumps}:
\begin{align*}
T^{\xi} \big|F_{T}(x)^{2}-F_{n,T}(x)^{2} \big| &=T^{ \xi} \big|F_{T}(x) - F_{n,T}(x) \big| \big|F_{T}(x)+F_{n,T}(x) \big| \\[0.10cm]
&\leq 2T^{\xi}\big|F_{T}(x)-F_{n,T}(x)\big| \xrightarrow{\mathbb{P}} 0.
\end{align*}
Next,
\begin{equation*}
F_{t,T}(x-\eta_{n,T}) \leq F_{t,n,T}(x) \leq F_{t,T}(x+\eta_{n,T}),
\end{equation*}
hence $\big|F_{t,n,T}(x)-F_{t,T}(x) \big| \leq F_{t,T}(x+\eta_{n,T})-F_{t,T}(x-\eta_{n,T})$. By definition of $F_{t,T}$, it follows that for any $z,y \in \mathbb{R}_{+}$ with $z > y$:
\begin{align*}
F_{t,T}(z) - F_{t,T}(y) &= \frac{1}{T} \int_0^{T-T^{\xi}}1_{ \{y< \sigma_{s+t}^{2} \leq z,y<\sigma_{s}^{2}\leq z\}}ds \\[0.10cm]
&\leq \frac{1}{2T} \int_0^{T-T^{\xi}}1_{\{y<\sigma_{s+t}^{2}\leq z\}}+ \frac{1}{2T}\int_0^{T-T^{\xi}}1_{\{y<\sigma_{s}^{2}\leq z\}}ds \\[0.10cm]
&= F_{T}(z)-F_{T}(y) + O \big(T^{\xi-1} \big).
\end{align*}
Taking $z=x+\eta_{n,T}$ and $y=x-\eta_{n,T}$:
\begin{equation*}
\int_0^{T^{\xi}} \big|F_{t,n,T}(x)-F_{t,T}(x) \big|dt = T^{\xi}\big(F_{T}(x+\eta_{n,T})-F_{T}(x-\eta_{n,T})\big)+O \big(T^{2\xi-1} \big).
\end{equation*}
As in the proof of Lemma \ref{lemma:rate-of-convergence} -- \ref{lemma:rate-of-convergence-jumps}, we can show that
\begin{equation*}
\int_0^{T^{\xi}} \big|F_{t,n,T}(x)-F_{t,T}(x) \big|dt \xrightarrow{\mathbb{P}} 0.
\end{equation*}
Finally, for the remaining term in \eqref{SigmaSplit}, we first note that stationarity of $(V_{t})_{t \geq0}$ implies that $\mathbb{E} \big[F_{t,T}(x) \big] = \mathbb{E} \big[F_{t}(x,x) \big] + O\big(T^{\xi-1}\big)$ and $\mathbb{E} \big[F_{T}(x) \big]=F(x)$. Hence,
\begin{align*}
\mathbb{E} \bigg[2 \int_0^{T^\xi} \big(F_{t,T}(x) - F_{T}(x)^{2} \big)dt \bigg]=2\int_0^{T^{\xi}} \big(F_{t}(x,x)-F(x)^{2}\big)dt-2T^{\xi} \text{var} \big(F_{T}(x) \big) + O\big(T^{2\xi-1} \big).
\end{align*}
Since $\alpha(t)=O(t^{-\gamma})$ and $\xi\in(0,1/3)$, Theorem 3.3 in \citet*{dehay:05a} implies that $T^{\xi} \text{var} \big(F_{T}(x) \big)\rightarrow 0$ as $T\rightarrow \infty$. Therefore,
\begin{align*}
\mathbb{E} \bigg[2\int_0^{T^\xi}\big(F_{t,T}(x)-F_{T}(x)^{2}\big)dt\bigg] \rightarrow \Sigma(x),
\end{align*}
as $T \rightarrow \infty$. It thus suffices to show that $2 \int_0^{T^\xi} \big(F_{t,T}(x)-F_{T}(x)^{2} \big)dt-\mathbb{E}\Big[2\int_0^{T^\xi}\big(F_{t,T}(x)-F_{T}(x)^{2}\big)dt\Big]\rightarrow 0$ in $L^{2}$, as $\Delta_{n} \rightarrow 0$ and $T\rightarrow \infty$, in order to conclude that $\Sigma_{n,T}(x) \xrightarrow{ \mathbb{P}} \Sigma(x)$. To this end, we decompose the difference as follows
\begin{align*}
2\int_0^{T^\xi} \big(F_{t,T}(x)-F_{T}(x)^{2} \big)dt - \mathbb{E} \bigg[2 \int_0^{T^\xi} \big(F_{t,T}(x)-F_{T}(x)^{2} \big)dt \bigg]&= 2 \int_0^{T^{\xi}} \big(F_{t,T}(x)-F_{t}(x,x) \big)dt \\[0.10cm]
&+T^{ \xi}\big(F_{T}(x)^{2} - F(x)^{2} \big) \\[0.10cm]
&+2T^{ \xi} \text{var} \big(F_{T}(x) \big) + O \big(T^{2 \xi - 1} \big).
\end{align*}
By Minkowski's inequality, it is enough to look at each term on the right-hand side separately. We define $\tilde{F}_{t,T}(x) = \frac{1}{T} \int_{0}^{T} 1_{ \left\{ V_{s+t} \leq x, V_{s} \leq x \right\}}ds$ and note that $\tilde{F}_{t,T}(x) = F_{t,T}(x) + O\big(T^{\xi-1} \big)$, $\mathbb{E} \big[ \tilde{F}_{t,T}(x) \big] = F_{t}(x,x)$ and
\begin{equation*}
2 \int_0^{T^{ \xi}} \big(F_{t,T}(x) - F_{t}(x,x) \big) dt = 2 \int_{0}^{T^{ \xi}} \big( \tilde{F}_{t,T}(x) - F_{t}(x,x) \big) dt + O \big(T^{2 \xi-1} \big).
\end{equation*}
The Cauchy-Schwarz inequality yields that
\begin{align*}
\mathbb{E} \left[ \bigg( \int_{0}^{T^{ \xi}} \big( \tilde{F}_{t,T}(x) - F_{t}(x,x) \big) dt \bigg)^{2} \right] &\leq T^{ \xi} \int_{0}^{T^{ \xi}} \text{var} \big( \tilde{F}_{t,T}(x) \big)dt \\[0.10cm]
&= O \big(T^{2 \xi-1} \big),
\end{align*}
where the last inequality follows from the calculations on p. 941 in \citet*{dehay:05a}. Therefore,
\begin{equation*}
2 \int_{0}^{T^{ \xi}} \big(F_{t,T}(x) - F_{t}(x,x) \big)dt \rightarrow 0
\end{equation*}
in $L^{2}$. As both $F_{T}$ and $F$ are bounded by 1, the bounded convergence theorem and Theorem 3.3 in \citet*{dehay:05a} imply that
\begin{align}
T^{ \xi} \left(F_{T}(x)^{2} - F(x)^{2} \right) \rightarrow 0
\end{align}
in $L^{2}$, as $T \rightarrow \infty$. Thus,
\begin{align*}
2 \int_{0}^{T^{ \xi}} \big(F_{t,T}(x) - F_{T}(x)^{2} \big)dt - \mathbb{E} \bigg[2 \int_{0}^{T^{ \xi}} \big(F_{t,T}(x) - F_{T}(x)^{2} \big)dt \bigg] \rightarrow 0
\end{align*}
in $L^{2}$, given the rate condition from Theorem 3.3. This implies $\Sigma_{n,T}(x) \xrightarrow{ \mathbb{P}} \Sigma(x)$. \qed

\bigskip

\noindent We now state a multivariate version of Theorem 3.3.
\begin{lemma} \label{lemma:multivariate}
Suppose that the conditions of Theorem 3.3 are fulfilled. Let $p \in \mathbb{N}$ be given. Then, for any $x_1, \ldots, x_p \in\mathbb{R}_{+}$,
\begin{align*}
\big(G_{n,T}(x_1),\ldots,G_{n,T}(x_p)\big) \xrightarrow{d}N\big(0,\Sigma(x_1,\ldots,x_p)\big),
\end{align*}
where $\Sigma(x_1,\ldots,x_p)$ is the $p\times p$ covariance matrix with elements of the form
\begin{align*}
\Sigma(x_1,\ldots,x_p)_{ij} = \int_0^{ \infty}\big(F_{t}(x_i,x_j)+F_{t}(x_i,x_j)-2F(x_i)F(x_j)\big)dt,
\end{align*}
for $i,j = 1, \ldots, p$. Furthermore, let $\xi\in(0,1/3)$ be given. Then,
\begin{align*}
\Sigma_{n,T}(x_i,x_j) = \int_0^{T^{ \xi}} \big(F_{t,n,T}(x_i,x_j) + F_{t,n,T}(x_j,x_i) - 2F_{n,T}(x_i)F_{n,T}(x_j)\big)dt \xrightarrow{ \mathbb{P}} \Sigma(x_{1}, \ldots, x_{p})_{ij}.
\end{align*}
\end{lemma}
\textbf{Proof.} The first part follows by a Cr\'{a}mer-Rao device. Also, proving consistency of the covariance estimator goes along the lines of the univariate case. We therefore skip it. \qed

\subsection*{Proof of Theorem \ref{theorem:functional-clt}}
In view of Lemma \ref{lemma:rate-of-convergence}-\ref{lemma:rate-of-convergence-jumps}, we obtain that $\sup_{x \in \mathbb{R}_{+}} \sqrt{T} \big(F_{n,T}(x) - F_{T}(x) \big) \xrightarrow{ \mathbb{P}} 0$. On the other hand, Theorem 5.2 in \citet*{dehay:05a} implies a functional central limit theorem for the process
$G_{T} =\big \{ \sqrt{T} \big(F_{T}(x)-F(x) \big), x \in \mathbb{R}_{+} \big \}$. \qed

\subsection*{Proof of Corollary \ref{corollary:weak-convergence}}

This follows from Theorem \ref{theorem:functional-clt} by noting that the sup and weighted $L^{2}$ norm are both continuous mappings from $\mathbb{D}(\mathbb{R}_+) \to \mathbb{R}$. \qed

\subsection*{Proof of Proposition \ref{proposition:distribution-free}}

From Corollary \ref{corollary:weak-convergence} the limiting distribution of $rL^{2}$ under $\mathcal{H}_{0}$ is $||G_{F}||^{2}_{w}$, where $G_{F}$ is a Gaussian process with covariance operator $\Sigma$. The distribution can therefore be written as $||G_{F}||_{w}^{2} = \sum_{i=1}^{ \infty} \lambda_{i} \Phi_{i}^{2}$, where $\Phi_{i}$ are independent standard normal random variables, and $\lambda_i$ are the eigenvalues of $\Sigma$ \citep*[see, e.g.,][]{kuo:75a}. Moreover, from \citet*[][chapter 1]{bosq:00a} we deduce that $\tilde{ \Sigma} = \sum_{i=1}^{ \infty} \lambda_{i} < \infty$, so that $\mathbb{E} \big[||G_{F}||_{w}^{2} \big] = \tilde{ \Sigma}$. Combining this result with the fact that $T(F)$ is a quadratic form of mean zero Gaussian random variables, it follows that for all $\alpha \in (0,0.215)$:
\begin{equation*}
\mathbb{P} \big(T(F) > \alpha \big) = 1 - \mathbb{P} \big( T(F) \leq \alpha \big) \leq 1 - \mathbb{P} \big( \chi^{2} \leq \alpha \big) = \mathbb{P} \big( \chi^{2} > \alpha \big),
\end{equation*}
where the inequality is due to \cite{szekely-bakirov:03a}. This concludes the proof of the first statement, while the second part follows upon observing that $||G_{F}||_{w}^{2} \xrightarrow{ \mathbb{P}} \infty$ under $\mathcal{H}_{a}$. \qed

\subsection*{Proof of Corollary \ref{corollary:function-clt}}

The proof follows from Theorem \ref{theorem:functional-clt} above and Theorem 19.23 in \citet*{vaart:98a}. \qed

\section{Simulation of critical values for goodness-of-fit testing} \label{appendix:critical-value}

To describe how critical values for our goodness-of-fit tests are generated, we again fix the sampling frequency at $\Delta_{n}$ and the time horizon at $[0,T]$.

The stochastic volatility model is assumed to be indexed by a parameter vector $\upsilon$. We discriminate between the case where $\upsilon$ is known (I), and the more realistic setup where $\upsilon$ is unknown, but a preliminary estimate $\hat{ \upsilon}$ is available along with an estimate of the noise variance $\hat{ \omega}$ (II). As a biproduct, we emphasize that the latter approach also accounts for the estimation error induced by the infill step, which is negligible in the limit but can be sizable in small samples.

We proceed as follows:

\noindent I. Without parameter estimation
\begin{enumerate}
\item Simulate a volatility path of length $T$ based on $\upsilon$.
\item Sample the path at frequency $\Delta_{n}$.
\item Construct the EDF, $F_{T}$.
\item Calculate the $t$-statistic with $F$ based on $\upsilon$.
\item Repeat step 1 -- 4 $B$ times.
\end{enumerate}

\noindent II. With parameter estimation
\begin{enumerate}
\item Simulate a volatility path of length $T$ based on $\hat{ \upsilon}$. Call it $( \sigma_{t})_{t \in [0,T]}$.
\item Construct noisy log-returns at sampling frequency $\Delta_{n}$ according to: $\Delta_{i}^{n} Z = \sigma_{(i-1) \Delta_{n}} \sqrt{ \Delta_{n}} \Phi_{1} + \hat{ \omega} \Phi_{2}$, where $\Phi_{1}$ and $\Phi_{2}$ are independent standard normal random variables.
\item Construct the REDF, $F_{n,T}$.
\item Retrieve a new estimate of the parameter vector, denoted by $\tilde{ \upsilon}$, from the artificial data.
\item Calculate the $t$-statistic with $F$ based on $\tilde{ \upsilon}$.
\item Repeat step 1 -- 5 $B$ times.
\end{enumerate}

In both I. and II., the critical values are subsequently found as the $\alpha$-quantile of the distribution of $t$-statistics based on the above $B$ repetitions of the simulation experiment.

\pagebreak


\renewcommand{\baselinestretch}{1.0}
\small
\bibliography{userref}

@ARTICLE{ait-sahalia:96a,
 AUTHOR = {Y. A\"{i}t-Sahalia},
 YEAR = {1996},
 TITLE = {{Testing continuous-time models of the spot interest rate}},
 JOURNAL = {Review of Financial Studies},
 VOLUME = {9},
 NUMBER = {2},
 PAGES = {385--426}
}

@BOOK{ait-sahalia-jacod:14a,
 AUTHOR = {Y. A\"{i}t-Sahalia and J. Jacod},
 YEAR = {2014},
 TITLE = {{High-frequency Financial Econometrics}},
 EDITION = {1st},
 PUBLISHER = {Princeton University Press},
 ADDRESS = {New Jersey}
}

@ARTICLE{ait-sahalia-jacod-li:12a,
 AUTHOR = {Y. A\"{i}t-Sahalia and J. Jacod and J. Li},
 YEAR = {2012},
 TITLE = {{Testing for jumps in noisy high frequency data}},
 JOURNAL = {Journal of Econometrics},
 VOLUME = {168},
 NUMBER = {2},
 PAGES = {207--222}
}

@ARTICLE{ait-sahalia-kimmel:07a,
 AUTHOR = {Y. A\"{i}t-Sahalia and R. Kimmel},
 YEAR = {2007},
 TITLE = {{Maximum likelihood estimation of stochastic volatility models}},
 JOURNAL = {Journal of Financial Economics},
 VOLUME = {83},
 NUMBER = {2},
 PAGES = {413--452}
}

@ARTICLE{ait-sahalia-xiu:16a,
 AUTHOR = {Y. A\"{i}t-Sahalia and D. Xiu},
 YEAR = {2016},
 TITLE = {{Increased correlation among asset classes: Are volatility or jumps to blame, or both?}},
 JOURNAL = {Journal of Econometrics},
 VOLUME = {194},
 NUMBER = {2},
 PAGES = {205--219}
}

@ARTICLE{andersen-benzoni-lund:02a,
 AUTHOR = {T. G. Andersen and L. Benzoni and J. Lund},
 YEAR = {2002},
 TITLE = {{An empirical investigation of continuous-time equity return models}},
 JOURNAL = {Journal of Finance},
 VOLUME = {57},
 NUMBER = {4},
 PAGES = {1239--1284}
}

@ARTICLE{andersen-sorensen:96a,
 AUTHOR = {T. G. Andersen and B. E. S{\o}rensen},
 YEAR = {1996},
 TITLE = {{GMM estimation of a stochastic volatility model: A Monte Carlo study}},
 JOURNAL = {Journal of Business and Economic Statistics},
 VOLUME = {14},
 NUMBER = {3},
 PAGES = {328--352}
}

@ARTICLE{andersen-thyrsgaard-todorov:19a,
 AUTHOR = {T. G. Andersen and M. Thyrsgaard and V. Todorov},
 YEAR = {2019},
 TITLE = {{Time-varying periodicity in intraday volatility}},
 JOURNAL = {Journal of the American Statistical Association},
 VOLUME = {114},
 NUMBER = {528},
 PAGES = {1695--1707}
}

@ARTICLE{baeumer-meerschaert:10a,
 AUTHOR = {B. Baeumer and M. M. Meerschaert},
 YEAR = {2010},
 TITLE = {{Tempered stable L\'{e}vy motion and transient super-diffusion}},
 JOURNAL = {Journal of Computational and Applied Mathematics},
 VOLUME = {233},
 NUMBER = {10},
 PAGES = {2438--2448}
}

@ARTICLE{bandi-russell:06a,
 AUTHOR = {F. M. Bandi and J. R. Russell},
 YEAR = {2006},
 TITLE = {{Separating microstructure noise from volatility}},
 JOURNAL = {Journal of Financial Economics},
 VOLUME = {79},
 NUMBER = {3},
 PAGES = {655--692}
}

@INCOLLECTION{barndorff-nielsen-graversen-jacod-podolskij-shephard:06a,
 AUTHOR = {O. E. Barndorff-Nielsen and S. E. Graversen and J. Jacod and M. Podolskij and N. Shephard},
 YEAR = {2006},
 TITLE = {{A central limit theorem for realized power and bipower variations of continuous semimartingales}},
 BOOKTITLE = {From Stochastic Calculus to Mathematical Finance: The Shiryaev Festschrift},
 EDITOR = {Y. Kabanov and R. Lipster and J. Stoyanov},
 PAGES = {33--68},
 PUBLISHER = {Springer},
 ADDRESS = {Heidelberg}
}

@ARTICLE{barndorff-nielsen-hansen-lunde-shephard:08a,
 AUTHOR = {O. E. Barndorff-Nielsen and P. R. Hansen and A. Lunde and N. Shephard},
 YEAR = {2008},
 TITLE = {{Designing realized kernels to measure the ex post variation of equity prices in the presence of noise}},
 JOURNAL = {Econometrica},
 VOLUME = {76},
 NUMBER = {6},
 PAGES = {1481--1536}
}

@ARTICLE{barndorff-nielsen-shephard:01a,
 AUTHOR = {O. E. Barndorff-Nielsen and N. Shephard},
 YEAR = {2001},
 TITLE = {{Non-Gaussian Orstein-Uhlenbeck-based models and some of their uses in financial economics}},
 JOURNAL = {Journal of the Royal Statistical Society: Series B},
 VOLUME = {63},
 NUMBER = {2},
 PAGES = {167--241}
}

@ARTICLE{barndorff-nielsen-shephard:02a,
 AUTHOR = {O. E. Barndorff-Nielsen and N. Shephard},
 YEAR = {2002},
 TITLE = {{Econometric analysis of realized volatility and its use in estimating stochastic volatility models}},
 JOURNAL = {Journal of the Royal Statistical Society: Series B},
 VOLUME = {64},
 NUMBER = {2},
 PAGES = {253--280}
}

@ARTICLE{bibby-skovgaard-sorensen:05a,
 AUTHOR = {B. M. Bibby and I. M. Skovgaard and M. S{\o}rensen},
 YEAR = {2005},
 TITLE = {{Diffusion-type models with given marginal distribution and autocorrelation function}},
 JOURNAL = {Bernoulli},
 VOLUME = {11},
 NUMBER = {2},
 PAGES = {191--220}
}

@ARTICLE{bibinger-winkelmann:18a,
 AUTHOR = {M. Bibinger and L. Winkelmann},
 YEAR = {2018},
 TITLE = {{Common price and volatility jumps in noisy high-frequency data}},
 JOURNAL = {Electronic Journal of Statistics},
 VOLUME = {12},
 NUMBER = {1},
 PAGES = {2018--2073}
}

@BOOK{bickel-klaassen-ritov-wellner:98a,
 AUTHOR = {P. J. Bickel and C. A. J. Klaassen and Y. Ritov and J. A. Wellner},
 YEAR = {1998},
 TITLE = {{Efficient and Adaptive Estimation for Semiparametric Models}},
 EDITION = {1st},
 PUBLISHER = {Springer},
 ADDRESS = {Berlin}
}

@ARTICLE{bollerslev-zhou:02a,
 AUTHOR = {T. Bollerslev and H. Zhou},
 YEAR = {2002},
 TITLE = {{Estimating stochastic volatility diffusion using conditional moments of integrated volatility}},
 JOURNAL = {Journal of Econometrics},
 VOLUME = {109},
 NUMBER = {1},
 PAGES = {33--65}
}

@BOOK{bosq:00a,
 AUTHOR = {D. Bosq},
 YEAR = {2000},
 TITLE = {{Linear Processes in Function Spaces: Theory and Applications}},
 EDITION = {1st},
 PUBLISHER = {Springer},
 ADDRESS = {Berlin}
}

@ARTICLE{bull:17a,
 AUTHOR = {A. D. Bull},
 YEAR = {2017},
 TITLE = {{Semimartingale detection and goodness-of-fit tests}},
 JOURNAL = {Annals of Statistics},
 VOLUME = {45},
 NUMBER = {3},
 PAGES = {1254--1283}
}

@ARTICLE{chernov-gallant-ghysels-tauchen:03a,
 AUTHOR = {M. Chernov and A. R. Gallant and E. Ghysels and G. Tauchen},
 YEAR = {2003},
 TITLE = {{Alternative models for stock price dynamics}},
 JOURNAL = {Journal of Econometrics},
 VOLUME = {116},
 NUMBER = {1--2},
 PAGES = {225--257}
}

@ARTICLE{christensen-oomen-podolskij:14a,
 AUTHOR = {K. Christensen and R. C. A. Oomen and M. Podolskij},
 YEAR = {2014},
 TITLE = {{Fact or friction: Jumps at ultra high frequency}},
 JOURNAL = {Journal of Financial Economics},
 VOLUME = {114},
 NUMBER = {3},
 PAGES = {576--599}
}

@ARTICLE{christoffersen-jacobs-mimouni:10a,
 AUTHOR = {P. F. Christoffersen and K. Jacobs and K. Mimouni},
 YEAR = {2010},
 TITLE = {{Volatility dynamics for the S\&P500: Evidence from realized volatility, daily returns, and option prices}},
 JOURNAL = {Review of Financial Studies},
 VOLUME = {23},
 NUMBER = {8},
 PAGES = {3141--3189}
}

@ARTICLE{comte-renault:98a,
 AUTHOR = {F. Comte and E. Renault},
 YEAR = {1998},
 TITLE = {{Long memory in continuous-time stochastic volatility models}},
 JOURNAL = {Mathematical Finance},
 VOLUME = {8},
 NUMBER = {4},
 PAGES = {291--323}
}

@ARTICLE{corradi-distaso:06a,
 AUTHOR = {V. Corradi and W. Distaso},
 YEAR = {2006},
 TITLE = {{Semi-parametric comparison of stochastic volatility models using realized measures}},
 JOURNAL = {Review of Economic Studies},
 VOLUME = {73},
 NUMBER = {3},
 PAGES = {635--667}
}

@ARTICLE{cox-ingersoll-ross:85a,
 AUTHOR = {J. C. Cox and J. E. Ingersoll and S. A. Ross},
 YEAR = {1985},
 TITLE = {{A theory of the term structure of interest rates}},
 JOURNAL = {Econometrica},
 VOLUME = {53},
 NUMBER = {2},
 PAGES = {385--407}
}

@ARTICLE{david-johnson:48a,
 AUTHOR = {F. N. David and N. L. Johnson},
 YEAR = {1948},
 TITLE = {{The probability integral transformation when parameters are estimated from the sample}},
 JOURNAL = {Biometrika},
 VOLUME = {35},
 NUMBER = {1--2},
 PAGES = {182-190}
}

@ARTICLE{dehay:05a,
 AUTHOR = {D. Dehay},
 YEAR = {2005},
 TITLE = {{On invariant distribution function estimation for continuous-time stationary processes}},
 JOURNAL = {Bernoulli},
 VOLUME = {11},
 NUMBER = {5},
 PAGES = {933--948}
}

@ARTICLE{delbaen-schachermayer:94a,
 AUTHOR = {F. Delbaen and W. Schachermayer},
 YEAR = {1994},
 TITLE = {{A general version of the fundamental theorem of asset pricing}},
 JOURNAL = {Mathematische Annalen},
 VOLUME = {300},
 NUMBER = {1},
 PAGES = {463--520}
}

@ARTICLE{dette-podolskij:08a,
 AUTHOR = {H. Dette and M. Podolskij},
 YEAR = {2008},
 TITLE = {{Testing the parametric form of the volatility in continuous time diffusion models--A stochastic process approach}},
 JOURNAL = {Journal of Econometrics},
 VOLUME = {143},
 NUMBER = {1},
 PAGES = {56--73}
}

@ARTICLE{dette-podolskij-vetter:06a,
 AUTHOR = {H. Dette and M. Podolskij and M. Vetter},
 YEAR = {2006},
 TITLE = {{Estimation of integrated volatility in continuous-time financial models with applications to goodness-of-fit testing}},
 JOURNAL = {Scandinavian Journal of Statistics},
 VOLUME = {33},
 NUMBER = {2},
 PAGES = {259--278}
}

@ARTICLE{diebold-strasser:13a,
 AUTHOR = {F. X. Diebold and G. H. Strasser},
 YEAR = {2013},
 TITLE = {{On the correlation structure of microstructure noise: A financial economic approach}},
 JOURNAL = {Review of Economic Studies},
 VOLUME = {80},
 NUMBER = {4},
 PAGES = {1304--1337}
}

@ARTICLE{foster-nelson:96a,
 AUTHOR = {D. P. Foster and D. B. Nelson},
 YEAR = {1996},
 TITLE = {{Continuous record asymptotics for rolling sample variance estimators}},
 JOURNAL = {Econometrica},
 VOLUME = {64},
 NUMBER = {1},
 PAGES = {139--174}
}

@ARTICLE{gallant-hsieh-tauchen:97a,
 AUTHOR = {A. R. Gallant and D. A. Hsieh and G. E. Tauchen},
 YEAR = {1997},
 TITLE = {{Estimation of stochastic volatility with diagnostics}},
 JOURNAL = {Journal of Econometrics},
 VOLUME = {81},
 NUMBER = {1},
 PAGES = {159--192}
}

@ARTICLE{gatheral-jaisson-rosenbaum:18a,
 AUTHOR = {J. Gatheral and T. Jaisson and M. Rosenbaum},
 YEAR = {2018},
 TITLE = {{Volatility is rough}},
 JOURNAL = {Quantitative Finance},
 VOLUME = {18},
 NUMBER = {6},
 PAGES = {933--949}
}

@ARTICLE{gatheral-oomen:10a,
 AUTHOR = {J. Gatheral and R. C. A. Oomen},
 YEAR = {2010},
 TITLE = {{Zero-intelligence realized variance estimation}},
 JOURNAL = {Finance and Stochastics},
 VOLUME = {14},
 NUMBER = {2},
 PAGES = {249--283}
}

@ARTICLE{geman-horowitz:80a,
 AUTHOR = {D. Geman and J. Horowitz},
 YEAR = {1980},
 TITLE = {{Occupation densities}},
 JOURNAL = {Annals of Probability},
 VOLUME = {8},
 NUMBER = {1},
 PAGES = {1--67}
}

@ARTICLE{hansen-lunde:06b,
 AUTHOR = {P. R. Hansen and A. Lunde},
 YEAR = {2006},
 TITLE = {{Realized variance and market microstructure noise}},
 JOURNAL = {Journal of Business and Economic Statistics},
 VOLUME = {24},
 NUMBER = {2},
 PAGES = {127--161}
}

@ARTICLE{heston:93a,
 AUTHOR = {S. L. Heston},
 YEAR = {1993},
 TITLE = {{A closed-form solution for options with stochastic volatility with applications to bond and currency options}},
 JOURNAL = {Review of Financial Studies},
 VOLUME = {6},
 NUMBER = {2},
 PAGES = {327--343}
}

@ARTICLE{hull-white:87a,
 AUTHOR = {J. Hull and A. White},
 YEAR = {1987},
 TITLE = {{The pricing of options on assets with stochastic volatilities}},
 JOURNAL = {Journal of Finance},
 VOLUME = {42},
 NUMBER = {2},
 PAGES = {281--300}
}

@ARTICLE{jacod-li-mykland-podolskij-vetter:09a,
 AUTHOR = {J. Jacod and Y. Li and P. A. Mykland and M. Podolskij and M. Vetter},
 YEAR = {2009},
 TITLE = {{Microstructure noise in the continuous case: The pre-averaging approach}},
 JOURNAL = {Stochastic Processes and their Applications},
 VOLUME = {119},
 NUMBER = {7},
 PAGES = {2249--2276}
}

@BOOK{jacod-protter:12a,
 AUTHOR = {J. Jacod and P. E. Protter},
 YEAR = {2012},
 TITLE = {{Discretization of Processes}},
 EDITION = {2nd},
 PUBLISHER = {Springer},
 ADDRESS = {Berlin}
}

@ARTICLE{jacod-rosenbaum:13a,
 AUTHOR = {J. Jacod and M. Rosenbaum},
 YEAR = {2013},
 TITLE = {{Quarticity and other functionals of volatility: Efficient estimation}},
 JOURNAL = {Annals of Statistics},
 VOLUME = {41},
 NUMBER = {3},
 PAGES = {1462--1484}
}

@ARTICLE{jongbloed-meulen-vaart:05a,
 AUTHOR = {G. Jongbloed and F. H. van der Meulen and A. W. van der Vaart},
 YEAR = {2005},
 TITLE = {{Nonparametric inference for L\'{e}vy-driven Ornstein-Uhlenbeck processes}},
 JOURNAL = {Bernoulli},
 VOLUME = {11},
 NUMBER = {5},
 PAGES = {759--791}
}

@BOOK{kuo:75a,
 AUTHOR = {H.-H. Kuo},
 YEAR = {1975},
 TITLE = {{Gaussian Measures in Banach Spaces}},
 EDITION = {1st},
 PUBLISHER = {Springer},
 ADDRESS = {Berlin}
}

@ARTICLE{li-todorov-tauchen:13a,
 AUTHOR = {J. Li and V. Todorov and G. Tauchen},
 YEAR = {2013},
 TITLE = {{Volatility occupation times}},
 JOURNAL = {Annals of Statistics},
 VOLUME = {41},
 NUMBER = {4},
 PAGES = {1865--1891}
}

@ARTICLE{li-todorov-tauchen:16a,
 AUTHOR = {J. Li and V. Todorov and G. Tauchen},
 YEAR = {2016},
 TITLE = {{Estimating the volatility occupation time via regularized Laplace inversion}},
 JOURNAL = {Econometric Theory},
 VOLUME = {32},
 NUMBER = {5},
 PAGES = {1253--1288}
}

@ARTICLE{li-todorov-tauchen:17a,
 AUTHOR = {J. Li and V. Todorov and G. Tauchen},
 YEAR = {2017},
 TITLE = {{Adaptive estimation of continuous-time regression models using high-frequency data}},
 JOURNAL = {Journal of Econometrics},
 VOLUME = {200},
 NUMBER = {1},
 PAGES = {36--47}
}

@ARTICLE{lilliefors:67a,
 AUTHOR = {H. W. Lilliefors},
 YEAR = {1967},
 TITLE = {{On the Kolmogorov-Smirnov test for normality with mean and variance unknown}},
 JOURNAL = {Journal of the American Statistical Association},
 VOLUME = {62},
 NUMBER = {318},
 PAGES = {399--402}
}

@ARTICLE{lilliefors:69a,
 AUTHOR = {H. W. Lilliefors},
 YEAR = {1969},
 TITLE = {{On the Kolmogorov-Smirnov test for the exponential distribution with mean unknown}},
 JOURNAL = {Journal of the American Statistical Association},
 VOLUME = {64},
 NUMBER = {325},
 PAGES = {387--389}
}

@ARTICLE{lin-lee-guo:13a,
 AUTHOR = {L.-C. Lin and S. Lee and M. Guo},
 YEAR = {2013},
 TITLE = {{Goodness-of-fit test for stochastic volatility models}},
 JOURNAL = {Journal of Multivariate Analysis},
 VOLUME = {116},
 NUMBER = {1},
 PAGES = {473--498}
}

@ARTICLE{lin-lee-guo:16a,
 AUTHOR = {L.-C. Lin and S. Lee and M. Guo},
 YEAR = {2016},
 TITLE = {{Goodness-of-fit test for the SVM based on noisy observations}},
 JOURNAL = {Statistica Sinica},
 VOLUME = {26},
 NUMBER = {3},
 PAGES = {1305--1329}
}

@ARTICLE{magdziarz-weron:11a,
 AUTHOR = {M. Magdziarz and A. Weron},
 YEAR = {2011},
 TITLE = {{Ergodic properties of anomalous diffusion processes}},
 JOURNAL = {Annals of Physics},
 VOLUME = {326},
 NUMBER = {9},
 PAGES = {2431--2443}
}

@ARTICLE{meddahi:03a,
 AUTHOR = {N. Meddahi},
 YEAR = {2003},
 TITLE = {{ARMA representation of integrated and realized variances}},
 JOURNAL = {Econometrics Journal},
 VOLUME = {6},
 NUMBER = {2},
 PAGES = {335--356}
}

@ARTICLE{merton:80a,
 AUTHOR = {R. C. Merton},
 YEAR = {1980},
 TITLE = {{On estimating the expected return on the market: An exploratory investigation}},
 JOURNAL = {Journal of Financial Economics},
 VOLUME = {8},
 NUMBER = {4},
 PAGES = {323--361}
}

@ARTICLE{newey-west:94a,
 AUTHOR = {W. K. Newey and K. D. West},
 YEAR = {1994},
 TITLE = {{Automatic lag selection in covariance matrix estimation}},
 JOURNAL = {Review of Economic Studies},
 VOLUME = {61},
 NUMBER = {4},
 PAGES = {631--653}
}

@ARTICLE{oomen:06a,
 AUTHOR = {R. C. A. Oomen},
 YEAR = {2006},
 TITLE = {{Comment on 2005 JBES invited address ``Realized variance and market microstructure noise'' by Peter R. Hansen and Asger Lunde}},
 JOURNAL = {Journal of Business and Economic Statistics},
 VOLUME = {24},
 NUMBER = {2},
 PAGES = {195--202}
}

@ARTICLE{podolskij-vetter:09a,
 AUTHOR = {M. Podolskij and M. Vetter},
 YEAR = {2009},
 TITLE = {{Bipower-type estimation in a noisy diffusion setting}},
 JOURNAL = {Stochastic Processes and their Applications},
 VOLUME = {119},
 NUMBER = {9},
 PAGES = {2803--2831}
}

@ARTICLE{podolskij-vetter:09b,
 AUTHOR = {M. Podolskij and M. Vetter},
 YEAR = {2009},
 TITLE = {{Estimation of volatility functionals in the simultaneous presence of microstructure noise and jumps}},
 JOURNAL = {Bernoulli},
 VOLUME = {15},
 NUMBER = {3},
 PAGES = {634--658}
}

@BOOK{resnick:98a,
 AUTHOR = {S. I. Resnick},
 YEAR = {1998},
 TITLE = {{A probability path}},
 EDITION = {1st},
 PUBLISHER = {Birkh\"{a}user},
 ADDRESS = {Basel}
}

@ARTICLE{rosinski:07a,
 AUTHOR = {J. Rosi\'{n}ski},
 YEAR = {2007},
 TITLE = {{Tempering stable processes}},
 JOURNAL = {Stochastic Processes and their Applications},
 VOLUME = {117},
 NUMBER = {6},
 PAGES = {677--707}
}

@ARTICLE{szekely-bakirov:03a,
 AUTHOR = {G. J. Sz\'{e}kely and N. K. Bakirov},
 YEAR = {2003},
 TITLE = {{Extremal probabilities for Gaussian quadratic forms}},
 JOURNAL = {Probability Theory and Related Fields},
 VOLUME = {126},
 NUMBER = {2},
 PAGES = {184--202}
}

@ARTICLE{todorov:09a,
 AUTHOR = {V. Todorov},
 YEAR = {2009},
 TITLE = {{Estimation of continuous-time stochastic volatility models with jumps using high-frequency data}},
 JOURNAL = {Journal of Econometrics},
 VOLUME = {148},
 NUMBER = {2},
 PAGES = {131--148}
}

@ARTICLE{todorov-tauchen:12a,
 AUTHOR = {V. Todorov and G. Tauchen},
 YEAR = {2012},
 TITLE = {{The realized Laplace transform of volatility}},
 JOURNAL = {Econometrica},
 VOLUME = {80},
 NUMBER = {3},
 PAGES = {1105--1127}
}

@ARTICLE{todorov-tauchen-grynkiv:11a,
 AUTHOR = {V. Todorov and G. Tauchen and I. Grynkiv},
 YEAR = {2011},
 TITLE = {{Realized Laplace transforms for estimation of jump diffusive volatility models}},
 JOURNAL = {Journal of Econometrics},
 VOLUME = {164},
 NUMBER = {2},
 PAGES = {367--381}
}

@ARTICLE{todorov-tauchen-grynkiv:14a,
 AUTHOR = {V. Todorov and G. Tauchen and I. Grynkiv},
 YEAR = {2014},
 TITLE = {{Volatility activity: Specification and estimation}},
 JOURNAL = {Journal of Econometrics},
 VOLUME = {178},
 NUMBER = {1},
 PAGES = {180--193}
}

@BOOK{vaart:98a,
 AUTHOR = {A. W. van der Vaart},
 YEAR = {1998},
 TITLE = {{Asymptotic Statistics}},
 EDITION = {1st},
 PUBLISHER = {Cambridge University Press},
 ADDRESS = {Cambridge}
}

@ARTICLE{vaart-wellner:07a,
 AUTHOR = {A. W. van der Vaart and J. A. Wellner},
 YEAR = {2007},
 TITLE = {{Empirical processes indexed by estimated functions}},
 JOURNAL = {Lecture Notes-Monograph Series},
 VOLUME = {55},
 NUMBER = {},
 PAGES = {234--252}
}

@ARTICLE{veretennikov:88a,
 AUTHOR = {Y. A. Veretennikov},
 YEAR = {1988},
 TITLE = {{Bounds for the mixing rate in the theory of stochastic equations}},
 JOURNAL = {Theory of Probability and Its Applications},
 VOLUME = {32},
 NUMBER = {2},
 PAGES = {273--281}
}

@ARTICLE{vetter-dette:12a,
 AUTHOR = {M. Vetter and H. Dette},
 YEAR = {2012},
 TITLE = {{Model checks for the volatility under microstructure noise}},
 JOURNAL = {Bernoulli},
 VOLUME = {18},
 NUMBER = {4},
 PAGES = {1421--1447}
}

@ARTICLE{zhang-mykland-ait-sahalia:05a,
 AUTHOR = {L. Zhang and P. A. Mykland and Y. A\"{i}t-Sahalia},
 YEAR = {2005},
 TITLE = {{A tale of two time scales: determining integrated volatility with noisy high-frequency data}},
 JOURNAL = {Journal of the American Statistical Association},
 VOLUME = {100},
 NUMBER = {472},
 PAGES = {1394--1411}
}

@ARTICLE{zu:15a,
 AUTHOR = {Y. Zu},
 YEAR = {2015},
 TITLE = {{Nonparametric specification tests for stochastic volatility models based on volatility density}},
 JOURNAL = {Journal of Econometrics},
 VOLUME = {187},
 NUMBER = {1},
 PAGES = {323--344}
}

@ARTICLE{zu-boswijk:14a,
 AUTHOR = {Y. Zu and H. P. Boswijk},
 YEAR = {2014},
 TITLE = {{Estimating spot volatility with high-frequency financial data}},
 JOURNAL = {Journal of Econometrics},
 VOLUME = {181},
 NUMBER = {2},
 PAGES = {117--135}
}
\end{document}